\documentclass[traditabstract]{aa}

\usepackage{graphicx}
\usepackage{txfonts}

\begin{document}

\title{The long-term millimeter activity of active galactic nuclei\thanks{This study is based on observations carried out with the IRAM Plateau de Bure Interferometer. IRAM is supported by INSU/CNRS (France), MPG (Germany), and IGN (Spain).}}

\author{S. Trippe\inst{1} \and M. Krips\inst{2} \and V. Pi\'etu\inst{2} \and R. Neri\inst{2} \and J.M. Winters\inst{2} \and F. Gueth\inst{2}  \and M. Bremer\inst{2} \and P. Salome\inst{3} \and \\ R. Moreno\inst{4} \and J. Boissier\inst{5,6} \and F. Fontani\inst{6,7}}

\institute{
Seoul National University, Department of Physics and Astronomy, 599 Gwanak-ro, Gwanak-gu, Seoul 151-742, South Korea \\
e-mail: {\tt trippe@astro.snu.ac.kr} \and
Institut de Radioastronomie Millim\'etrique (IRAM), 300 rue de la Piscine, F-38406 Saint Martin d'H\`eres, France \and
LERMA, Observatoire de Paris, UMR 8112 du CNRS, 75014 Paris, France \and
Observatoire de Paris, LESIA, 92195 Meudon, France \and
Istituto di Radioastronomia -- INAF, Via Gobetti 101, Bologna, Italy \and
ESO, Karl Schwarzschild Str. 2, 85748 Garching bei M\"unchen, Germany \and
INAF -- Osservatorio Astrofisico di Arcetri, Largo E. Fermi 5, I-50125, Firenze, Italy
}

\date{Received 10 August 2010; accepted 22 July 2011}

\abstract{
We analyze the long-term evolution of the fluxes of six active galactic nuclei (AGN) -- 0923+392, 3C~111, 3C~273, 3C~345, 3C~454.3, and 3C~84 -- in the frequency range 80--267~GHz using archival calibration data of the IRAM Plateau de Bure Interferometer. Our dataset spans a long timeline of $\approx$14 years with 974 -- 3027 flux measurements per source. We find strong (factors $\approx$2--8) flux variability on timescales of years for all sources. The flux density distributions of five out of six sources show clear signatures of bi- or even multimodality. Our sources show mostly steep ($\alpha\approx0.5-1$), variable spectral indices that indicate outflow dominated emission; the variability is most probably due to optical depth variations. The power spectra globally correspond to red-noise spectra with five sources being located between the cases of white and flicker noise and one source (3C~111) being closer to the case of random walk noise. For three sources the low-frequency ends of their power spectra appear to be upscaled in spectral power by factors $\approx$2--3 with respect to the overall powerlaws. In two sources, 3C~454.3 and 3C~84, the 1.3-mm emission preceeds the 3-mm emission by $\approx$55 and $\approx$300 days, respectively, probably due to (combinations of) optical depth and emission region geometry effects. We conclude that the source emission cannot be described by uniform stochastic emission processes; instead, a distinction of ``quiescent'' and (maybe multiple) ``flare'' states of the source emission appears to be necessary.
}


\maketitle

\section{Introduction}

Active galactic nuclei (AGN) have been studied extensively in the wavelength range from cm-radio to $\gamma$-rays in the last decades (see, e.g., Kembhavi \& Narlikar \cite{kembhavi1999}, or Krolik \cite{krolik1999}, and references therein for a review). There is overwhelming observational evidence that their emission originates from accretion onto supermassive black holes (SMBH) with masses $M_{\bullet}\simeq10^{6...9}M_{\odot}$ (e.g., Ferrarese \& Ford \cite{ferrarese2005}, and references therein).

The viewing-angle unification standard model of AGN (Lawrence \cite{lawrence1987}; see also Urry \& Padovani \cite{urry1995}) has long been established, although recent studies emphasize the impact of galaxy mergers on AGN fueling and luminosity (e.g. Hopkins et al. \cite{hopkins2006}). However, the details of AGN activity are still not well understood. One example is the nuclear emission; proposals for its origin range from shocks in continuous (e.g., Marscher \& Gear \cite{marscher1985}) or discontinuous (e.g., Spada et al. \cite{spada2001}) jets to orbiting plasma ``hotspots'' (e.g., Abramowicz et al. \cite{abramowicz1991}) or plasma density waves (e.g., Kato \cite{kato2000}) in accretion disks.

One possibility to constrain the AGN physics is offered by the analysis of characteristic timescales and the noise properties of the emission. Several studies aimed at identifying characteristic timescales, including possible quasi-periodic oscillations, from tens of minutes (using X-ray observations; e.g. Benlloch et al. \cite{benlloch2001}, and using mm-radio observations; e.g. Sch\"odel et al. \cite{schoedel2007}) to tens of years (based on cm- and mm-radio monitoring data; Hovatta et al. \cite{hovatta2007,hovatta2008}). Other works focused on the power spectral statistics of the emission which was found to globally follow red-noise type powerlaws (e.g. Lawrence \& Papadakis \cite{lawrence1993}; Uttley et al. \cite{uttley2002}).

The studies by Hovatta et al. (\cite{hovatta2007,hovatta2008}) used radio data in the frequency range 22--87~GHz spanning timelines of $\approx$25 years. A key result of their study was that AGN show large-amplitude activity on timescales $\gtrsim$6 years. Therefore, analysing the long-term behaviour of AGN requires monitoring on timescales of $\gtrsim$10 years. Building up on this conclusion, we have analyzed archival flux monitoring data from the IRAM Plateau de Bure Interferometer (PdBI) of six AGN -- 0923+392, 3C~111, 3C~273, 3C~345, 3C~454.3, and 3C~84 -- in the (observatory frame) frequency range 80--267~GHz. Given the range of redshifts $z\approx0.02-0.86$ of our sample, this corresponds to a range of rest-frame frequencies of $\approx$82--497~GHz. Our dataset comprises several thousand measurements for each source, covering timelines of $\approx$14 years. We note that our study complements earlier AGN mm/radio surveys carried out with the IRAM Pico Veleta observatory (Steppe et al. \cite{steppe1988,steppe1992,steppe1993}; Reuter et al. \cite{reuter1997}).

\section{Observations and Data Processing}

\begin{table*}
\caption{Physical properties and observation journal for our six target AGN. J2000 coordinates, source types, and redshifts are taken from the NED, black hole masses $M_{\bullet}$ are from the references given. The kpc-scale morphology has been taken from the MOJAVE database (Lister et al. \cite{lister2009}). We also give the total monitoring time $T$ and the number of measurements $N$ for the 1.3-mm (here labeled 1-mm for brevity), 2-mm, and 3-mm bands, respectively.}
\label{tab_journal}
\centering
\begin{tabular}{l c c c c c c c c c c}
\hline\hline
Object & R.A. & Dec & Type$^{\mathrm{a}}$ & Redshift & $M_{\bullet} [10^8M_{\odot}]$ & kpc-morphology & $T$ [yr] & $N_{\rm 1mm}$ & $N_{\rm 2mm}$ & $N_{\rm 3mm}$ \\
\hline
0923+392 & 09:27:03 & +39:02:21 & FSRQ & 0.695 & 12$^{\mathrm{b}}$ & 1-sided halo & 13.7 & 574 & 42 & 1019 \\
3C 111 & 04:18:21 & +38:01:36 & Sy 1 & 0.049 & 36$^{\mathrm{c}}$ & 2-sided jet & 12.9 & 398 & 7 & 569 \\
3C 273 & 12:29:06 & +02:03:09 & FSRQ & 0.158 & 8$^{\mathrm{b}}$ & 1-sided jet & 13.8 & 1053 & 81 & 1893 \\
3C 345 & 16:42:58 & +39:48:37 & FSRQ & 0.593 & 18$^{\mathrm{b}}$ & 1-sided halo & 13.9 & 587 & 57 & 1099 \\
3C 454.3 & 22:53:57 & +16:08:54 & FSRQ & 0.859 & 7$^{\mathrm{b}}$ & 1-sided jet & 13.8 & 835 & 116 & 1677 \\
3C 84 & 03:19:48 & +41:30:42 & Sy 2 & 0.018 & 19$^{\mathrm{c}}$ & 2-sided halo & 13.9 & 501 & 56 & 990 \\
\hline

\end{tabular}

\begin{list}{}{}
\item[$^{\mathrm{a}}$] ``Sy'' = Seyfert galaxy, ``FSRQ'' = flat spectrum radio quasar
\item[$^{\mathrm{b}}$] Liu et al. (\cite{liu2006})
\item[$^{\mathrm{c}}$] Marchesini et al. (\cite{marchesini2004})
\end{list}

\end{table*}

The PdBI is able to observe in three atmospheric windows located around wavelengths of 1.3~mm, 2~mm, and 3~mm. Each of these bands covers a continuous range of frequencies that are available for observations; these ranges are
201--267~GHz for the 1.3~mm band,
129--174~GHz for the 2~mm band,
and 80--116~GHz for the 3~mm band (Winters \& Neri \cite{winters2010}).

Until the end of 2006, observations were made using dual-band, single-polarization receivers. One frequency in the 1.3-mm band and one frequency in the 3-mm band could be selected for simultaneous observations; the 2-mm band was not reachable. At the beginning of 2007, new single-band, dual-polarization receivers were installed. Since then it is possible to observe a single frequency located in a given band simultaneously in both linear polarizations. At the beginning of 2008, the 2-mm band became available for observations.

Flux monitoring data for AGN are available from October 1996 on as a by-product of regular observatory operations. The PdBI uses the most luminous ($S_{\nu}\gtrsim3$~Jy in the 3-mm band) active galactic nuclei as calibrators for the spectral response of the instrument. The sources we present here are the most luminous targets regularly available (especially in terms of declination) as spectral calibrators. Therefore at least one of these sources has been observed in almost any individual observing project, with each project lasting typically few hours. This leads to a dense sampling over the entire observation timeline of $\gtrsim$13 years; gaps in the lightcurves are due to technical works at the interferometer and seasonal sunavoidance constraints. For our final sample of sources to be analyzed we picked all objects for which we have at least 1000 measurements over the entire observing timeline without major ($\gtrsim$1 year) interruptions of the monitoring. These were 0923+392, 3C~111, 3C~273, 3C~345, 3C~454.3, and 3C~84.

Antenna temperatures are converted into physical flux densities by using empirical antenna efficiencies as conversion factors. Those factors are functions of frequencies and are located in the range from $\approx$22~Jy/K (for the 3-mm band) to $\approx$37~Jy/K (for the 1.3-mm band). In order to estimate the systematic uncertainties of our dataset, we analyzed observations of the radio continuum star MWC~349A (e.g. Tafoya et al. \cite{tafoya2004}) that were obtained and calibrated like the AGN measurements. Due to its well-known flux properties MWC~349 serves as a ``standard candle'' for the PdBI (Krips et al. \emph{in prep.}). From the systematic scatter of the MWC~349 lightcurves we conclude that the \emph{systematic} relative uncertainties of our AGN observations are $\approx$10\% for the 2-mm and 3-mm bands and $\approx$16\% for the 1.3-mm band. We note that the systematic accuracy of our data is limited by the automatic monitoring and calibration; when calibrating observations individually and interactively while using MWC~349 as reference, systematic errors can be reduced to $\lesssim$5\% (Krips et al. \emph{in prep.}).

Before analysis, we had to remove erroneous values from the flux data. Errors are caused by either of two effects: (1) poor signal-to-noise ratio (S/N), or (2) failure of the flux calibration. We caught effect (1) by rejecting data with statistical flux errors larger than 0.1~Jy or S/N$<$10. \emph{Statistical} errors of individual measurements used for analysis are thus $<$0.1~Jy. Effect (2) is more difficult to treat as it leads to a systematic deviation of the observed from the true flux values. Especially, it leads to a clustering of data points in the range 0--1.5~Jy (details depending on the source), well separate from, and systematically below, the long-term lightcurves. In order to remove this effect, we rejected flux measurements with fluxes located below the ``divides'' that separate the long-term lightcurves from the clusters of erroneous points. Additionally, we checked the statistical behaviour of the data. For this, we analyzed the differences between the lightcurves and smoothed versions of the lightcurves, using smoothing by running medians. We found that the distributions of the residuals usually follow Gaussian distributions with extended tails. We regarded the data points located in the tails as ``suspicious'' outliers and rejected all data located more than $2.5\sigma$ away from the centers of the distributions. This conservative approach rejected between 2\% and 16\% of the data of a given lightcurve.

The data processing as well as the subsequent analysis made use of the IRAM Grenoble data processing software GILDAS\footnote{Developed and maintained by the IRAM GILDAS team, see http://www.iram.fr/IRAMFR/GILDAS/} and the MPE Garching data processing software DPUSER\footnote{Developed and maintained by Thomas Ott (MPE Garching), see http://www.mpe.mpg.de/$\sim$ott/dpuser/index.html}.

After processing the data, we were left with large numbers of measurements ranging from 974 to 3027 per target, corresponding to one measurements every 1.5--4.3 days. Our flux data base covers a timeline from September 1996 (for the 2-mm band: February 2008) to August 2010. In Table~\ref{tab_journal} we provide an overview over our target AGN and individual observation statistics. We collected the source types and redshifts from the NASA/IPAC Extragalactic Database (NED) and the kilo-parsec scale morphology from the MOJAVE database (Lister et al. \cite{lister2009}).

\section{Analysis}

\subsection{Lightcurves}

\begin{figure*}
\centering
\includegraphics[height=8.5cm,angle=-90]{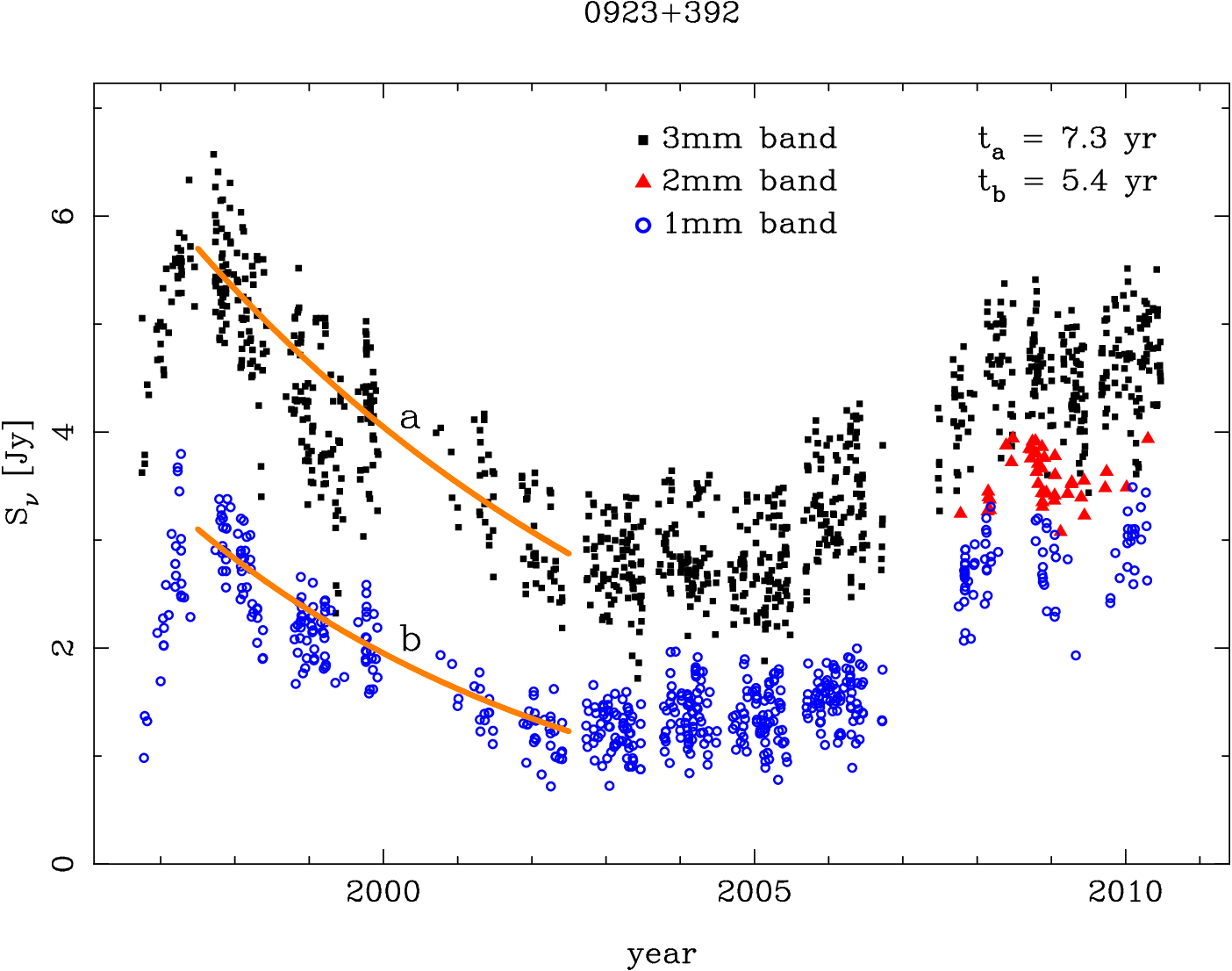}
\hspace{5mm}
\includegraphics[height=8.5cm,angle=-90]{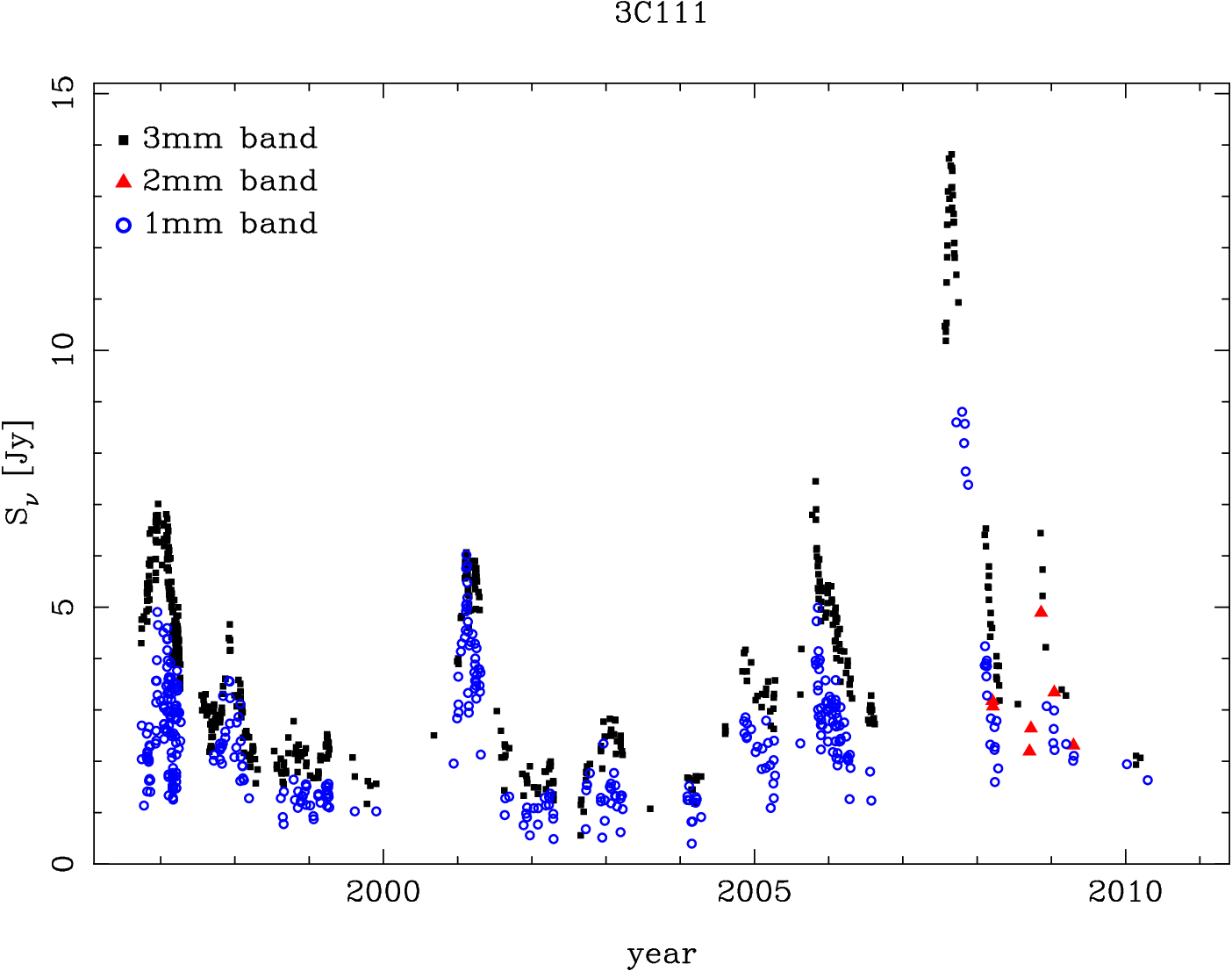} \\
\vspace{5mm}
\includegraphics[height=8.5cm,angle=-90]{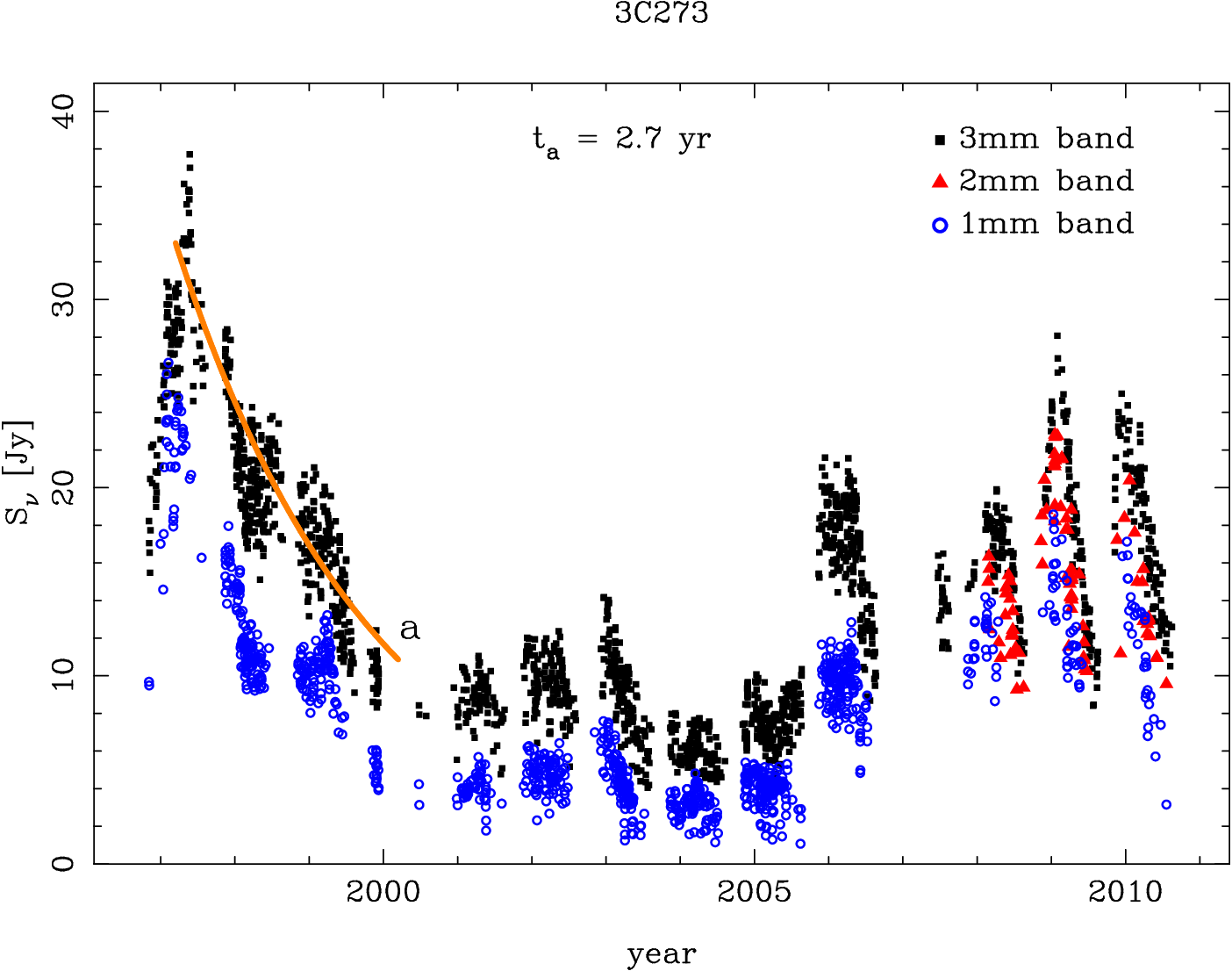}
\hspace{5mm}
\includegraphics[height=8.5cm,angle=-90]{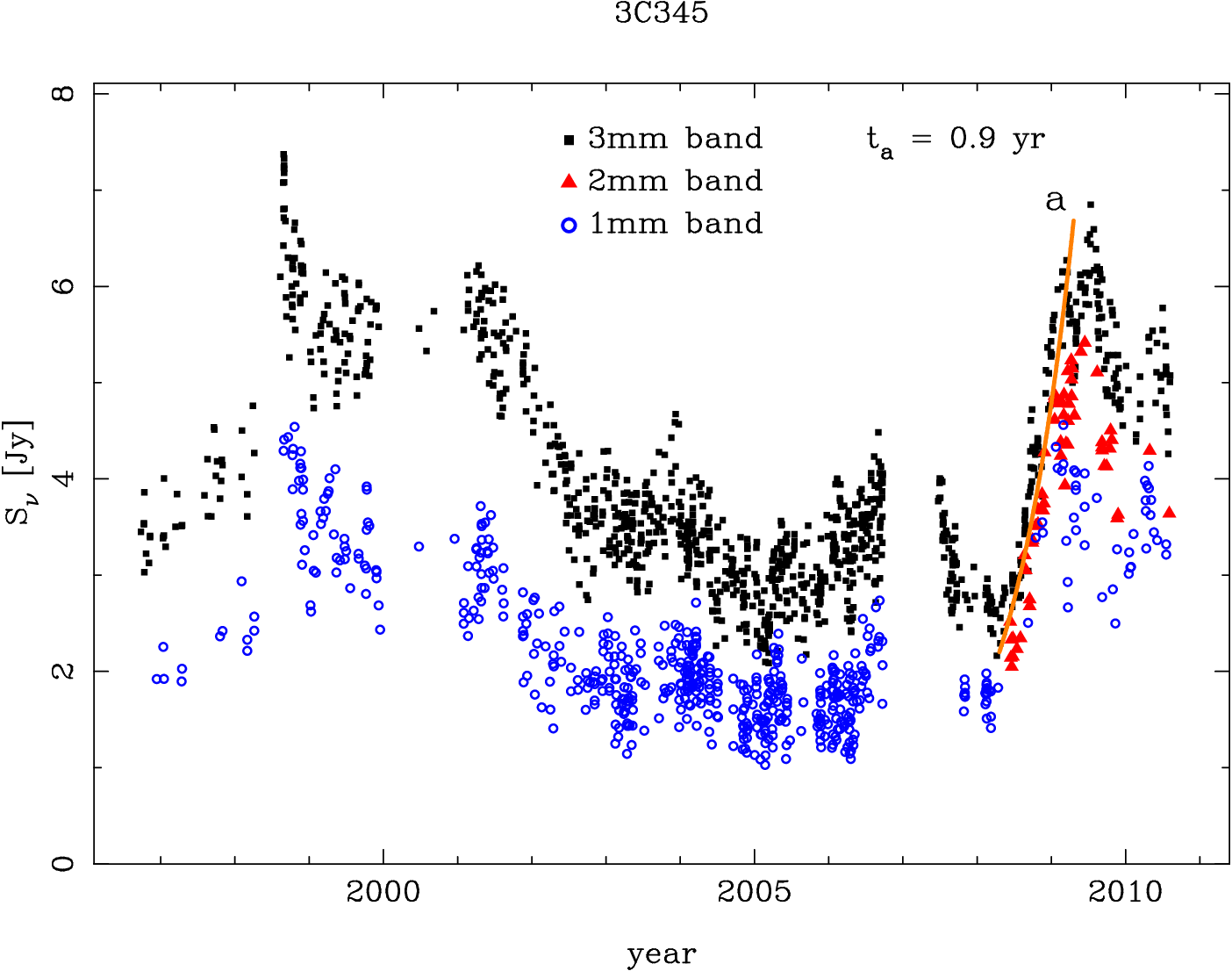} \\
\vspace{5mm}
\includegraphics[height=8.5cm,angle=-90]{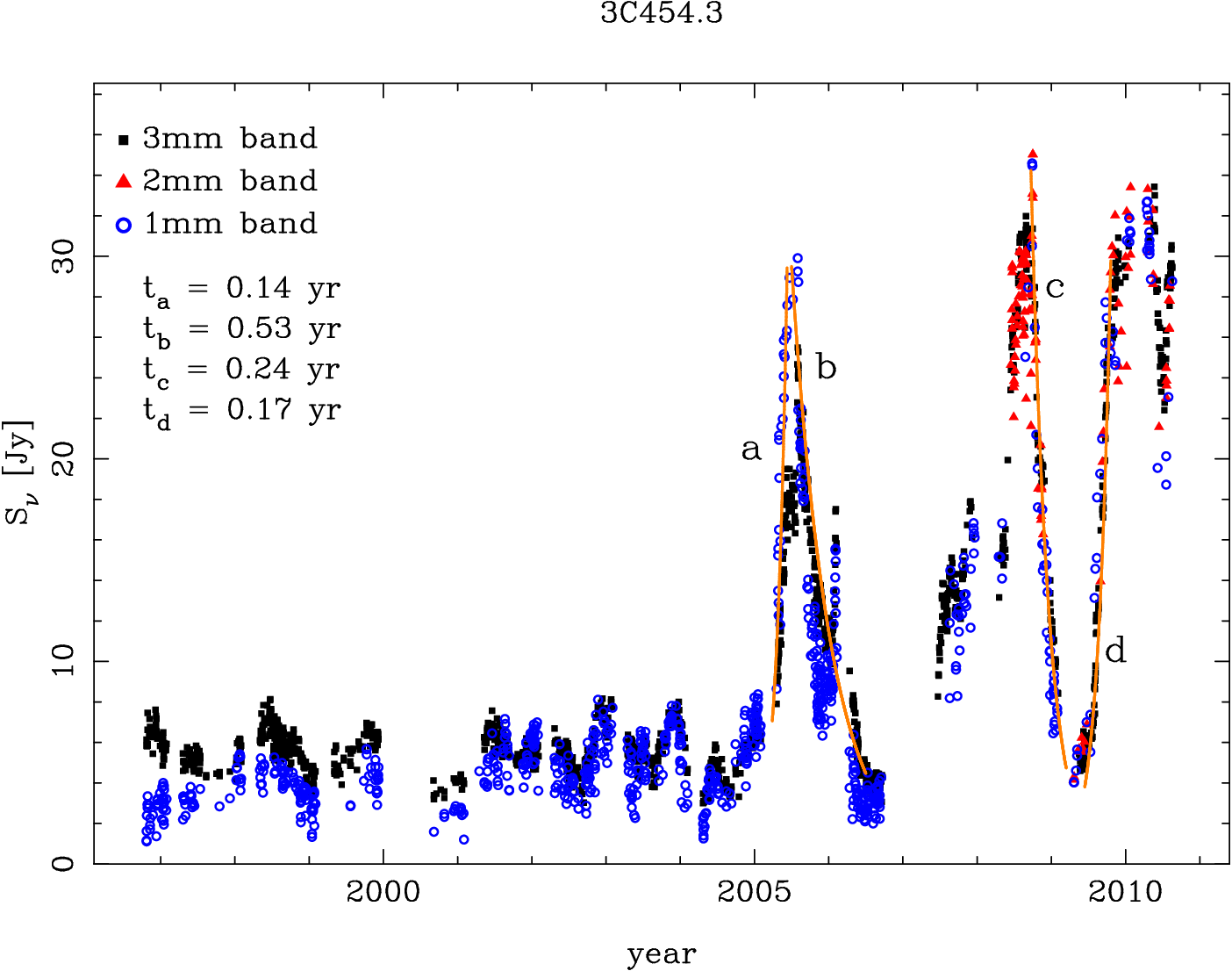}
\hspace{5mm}
\includegraphics[height=8.5cm,angle=-90]{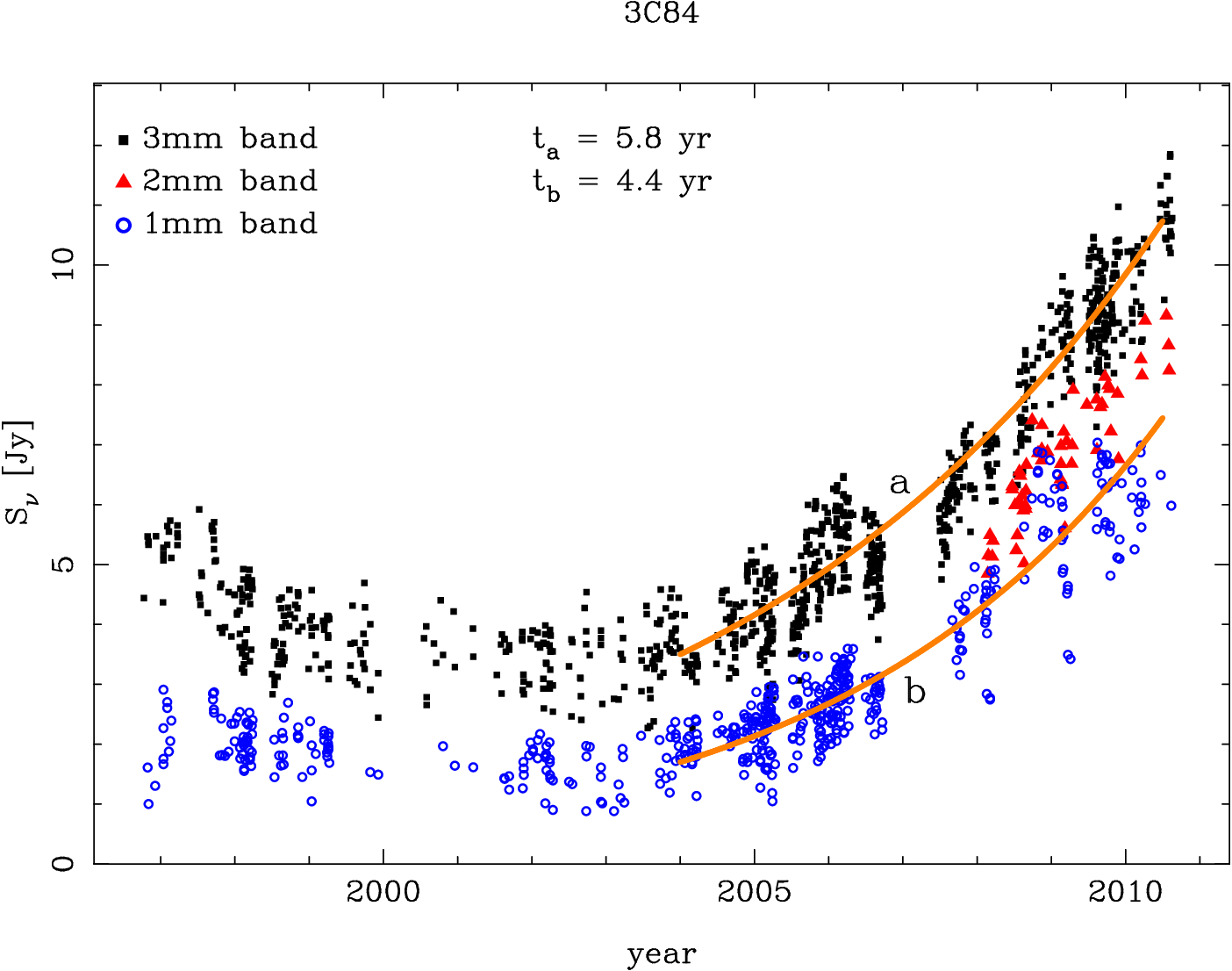}
\caption{The long-term lightcurves of our six target quasars spanning timelines from 1996 to 2010. Please note the different flux axis scales. For each source we give the fluxes vs. time separatly for the 1.3-mm (here labeled 1-mm for brevity), 2-mm, and 3-mm bands. For each source (except 3C~111) we approximate the most prominent, continuously sampled variations as exponential rises or decays according to $S_{\nu} \propto \exp(\pm t/t_x)$ (orange lines). Here $t$ is the time, $t_x$ the characteristic rise/decay timescale of the curve labeled $x~(x=a,b,c,...)$. The sign of $t$ is $+t$ for exponential rises and $-t$ for decays. All parameters are given in the corresponding diagrams.
}
\label{fig_lightcurve}
\end{figure*}

In Fig.~\ref{fig_lightcurve} we present the lightcurves of our six target AGN. For each source we give flux densities $S_{\nu}$ vs. time $t$ separately for the 1.3-mm, 2-mm, and 3-mm bands as far as available.

In order to quantify the variability timescales, we approximated the most prominent -- referring to smooth, continuously sampled variations in flux by factors $\gtrsim$2 -- variations of each lightcurve as exponential rises or decays like

\begin{equation}
S_{\nu} \propto {\rm e}^{\pm t/t_e}
\label{eq_risedecay}
\end{equation}

\noindent
with $t_e$ being the evolution timescale and the sign of the exponent being positive for exponential rises and negative for decays. We attempted those approximations only for parts of lightcurves that were sufficiently smooth for being described by a single analytical function and well sampled. These limitations prevented us from applying this approach to the lightcurve of 3C~111 which shows a combination of irregular sampling and fast variability. For the other sources, we give the results in the corresponding diagrams of Fig.~\ref{fig_lightcurve}.

\subsection{Flux Distributions}

\begin{figure*}
\centering
\includegraphics[height=8.5cm,angle=-90]{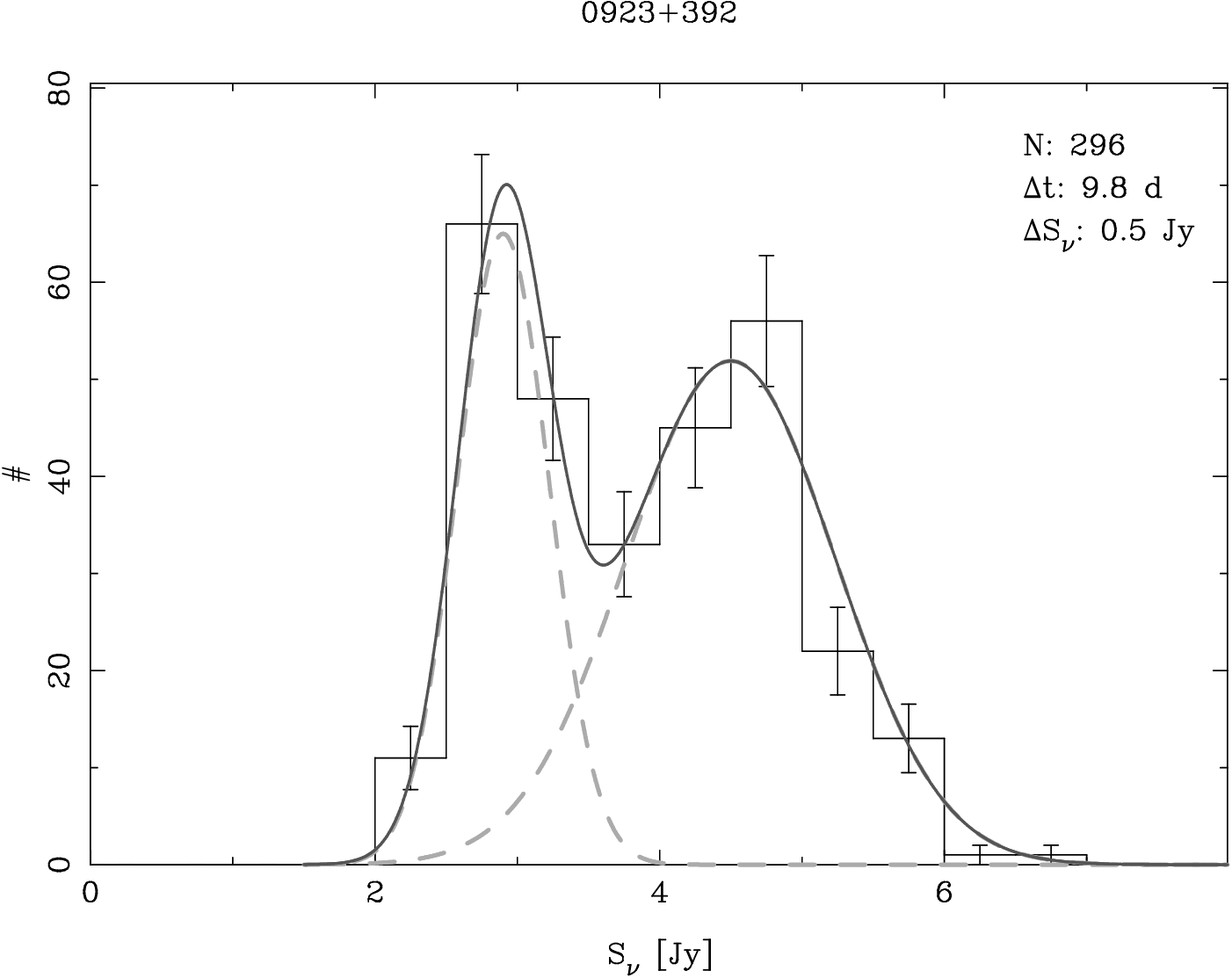}
\hspace{5mm}
\includegraphics[height=8.5cm,angle=-90]{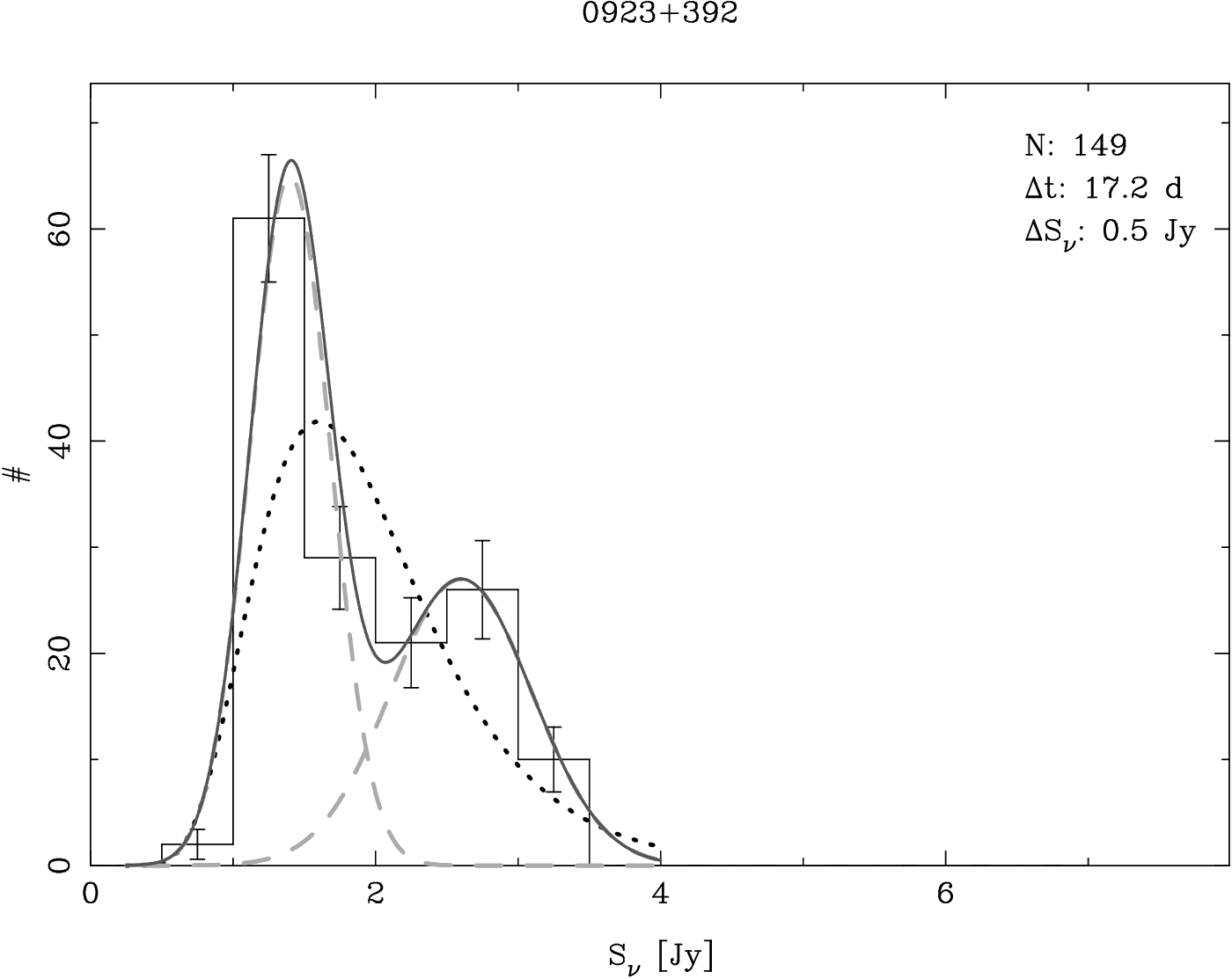} \\
\vspace{5mm}
\includegraphics[height=8.5cm,angle=-90]{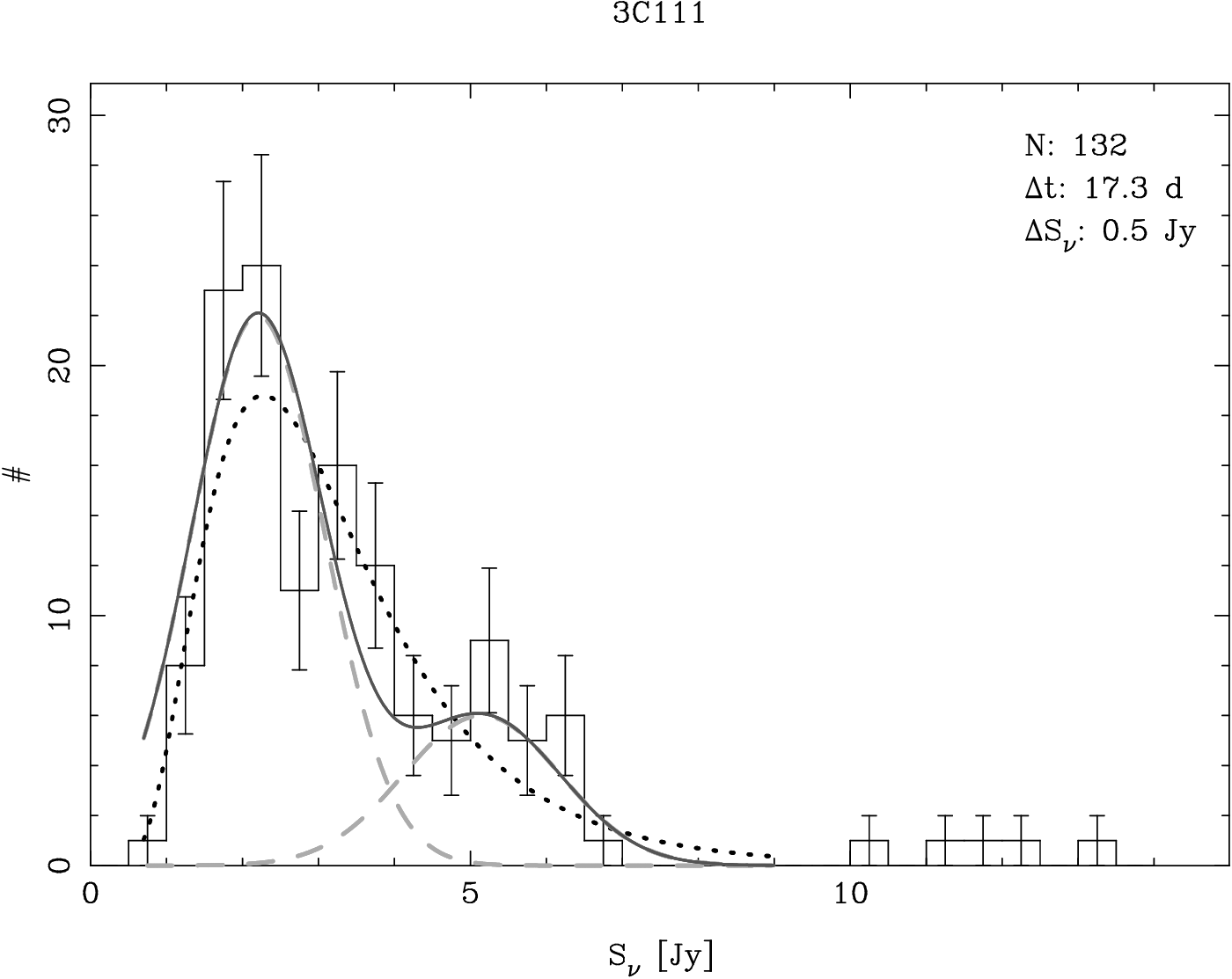}
\hspace{5mm}
\includegraphics[height=8.5cm,angle=-90]{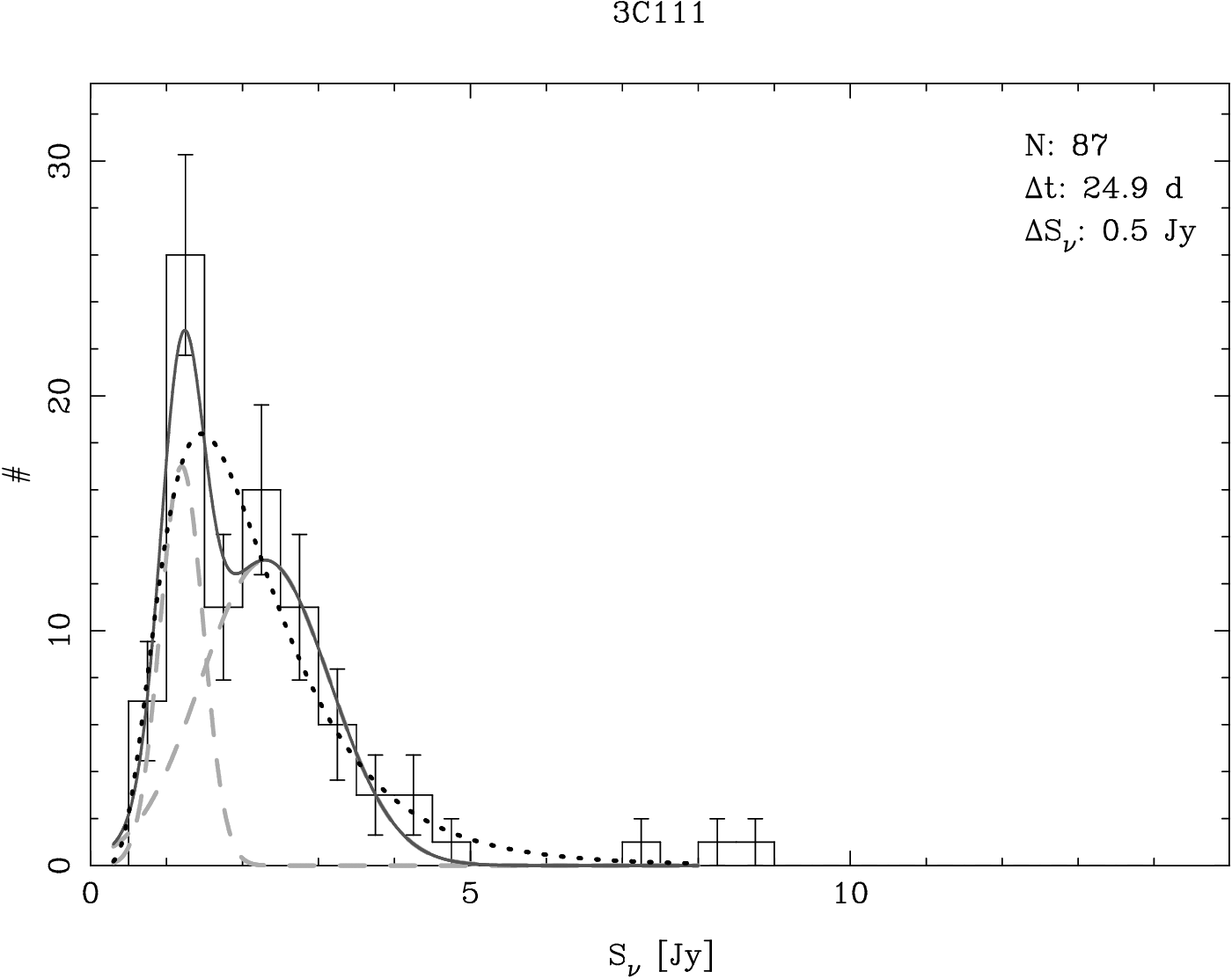} \\
\vspace{5mm}
\includegraphics[height=8.5cm,angle=-90]{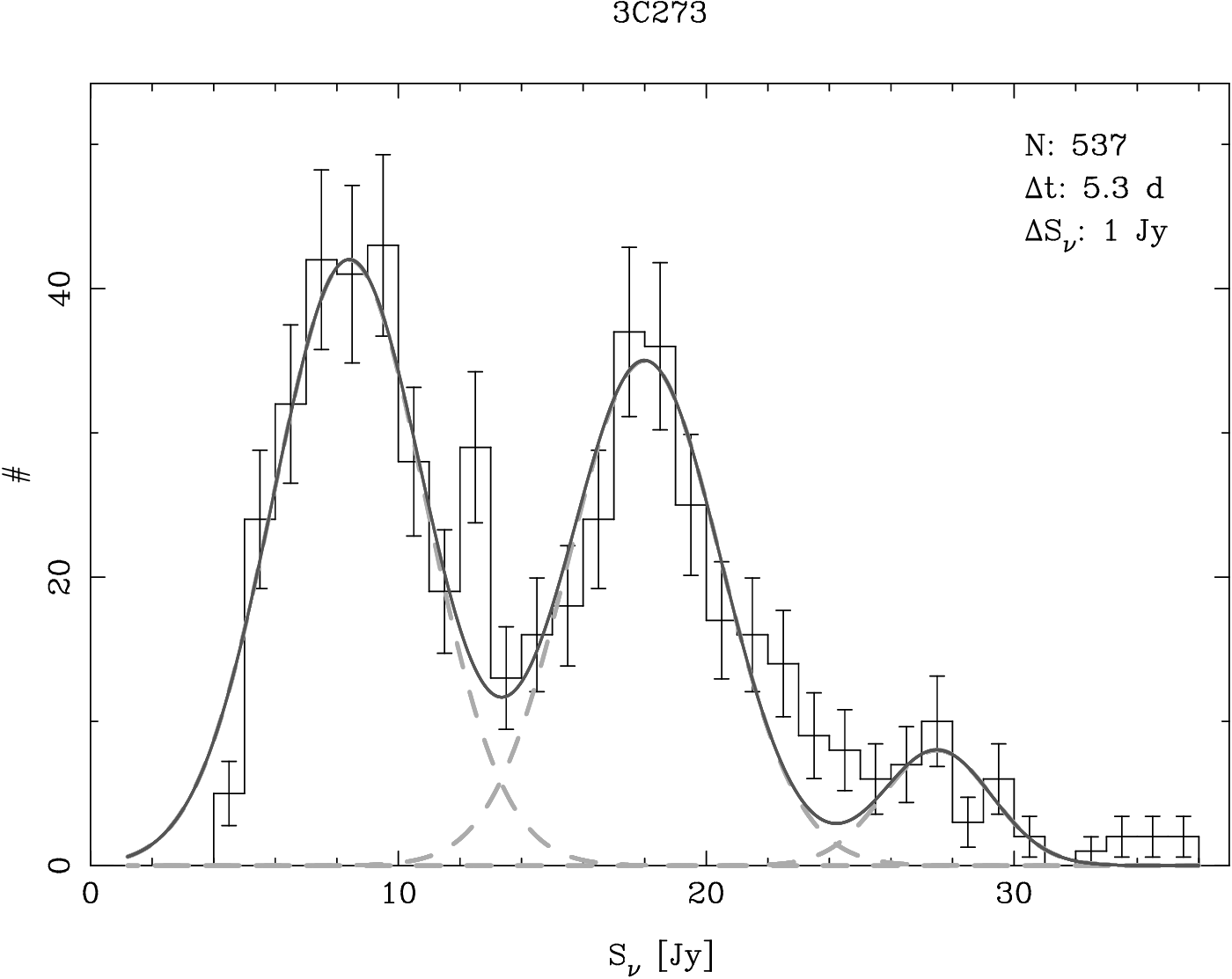}
\hspace{5mm}
\includegraphics[height=8.5cm,angle=-90]{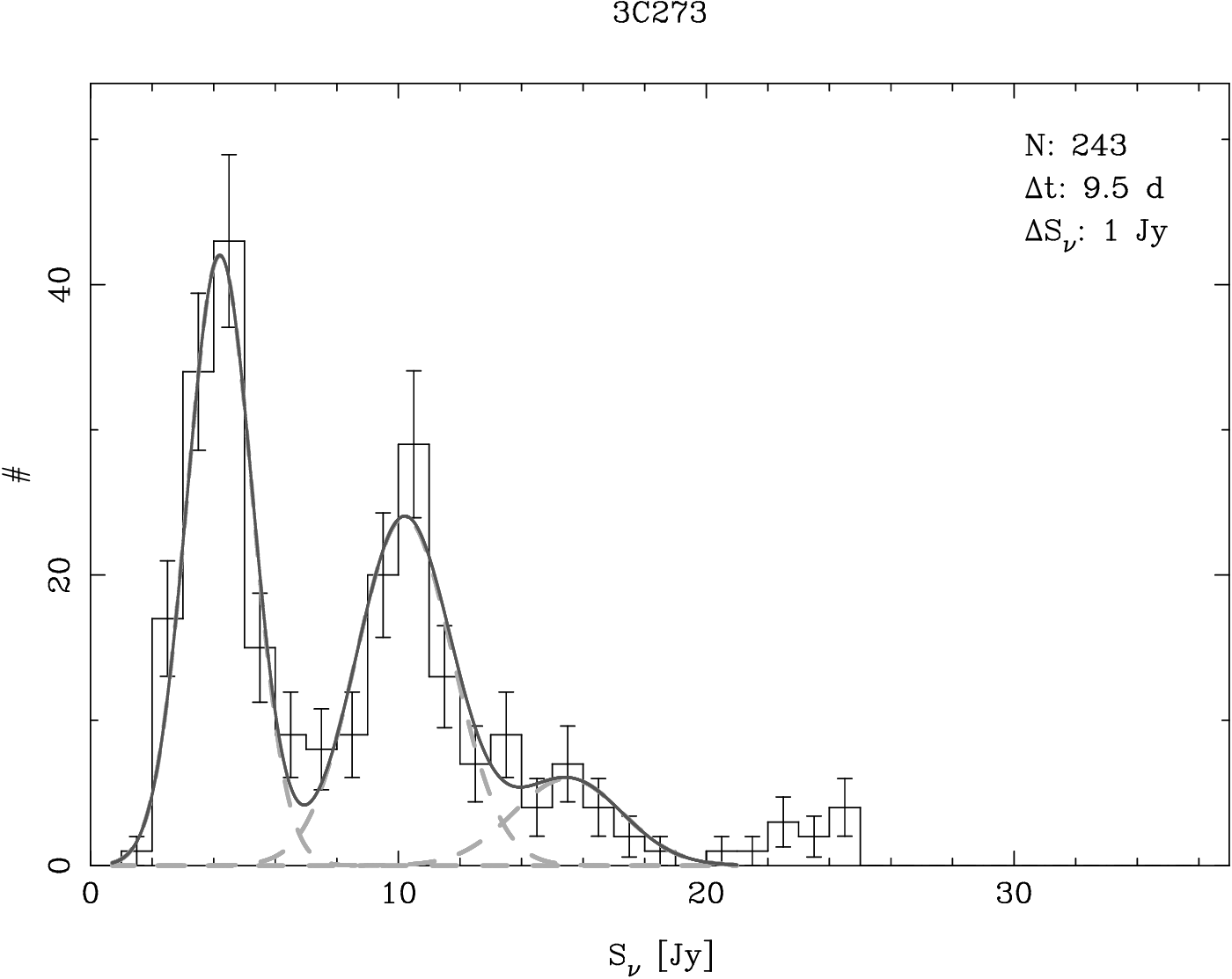}
\caption{Distributions of the 3-mm (left column) and 1.3-mm band (right column) fluxes of 0923+392, 3C~111, and 3C~273. Please note the different axis scales and bin sizes. In order to minimize the impact of irregular sampling, we binned all lightcurves (compare Fig.~\ref{fig_lightcurve}) in time using $N$ bins with sizes $\Delta t$. The parameters $N$, $\Delta t$, and the sizes of the flux bins $\Delta S_{\nu}$ are given in the respective diagrams. Errorbars indicate binomial errors. Continuous grey lines indicate approximations to the data composed of individual Gaussians (dashed grey lines). In three cases (0923+392 at 1.3~mm, 3C~111 at 1.3~mm and 3~mm) we also indicate the best-fitting log-normal distributions (dotted black curves).}
\label{fig_fluxdist1}
\end{figure*}

\begin{figure*}
\centering
\includegraphics[height=8.5cm,angle=-90]{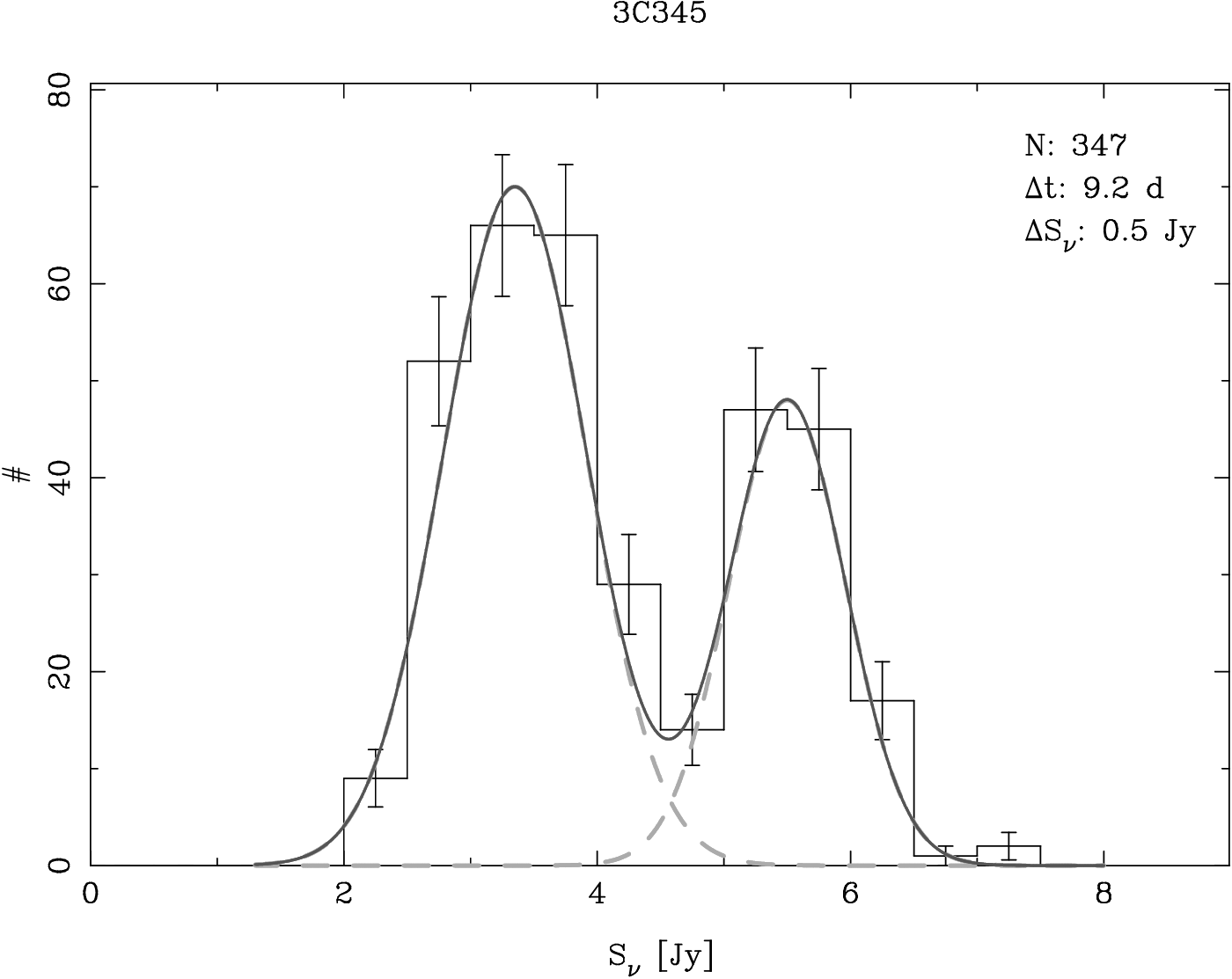}
\hspace{5mm}
\includegraphics[height=8.5cm,angle=-90]{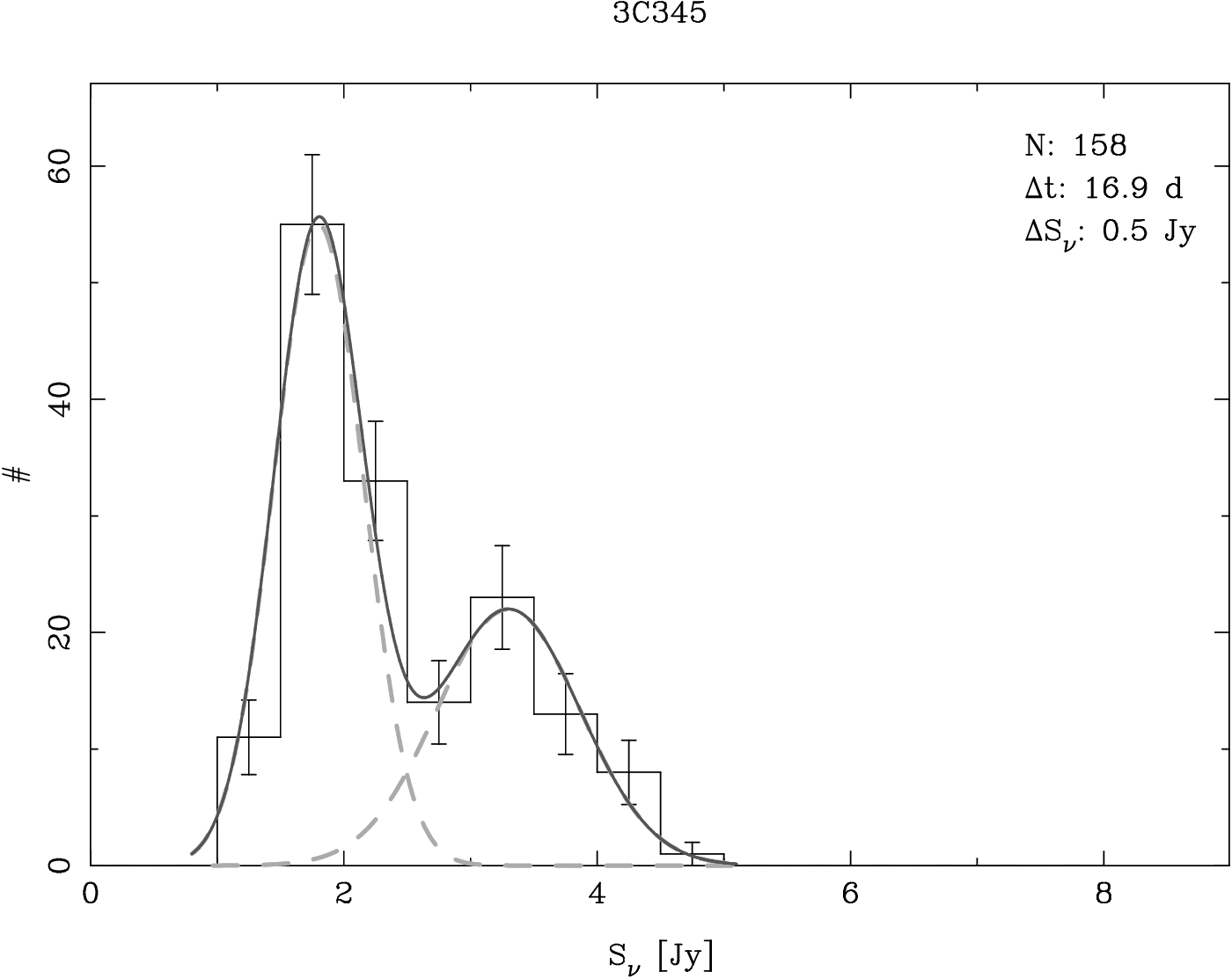} \\
\vspace{5mm}
\includegraphics[height=8.5cm,angle=-90]{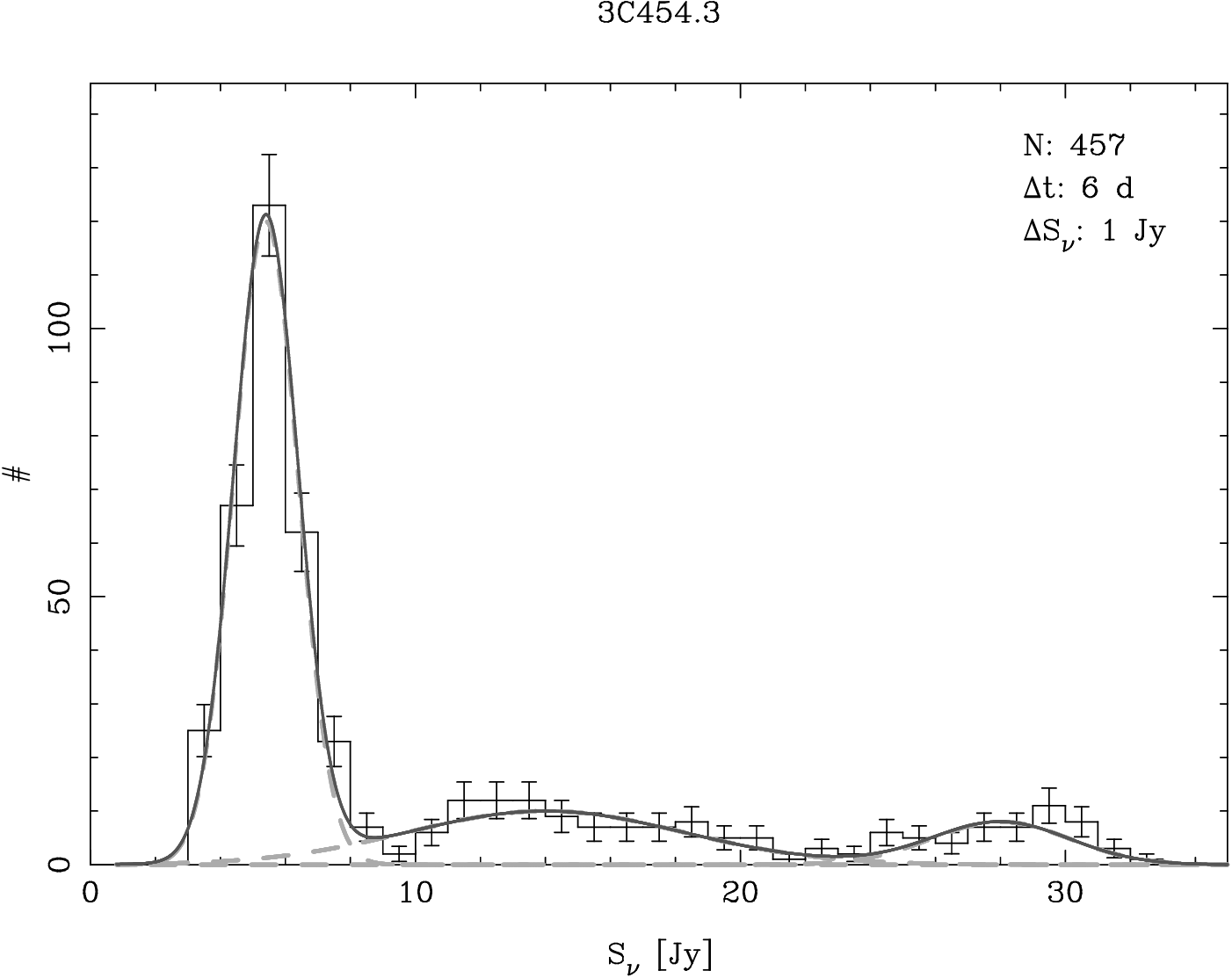}
\hspace{5mm}
\includegraphics[height=8.5cm,angle=-90]{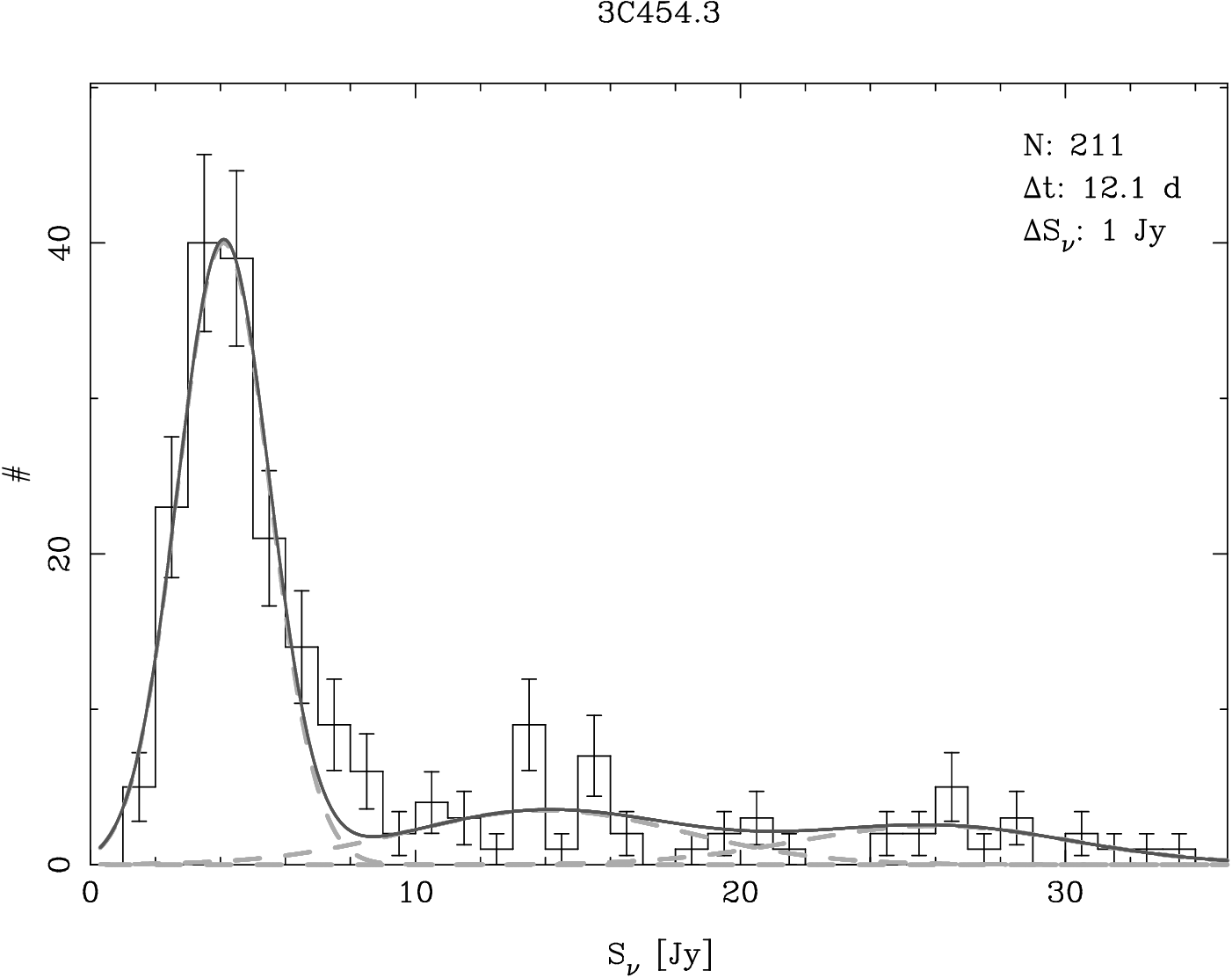} \\
\vspace{5mm}
\includegraphics[height=8.5cm,angle=-90]{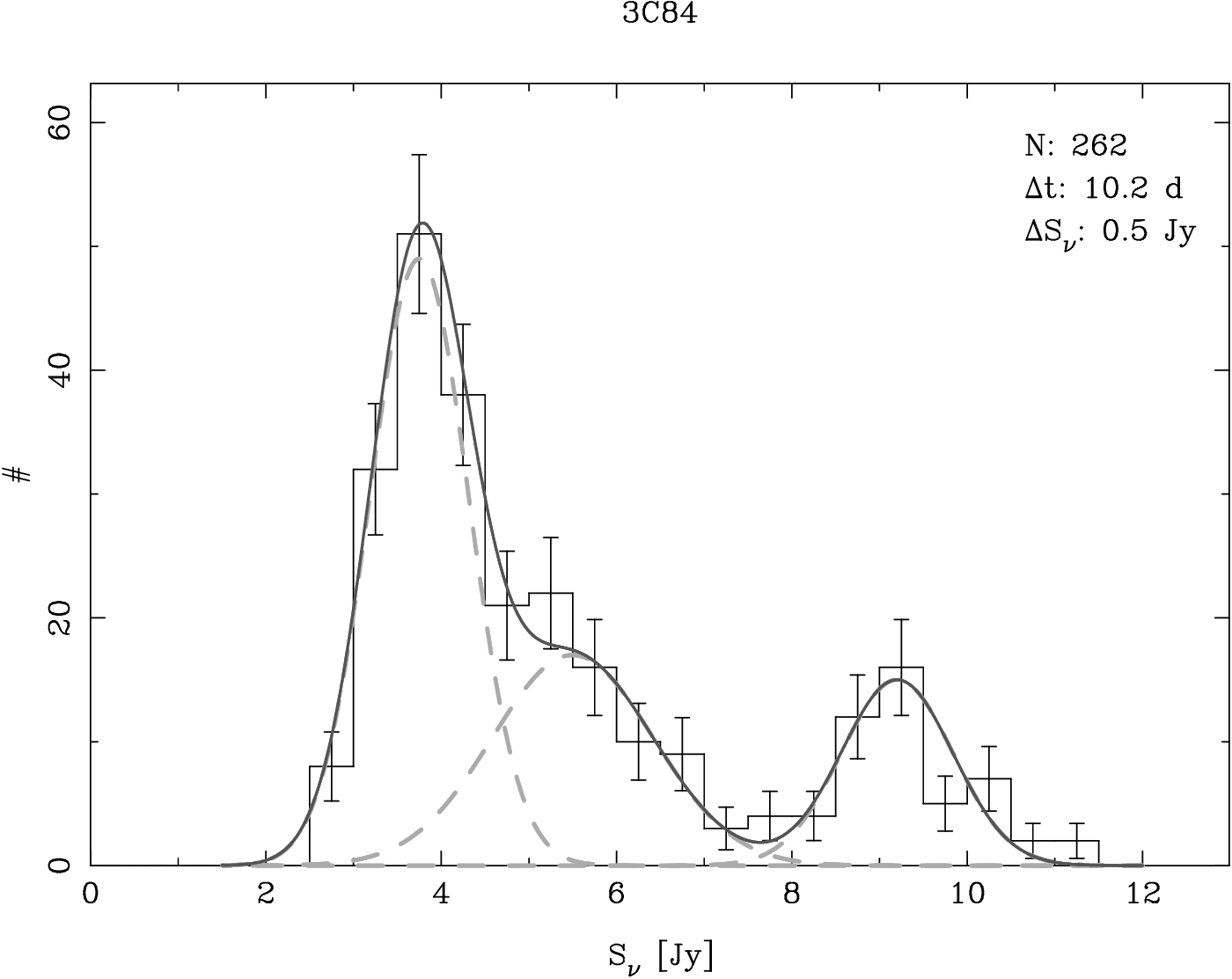}
\hspace{5mm}
\includegraphics[height=8.5cm,angle=-90]{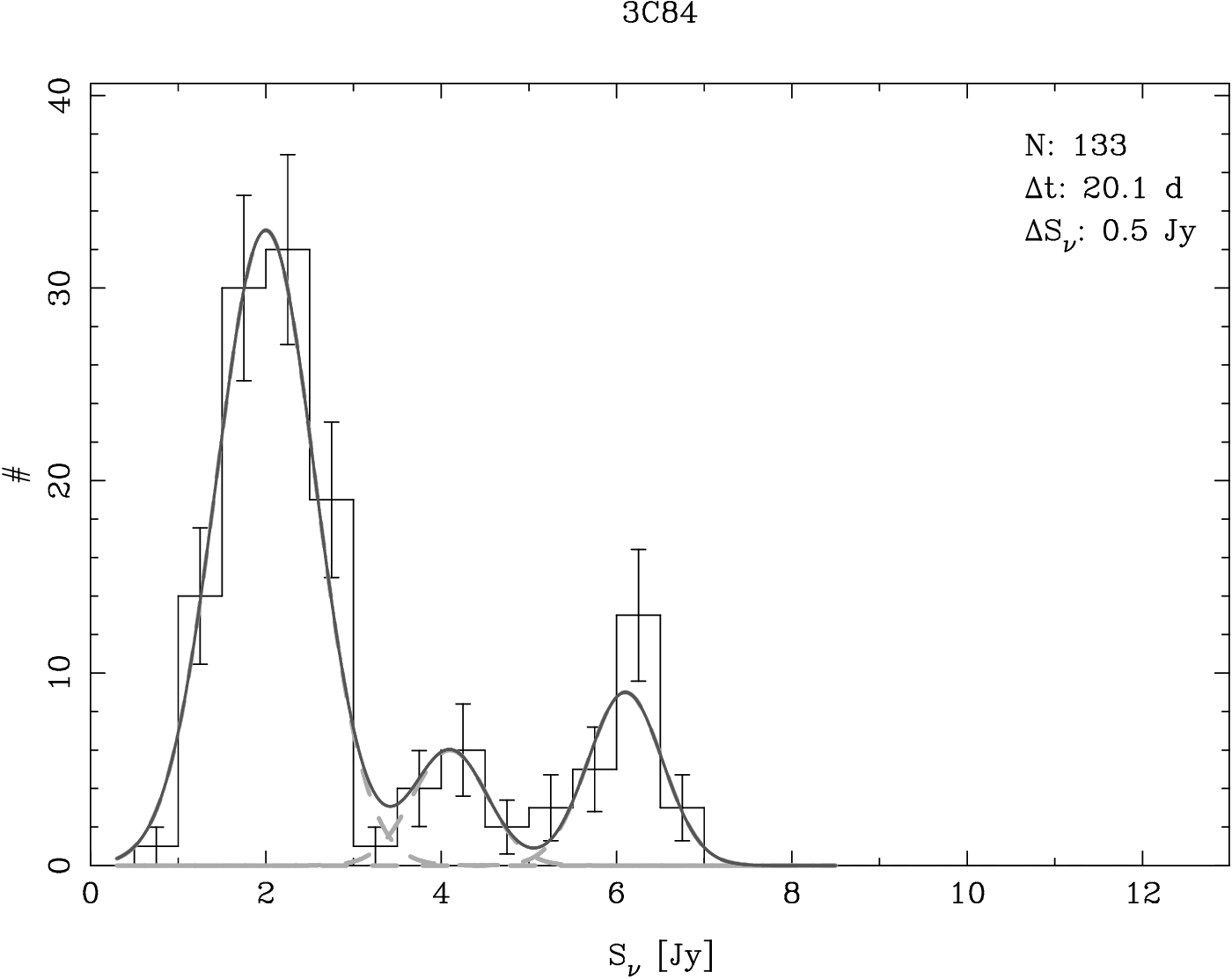}
\caption{Same as Fig.~\ref{fig_fluxdist1} for 3C~345, 3C~454.3, and 3C~84.}
\label{fig_fluxdist2}
\end{figure*}

In Fig. \ref{fig_fluxdist1} and \ref{fig_fluxdist2} we present for each source the flux distributions for the 1.3-mm and 3-mm bands. The 2-mm lightcurves do not contain enough data for calculating corresponding distributions. In order to prevent distortions introduced by irregular sampling, we binned our data in time. The size of the time bins for a specific lightcurve is given by

\begin{equation}
\Delta t = 2T / N
\label{eq_timebin}
\end{equation}

\noindent
Here $T$ is the total observation period, $N$ is the number of data points. For the special case of regular sampling, $\Delta t$ corresponds to the inverse of the Nyquist frequency.

As the resulting histograms show somewhat complicated morphologies, we investigated the minimum number of smooth components necessary for a reasonable description of their profiles. For this we approximated each distribution as a superposition of two (0923+392, 3C~111, 3C~345) or three (3C~273, 3C~454.3, 3C~84) Gaussian distributions

\begin{equation}
G(S) = a_0\exp[-(S-S_0)^2/\sigma^2]
\label{eq_gaussian}
\end{equation}

\noindent
with $S$ denoting the flux, $S_0$ being the center position, $a_0$ being the amplitude at $S_0$, and $\sigma^2$ denoting the variance. The choice of Gaussian distributions was governed by the fact that a priori such distributions are most likely to occur in case of uniform stochastic emission processes.

In three cases (0923+392 at 1.3~mm, 3C~111 at 1.3~mm and 3~mm) we also attempted to describe the flux histograms by log-normal distributions

\begin{equation}
\Lambda(S) = \frac{1}{\sqrt{2\pi}\sigma S}\exp[-({\rm ln}S-\mu)^2/2\sigma^2]
\label{eq_lognormal}
\end{equation}

\noindent
where ln denotes the natural logarithm and $\sigma$ and $\mu$ are the scale parameter and the location parameter, respectively. As log-normal distributions are characteristic for multiplicative stochastic processes\footnote{This is a simple consequence of the central limit theorem which states that sums of random numbers are normally distributed. As the logarithm of a product of random numbers equals the sum of the logarithms of the individual random numbers, it follows (under certain weak conditions) that the logarithms of products of random numbers are normally distributed (e.g., Montroll \& Shlesinger \cite{montroll1982}, and references therein).}, they may be expected in AGN flux distributions (compare, e.g., Dodds-Eden et al. \cite{doddseden2011}).

\subsection{Spectral Indices}

As we observed our target AGN over a range of frequencies located in three spectral bands, we were able to analyze their mm-spectroscopic properties. This analysis was however complicated by fast intrinsic variability as well as by the fact that the spectral information partially needed to be derived from non-simultaneous observations. We therefore inserted all measurements within selected time windows into flux--frequency diagrams and derived time-averaged spectral indices $\alpha$ via means of least-squares fits. Spectral indices were defined via the relation

\begin{equation}
S_{\nu}\propto\nu^{-\alpha}
\label{eq_index}
\end{equation}

\begin{figure*}
\centering
\includegraphics[height=8.5cm,angle=-90]{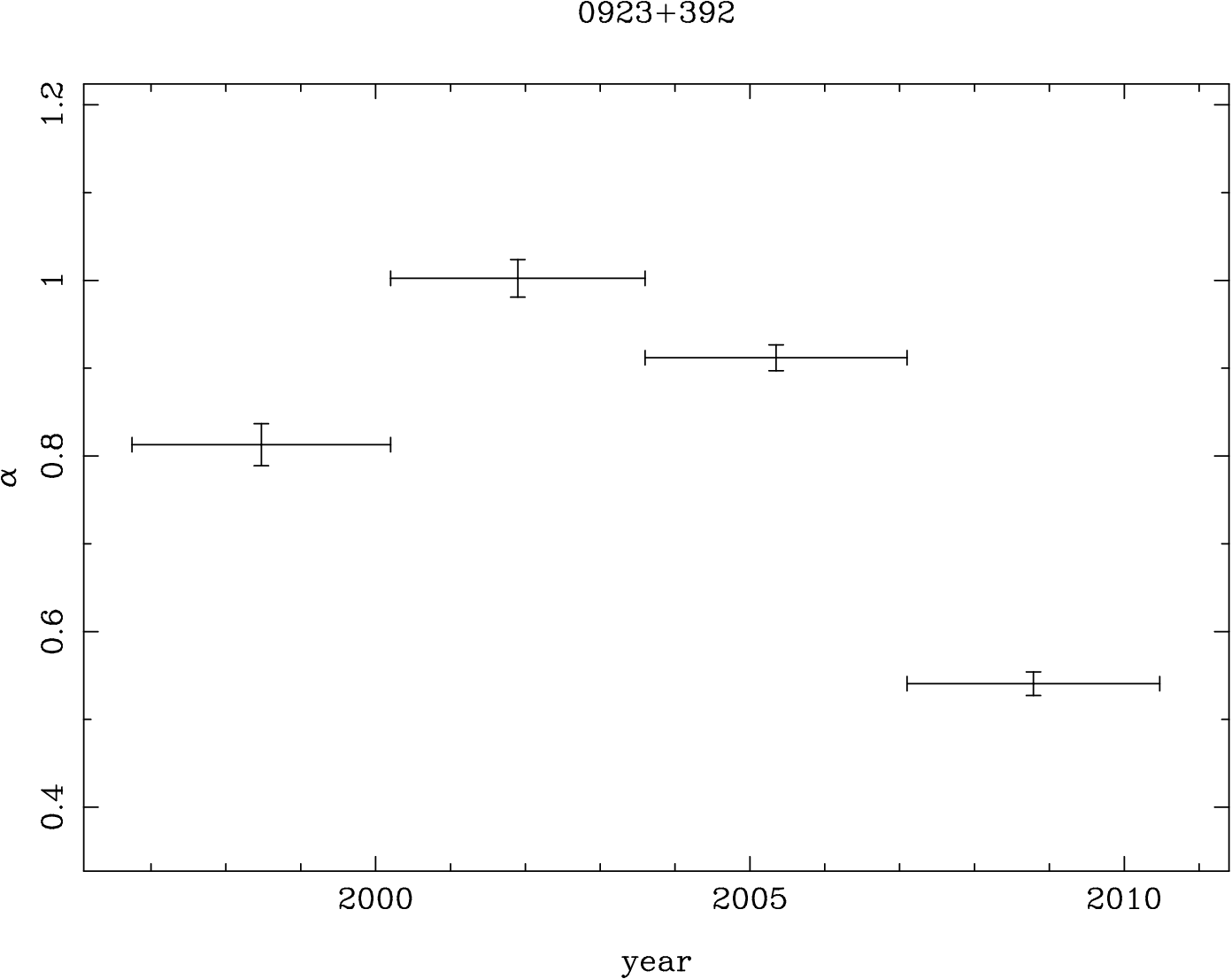}
\hspace{5mm}
\includegraphics[height=8.5cm,angle=-90]{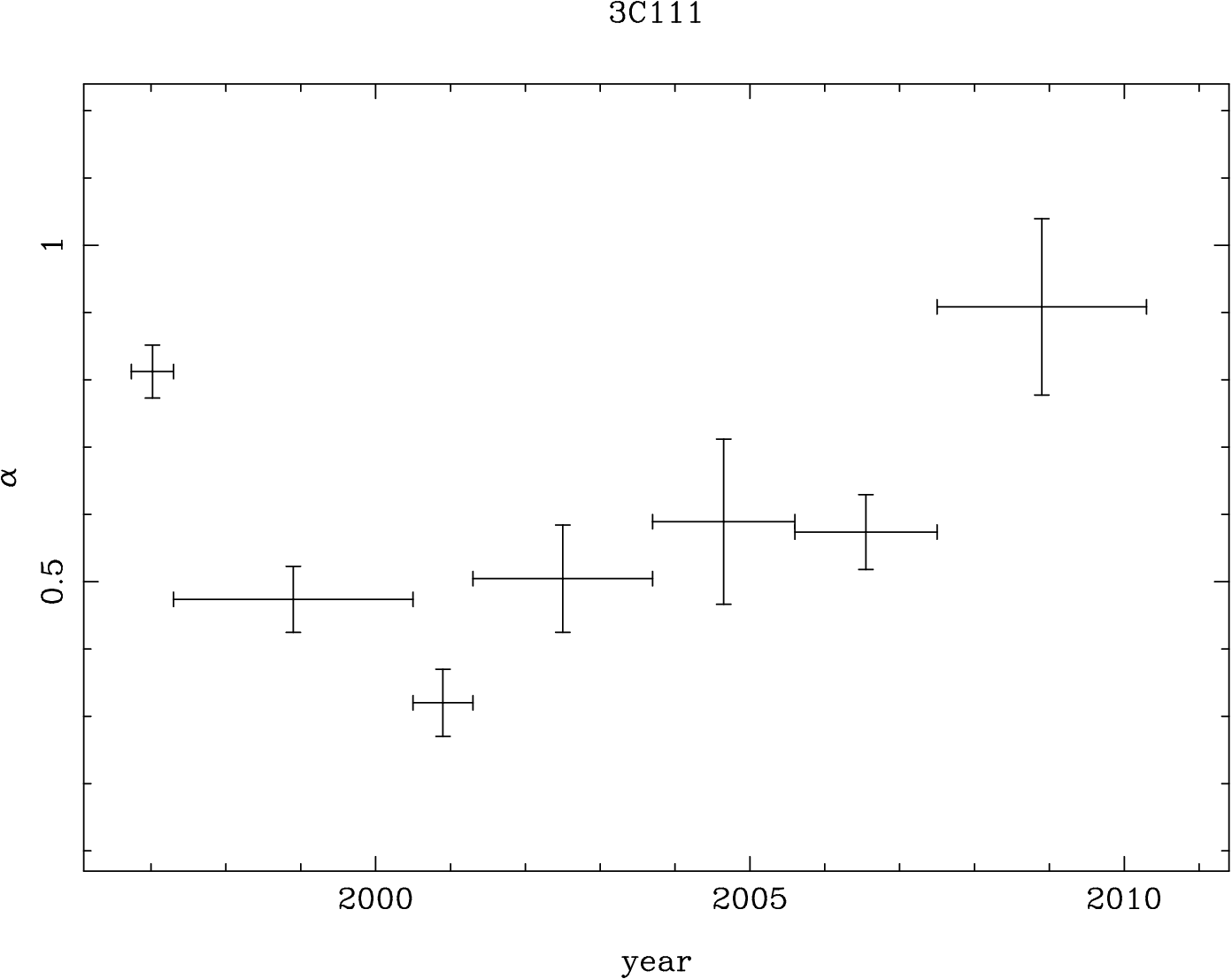} \\
\vspace{5mm}
\includegraphics[height=8.5cm,angle=-90]{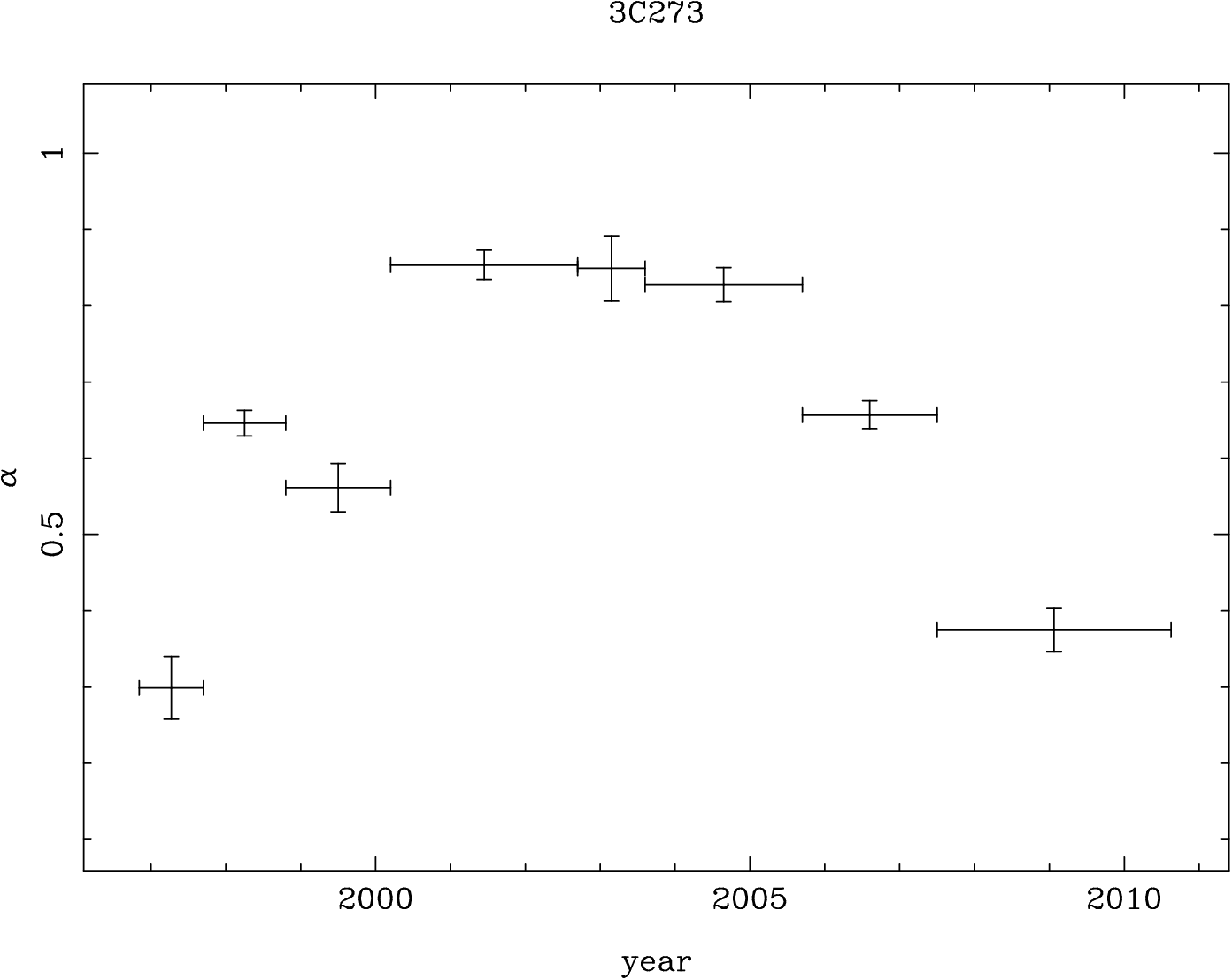}
\hspace{5mm}
\includegraphics[height=8.5cm,angle=-90]{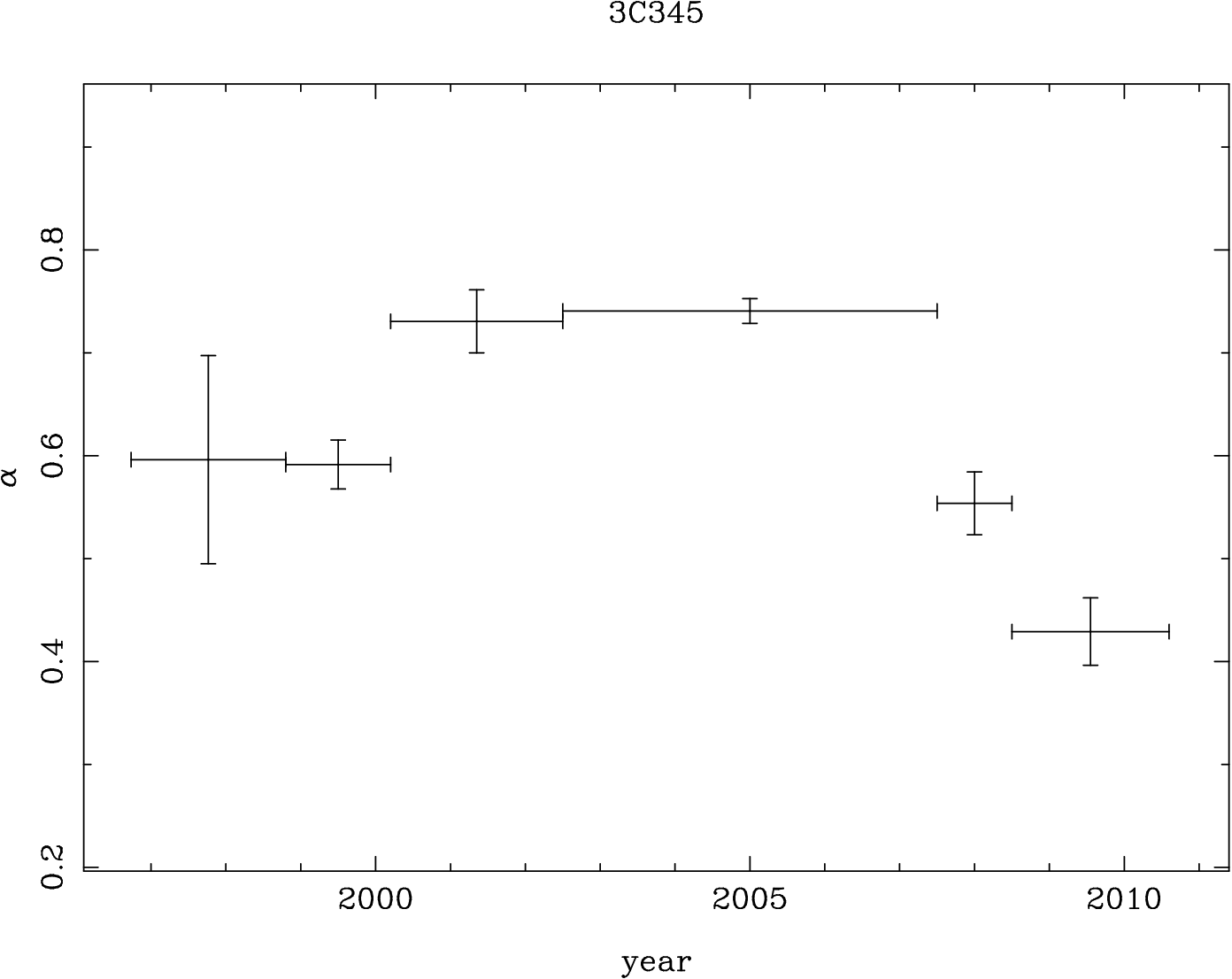} \\
\vspace{5mm}
\includegraphics[height=8.5cm,angle=-90]{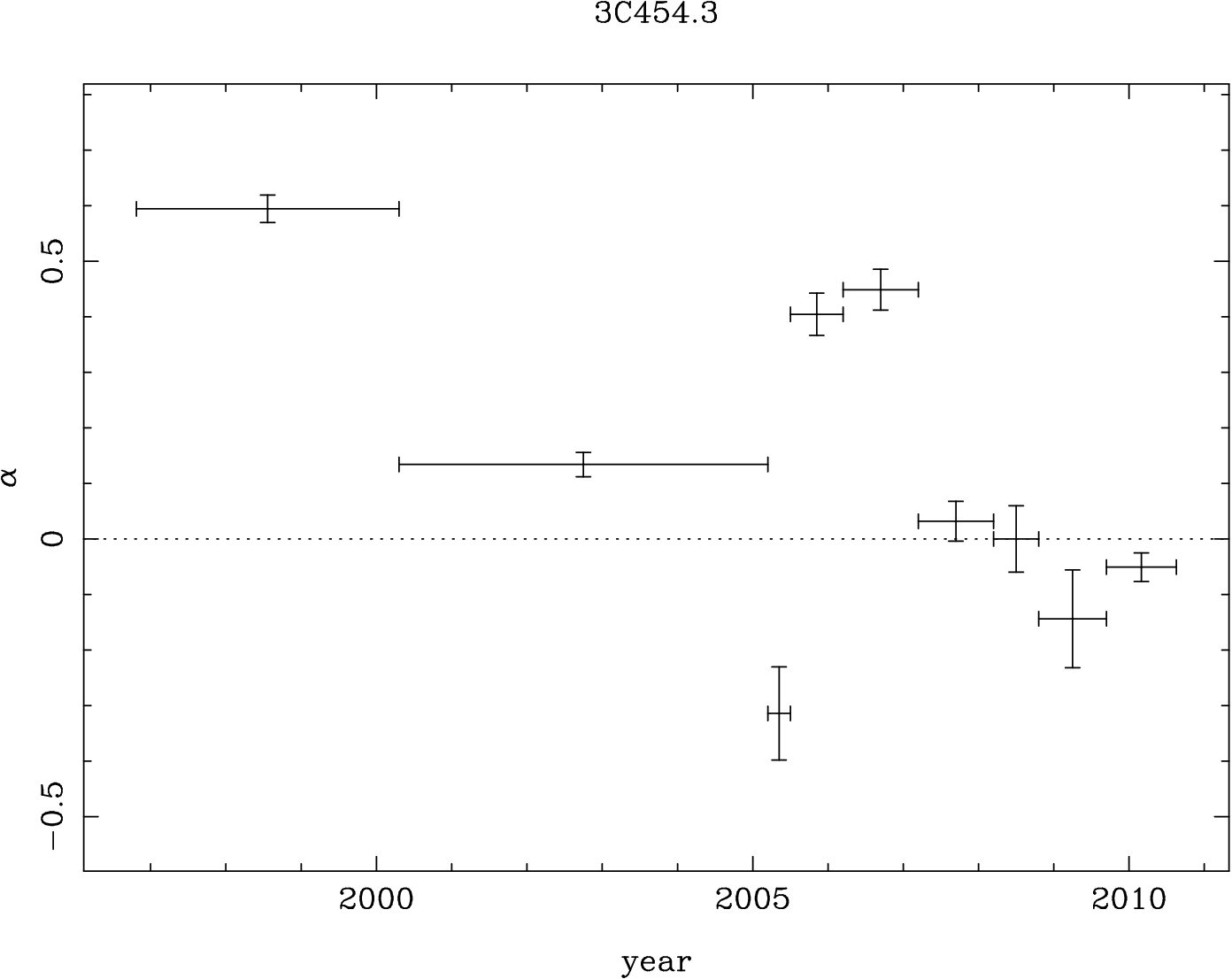}
\hspace{5mm}
\includegraphics[height=8.5cm,angle=-90]{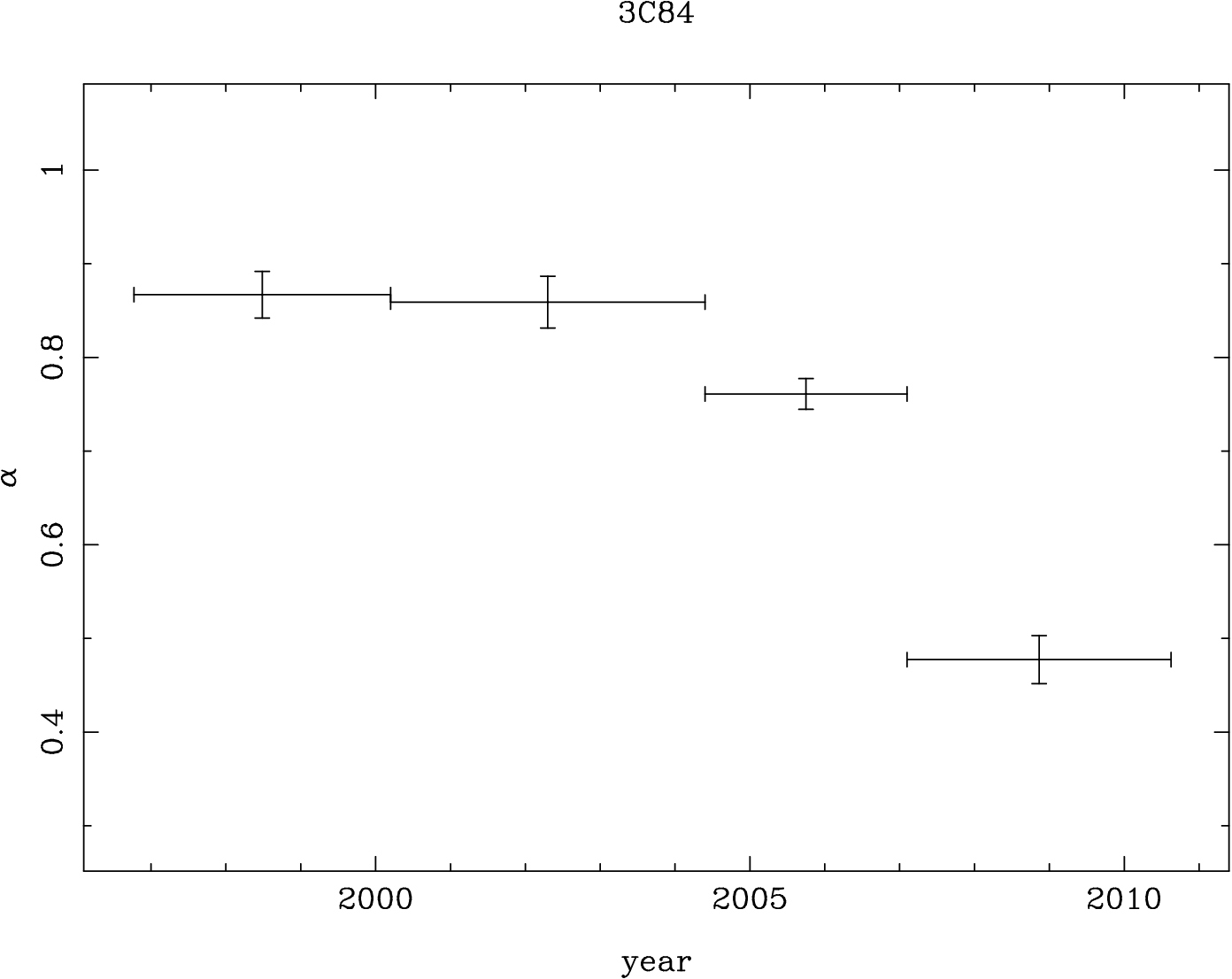}
\caption{Evolution of the spectral indices from 1996 to 2010. Please note the different axis scales. The spectral index $\alpha$ is defined via $S_{\nu}\propto\nu^{-\alpha}$. Errorbars along the time axes denote the bin sizes, error bars along the $\alpha$ axes denote the statistical $1\sigma$ errors. Horizontal dashed lines (where visible) mark the $\alpha=0$ lines. For each source we chose the bins such that each bin covers a phase of similar flux levels (quiescent phases, flares).}
\label{fig_specindex}
\end{figure*}

For each source we present the spectral index as function of time in Fig.~\ref{fig_specindex}. We selected the time windows such that a given window covers a phase of similar flux levels, meaning a ``quiet'' phase, an outburst, or a low-flux regime between two outbursts.

\begin{figure*}
\centering
\includegraphics[height=8.5cm,angle=-90]{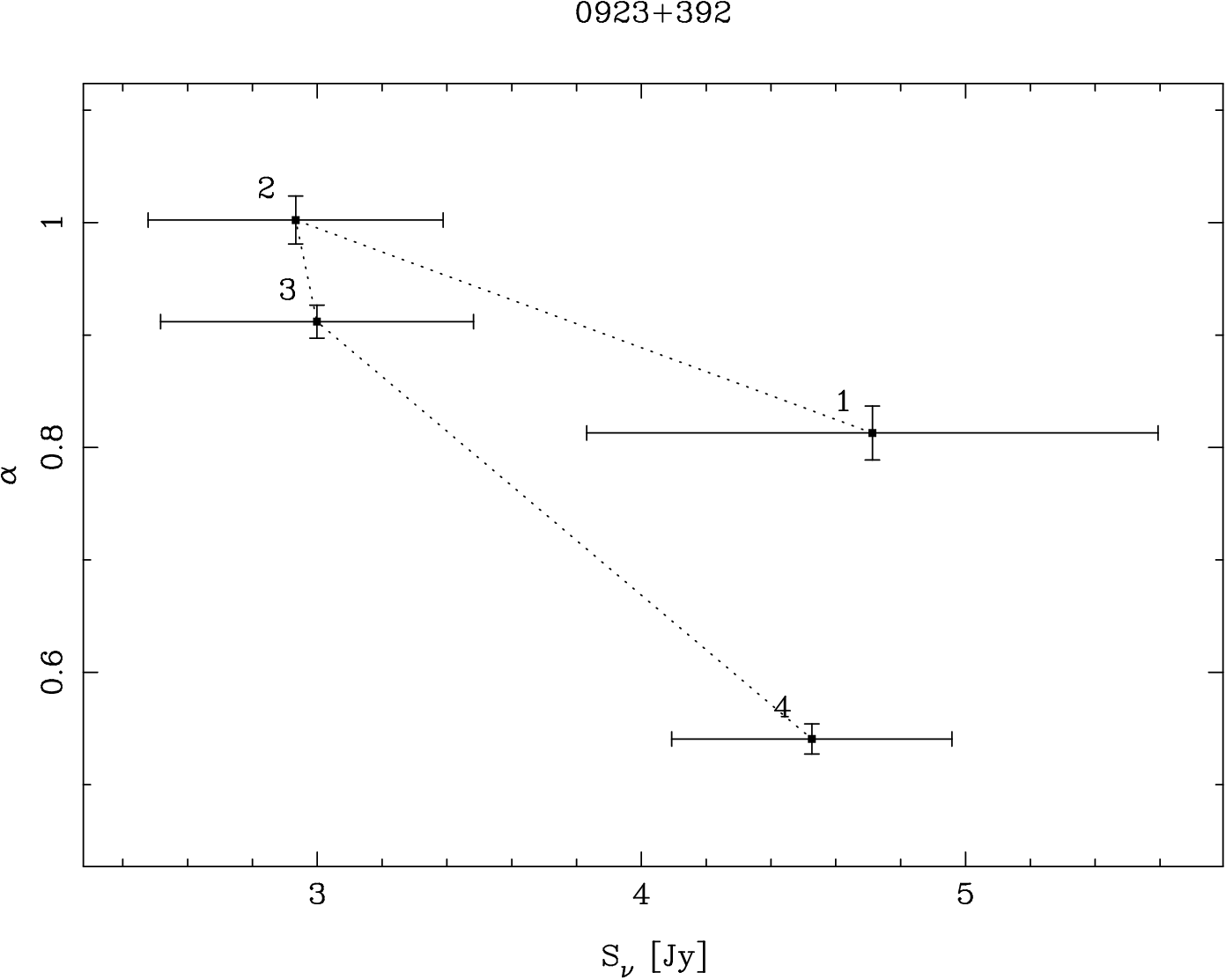}
\hspace{5mm}
\includegraphics[height=8.5cm,angle=-90]{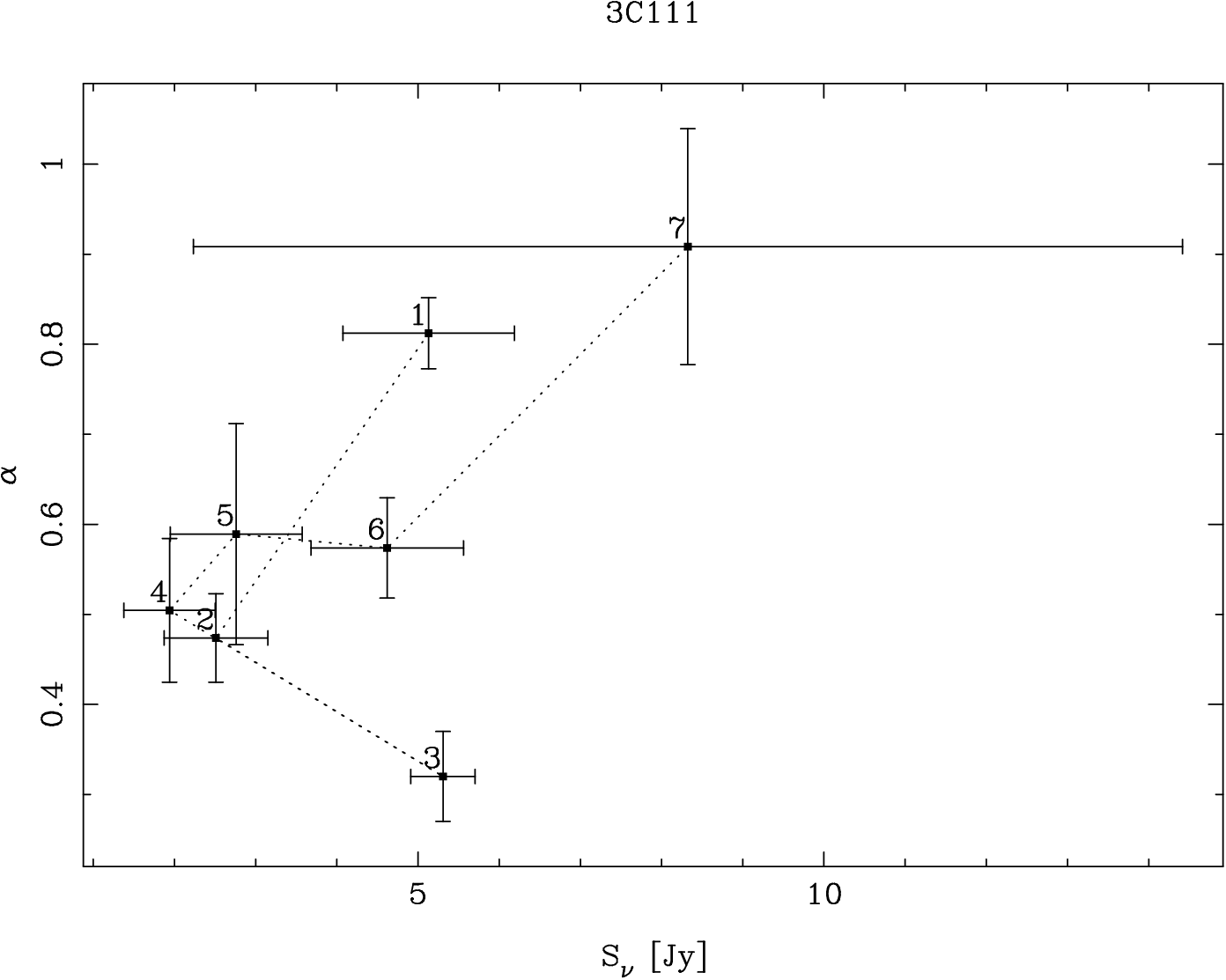} \\
\vspace{5mm}
\includegraphics[height=8.5cm,angle=-90]{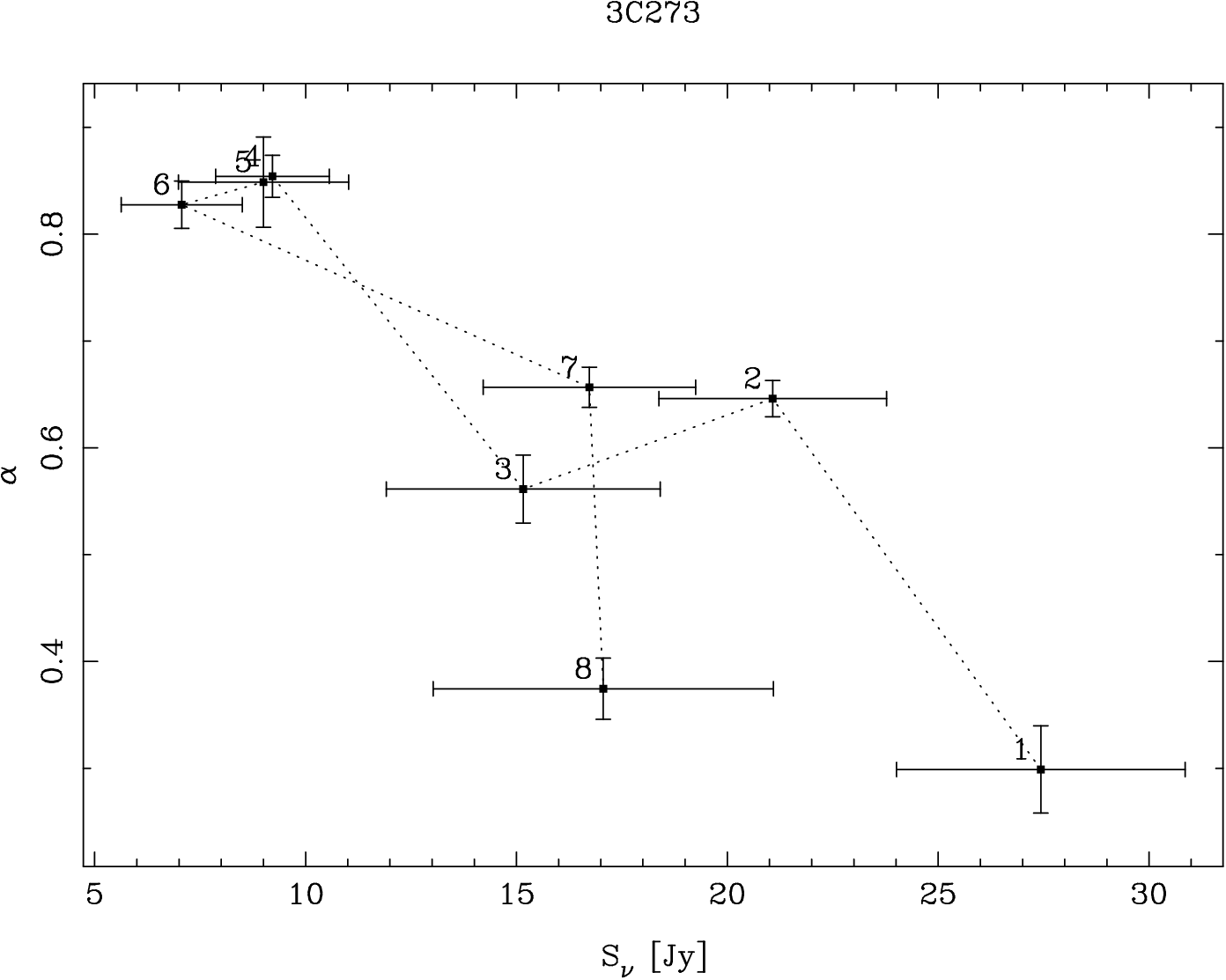}
\hspace{5mm}
\includegraphics[height=8.5cm,angle=-90]{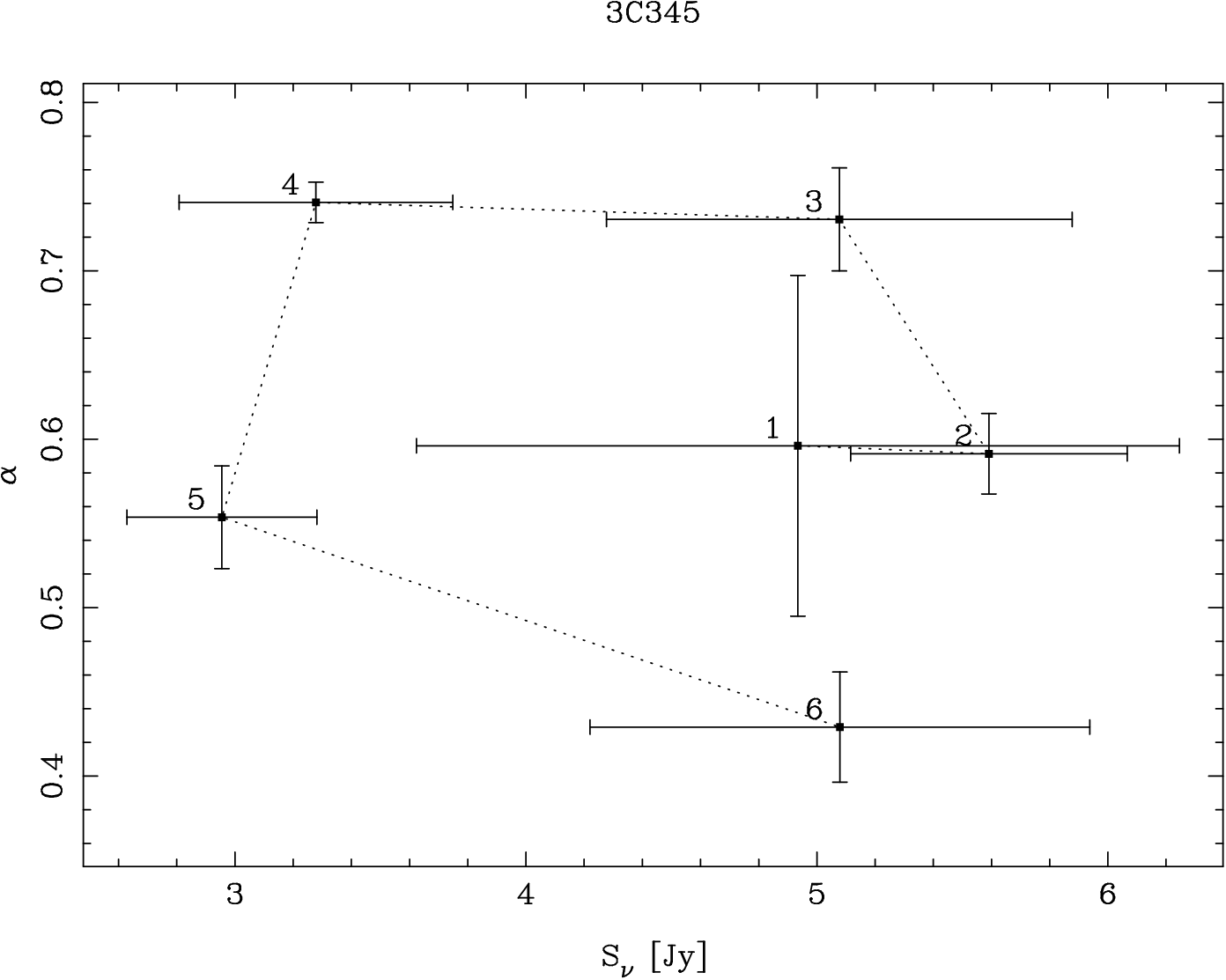} \\
\vspace{5mm}
\includegraphics[height=8.5cm,angle=-90]{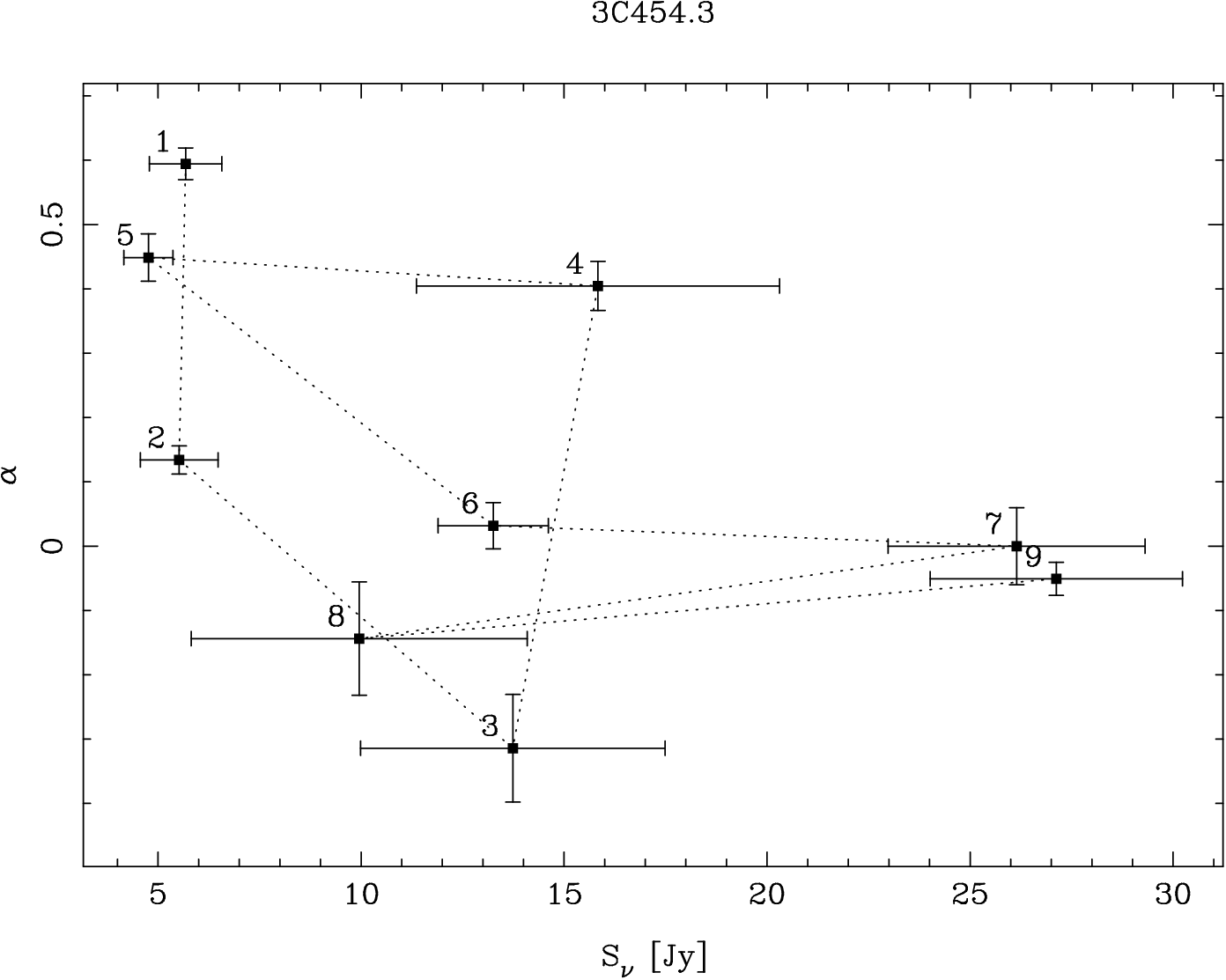}
\hspace{5mm}
\includegraphics[height=8.5cm,angle=-90]{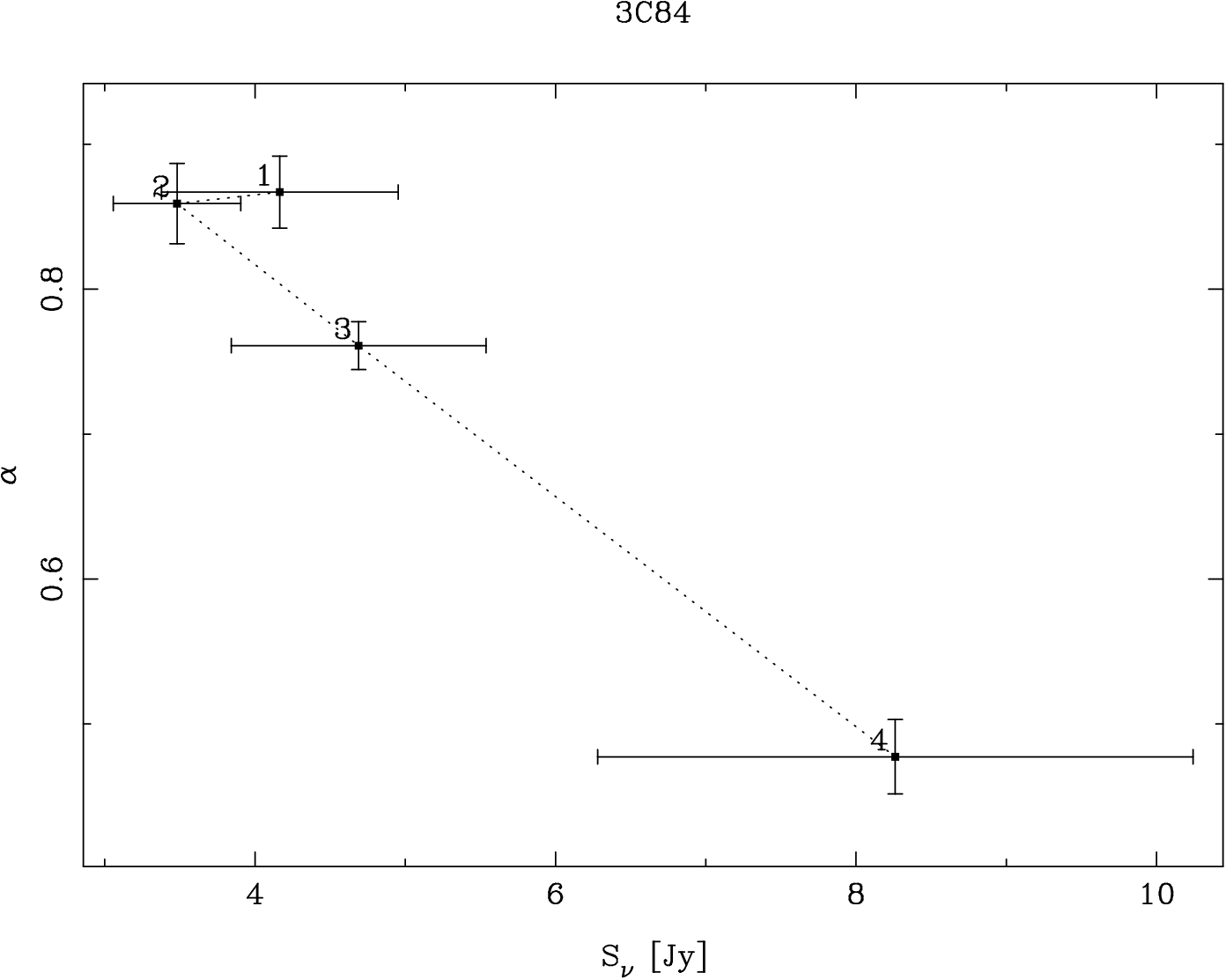}
\caption{Evolution of the spectral index $\alpha$ as function of the 3-mm flux. The data points are connected by dotted lines according to their order in time. Labels 1, 2, 3, ... mark the first, second, third, ... measurement in time. The time bins used here are those defined for Fig.~\ref{fig_specindex}. Error bars along the flux axis denote the flux variability (standard deviation), error bars along the $\alpha$ axis denote statistical $1\sigma$ errors.}
\label{fig_alpha_vs_flux}
\end{figure*}

In order to test possible correlations between spectral index and source flux, we present $\alpha$ as function of the 3-mm flux for each source in Fig.~\ref{fig_alpha_vs_flux}. For this diagram we use the same binning in time as for Fig.~\ref{fig_specindex}. In order to preserve some time information, we connect and label the data points according to their order in time.

\subsection{Power Spectra}

As the lightcurve of our sources show variability on many timescales, we performed a power spectrum analysis in order to quantify this variability. For each lightcurve we computed a normalized Scargle periodogram

\begin{equation}
A_{f} = \frac{1}{2\sigma^2}\cdot\left[\frac{\left(\sum_{j}S_j\cos 2\pi f t_j\right)^2}{\sum_{j}\cos^2 2\pi f t_j} + \frac{\left(\sum_{j}S_j\sin 2\pi f t_j\right)^2}{\sum_{j}\sin^2 2\pi f t_j}\right]
\label{eq_periodogram}
\end{equation}

\noindent
(Scargle \cite{scargle1982}) from our data. Here $A_f$ is the periodogram amplitude evaluated at frequency $f$, $S_j$ and $t_j$ are the flux value and the time of the $j$-th data point, respectively, and $\sigma^2$ denotes the variance of the data. The base frequencies are $f=f_{\rm min}, 2f_{\rm min}, 3f_{\rm min}, ..., f_{\rm max}$ with $f_{\rm min}=1/T$, $f_{\rm max}=N/(2T)$, $T$ being the total observation period, and $N$ being the number of data points. For the special case of regular sampling, $f_{\rm max}$ corresponds to the Nyquist frequency. The timelines and sampling of our observations (see Table~\ref{tab_journal}) lead to typical values of $f_{\rm min}\sim2\times10^{-4}$~days$^{-1}$ and $f_{\rm max}\sim0.2$~days$^{-1}$. Before calculating a periodogram we subtracted the average source flux from the data in order to remove the zero-frequency power. In order to exclude systematic errors due to spectral effects, we performed the periodogram analysis separately for the 1.3-mm and 3-mm bands of each source (for the 2-mm band the amount of data is insufficient for a comparison to the other bands).

The power spectra we find all show a global decrease of spectral power with increasing frequency, i.e. they correspond to red-noise spectra. In order to quantify this behaviour, we derived powerlaw indices by fitting straight lines to the data in log-space. The spectral index $\beta$ of the power spectra is defined via $A_f\propto f^{-\beta}$.

\begin{figure*}
\centering
\includegraphics[height=8.5cm,angle=-90]{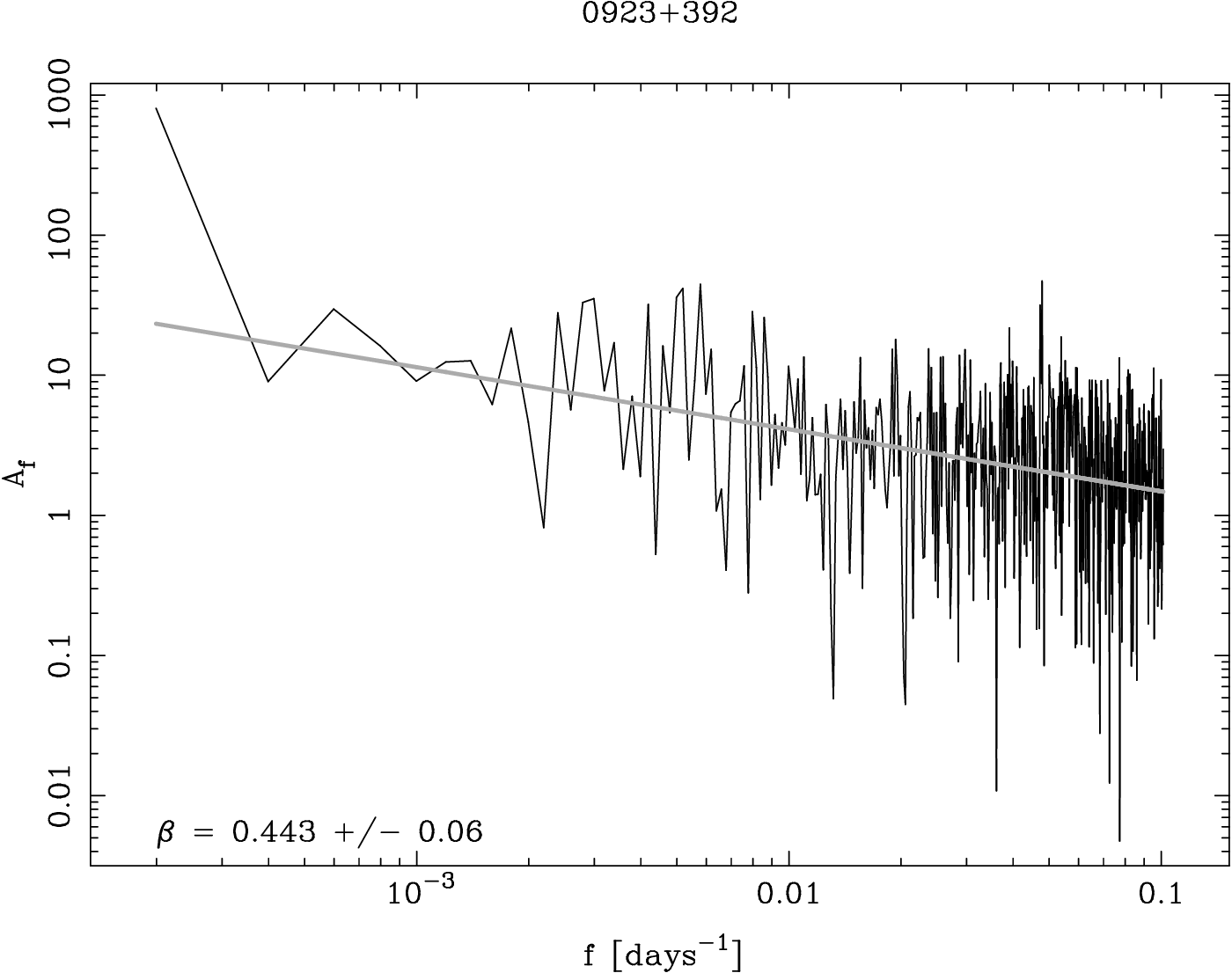}
\hspace{5mm}
\includegraphics[height=8.5cm,angle=-90]{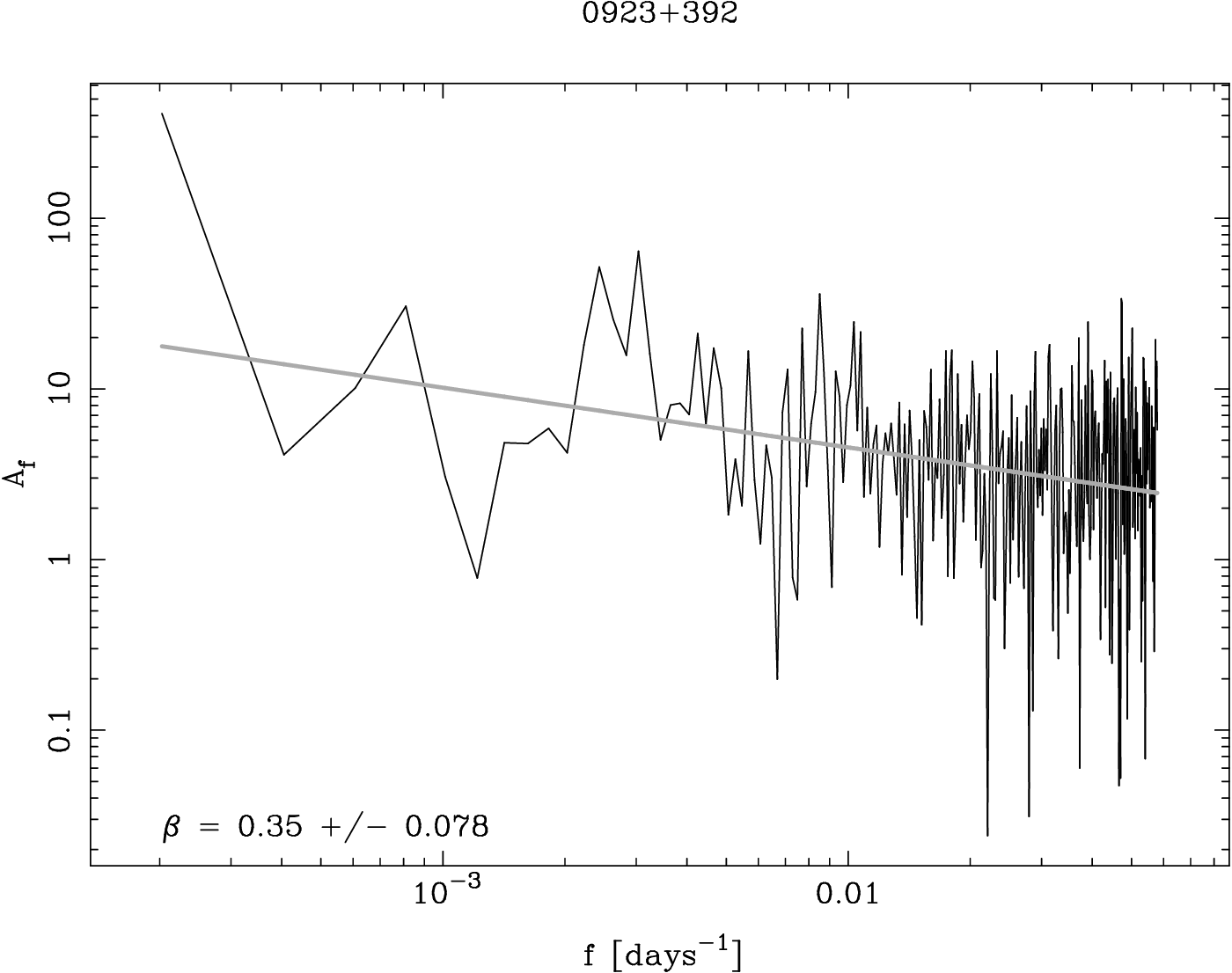} \\
\vspace{5mm}
\includegraphics[height=8.5cm,angle=-90]{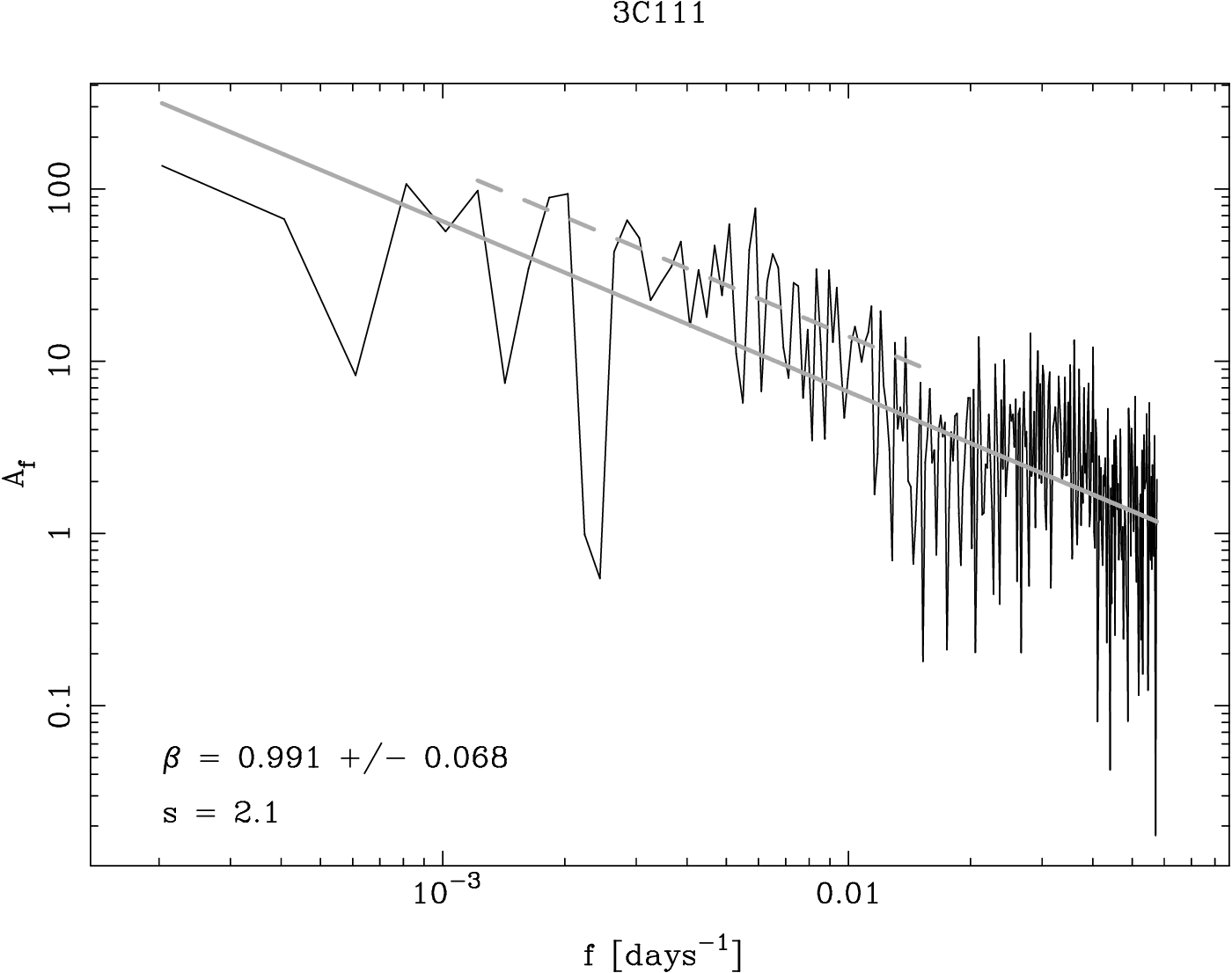}
\hspace{5mm}
\includegraphics[height=8.5cm,angle=-90]{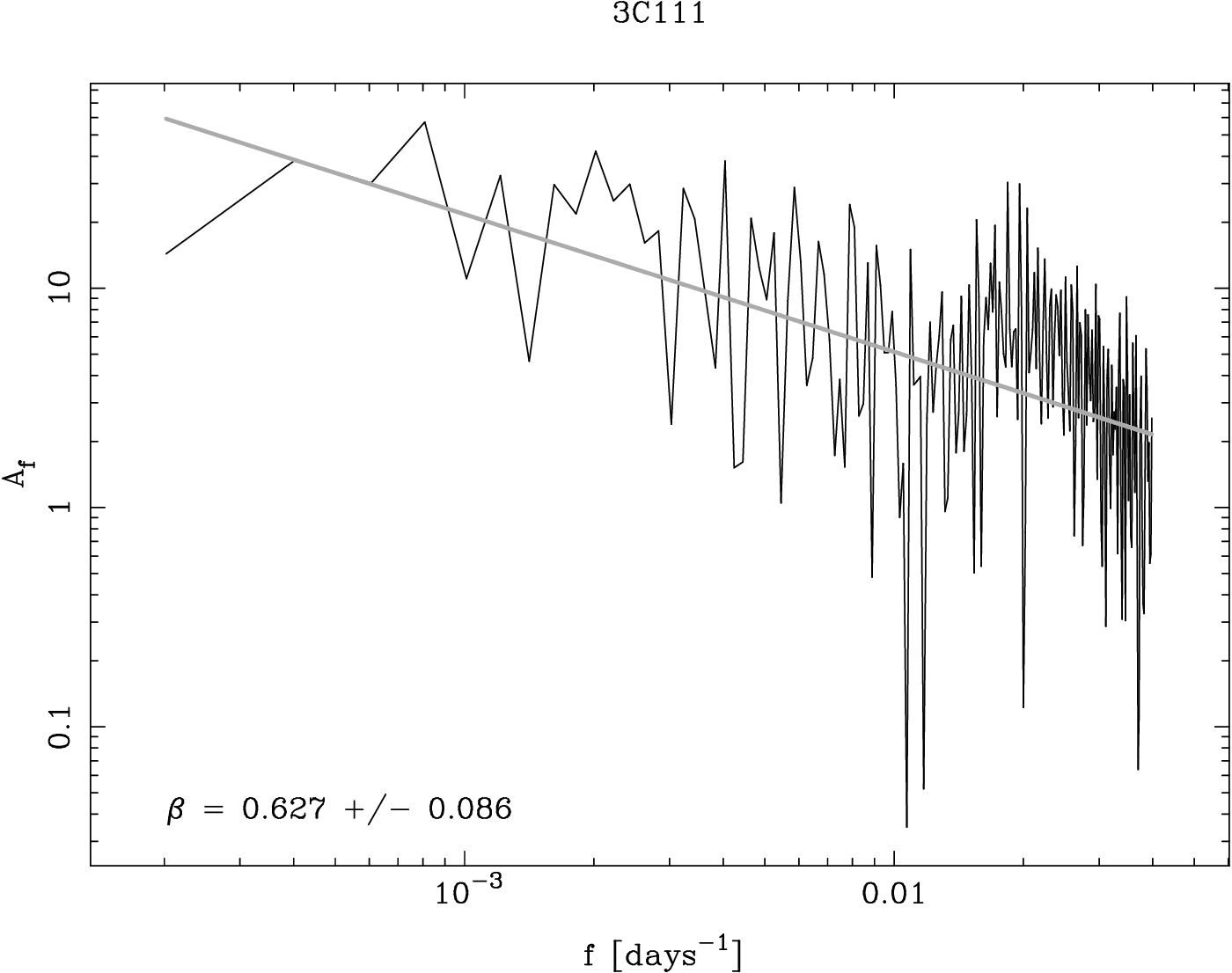} \\
\vspace{5mm}
\includegraphics[height=8.5cm,angle=-90]{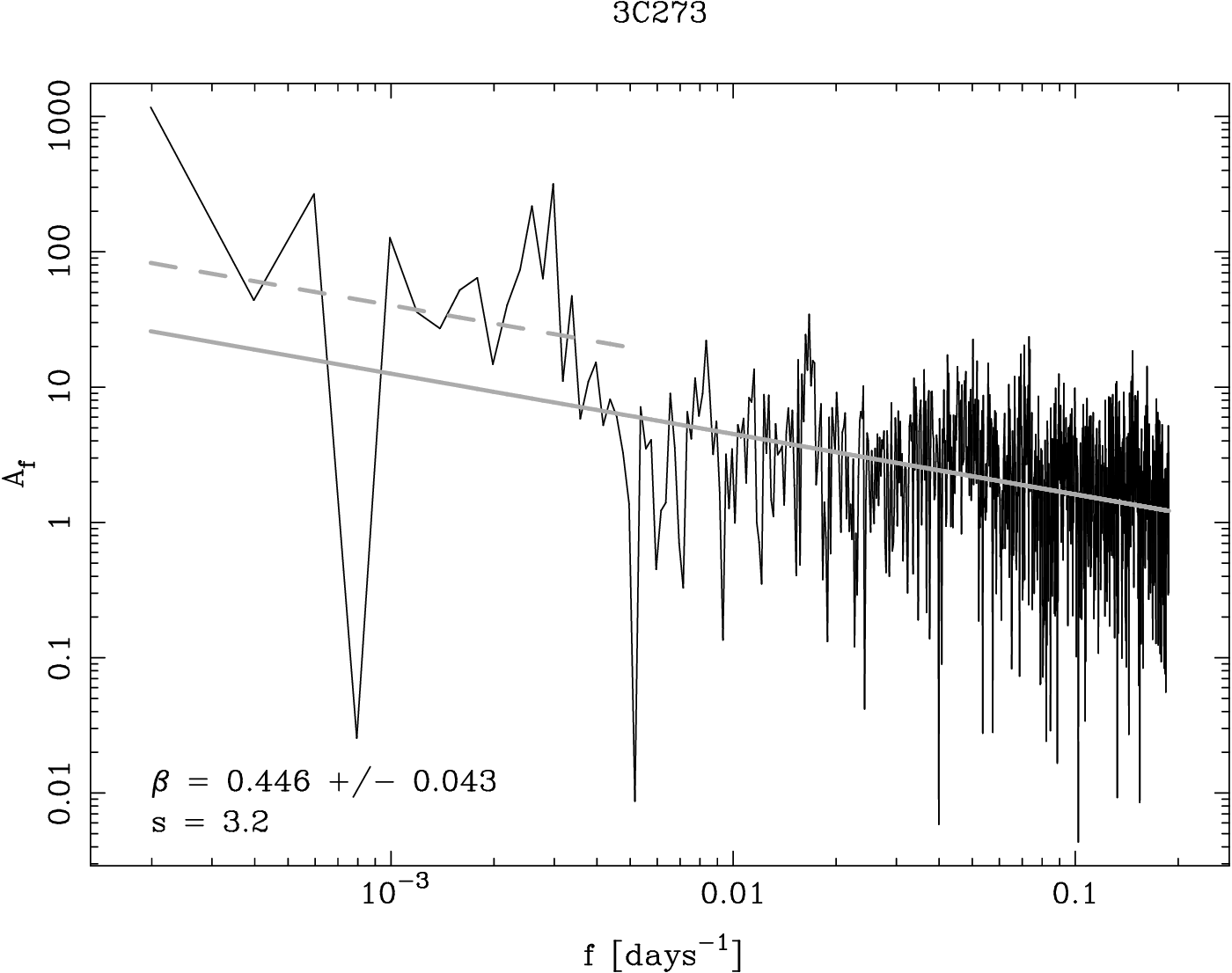}
\hspace{5mm}
\includegraphics[height=8.5cm,angle=-90]{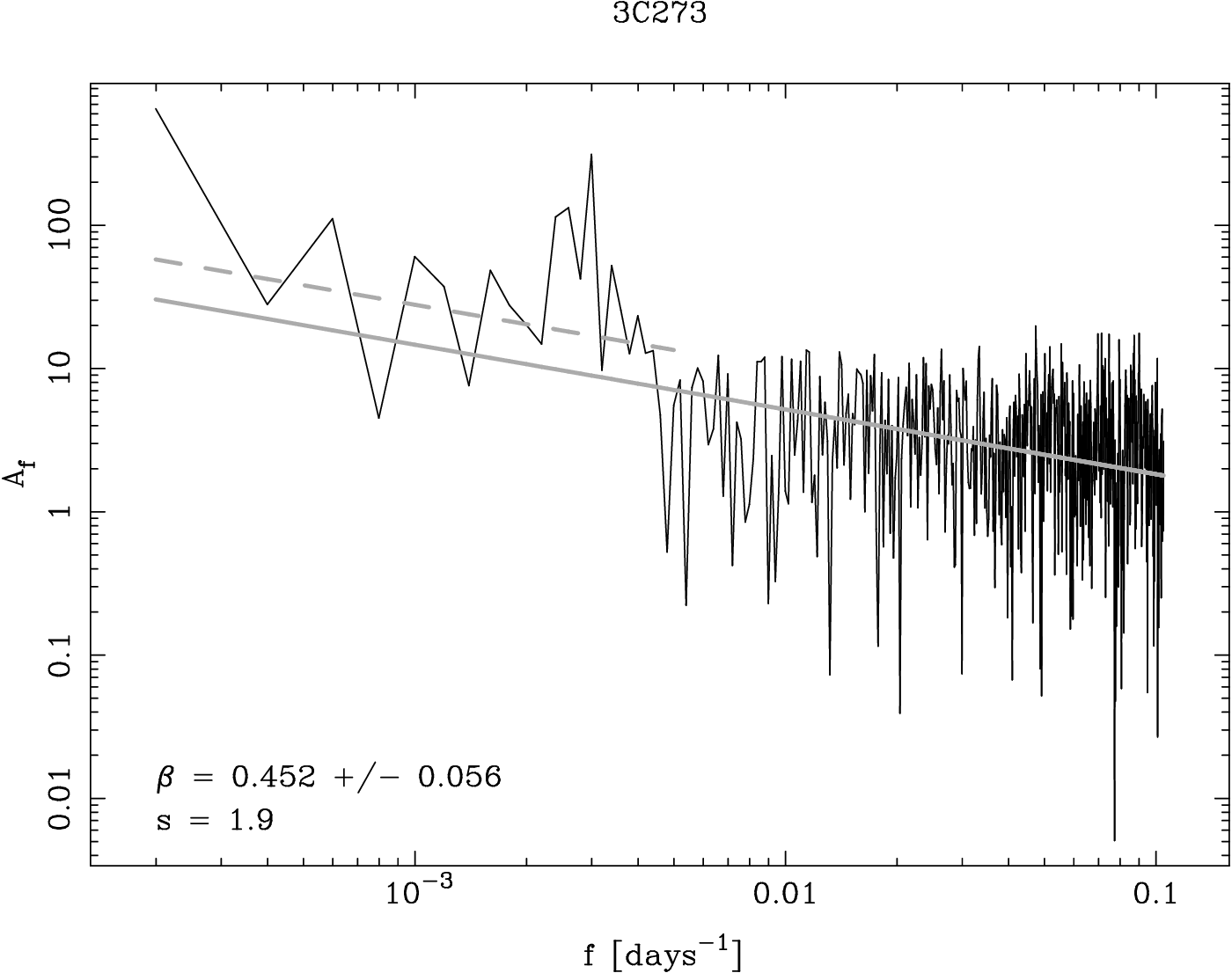}
\caption{Scargle periodograms (spectral power $A_f$ vs. frequency $f$) of the 3-mm (left column) and 1.3-mm band (right column) lightcurves of 0923+392, 3C~111, and 3C~273. Please note the different axis scales. The black curves denote the measured power spectra, the continuous grey lines marks the best-fitting powerlaws. The spectral index $\beta$ of the power spectra is defined via $A_f\propto f^{-\beta}$. Dashed grey lines (where present) correspond to the best-fiting powerlaws (continuous grey lines) upscaled by factors $s$ as given in the corresponding plots.}
\label{fig_scargle1}
\end{figure*}

\begin{figure*}
\centering
\includegraphics[height=8.5cm,angle=-90]{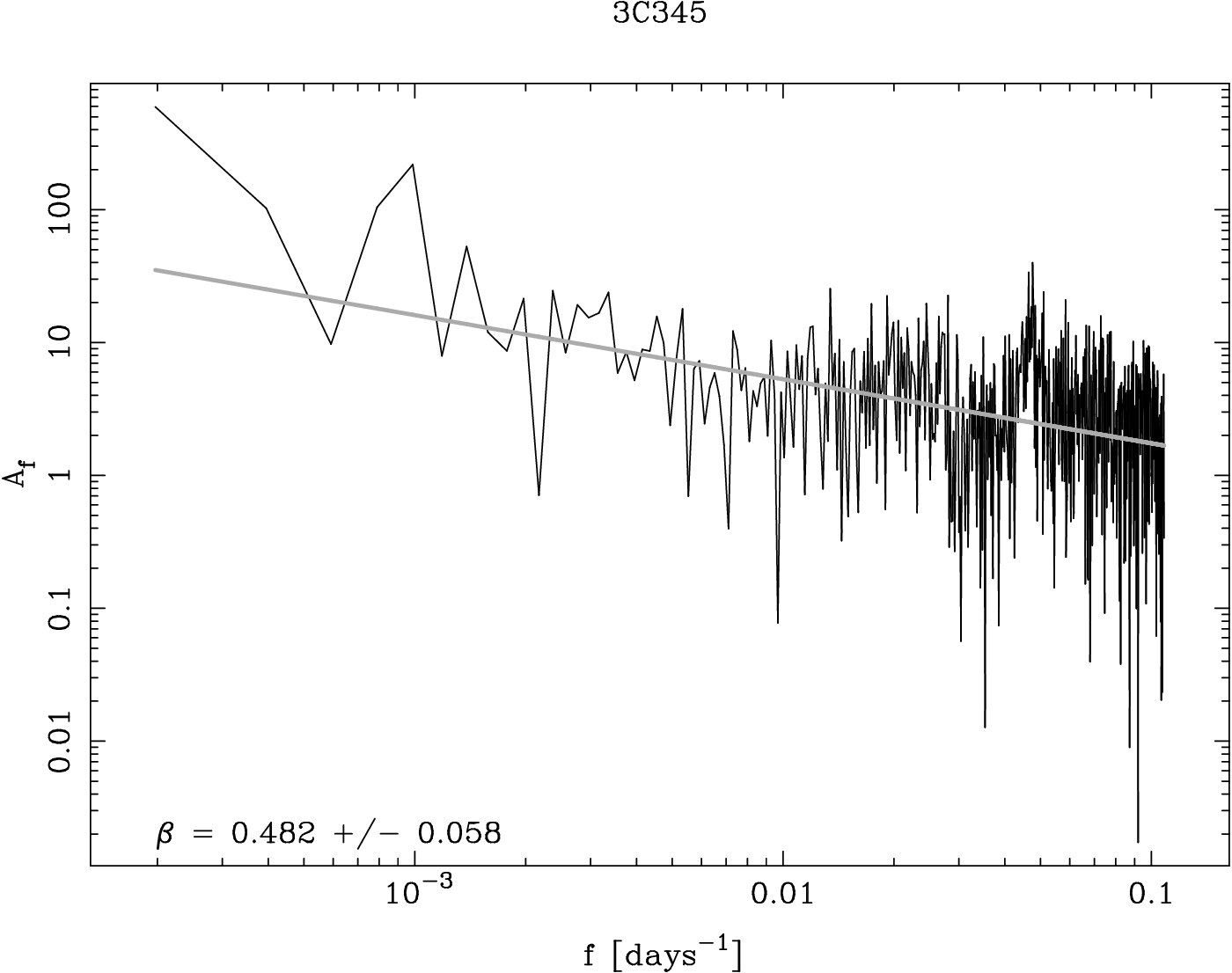}
\hspace{5mm}
\includegraphics[height=8.5cm,angle=-90]{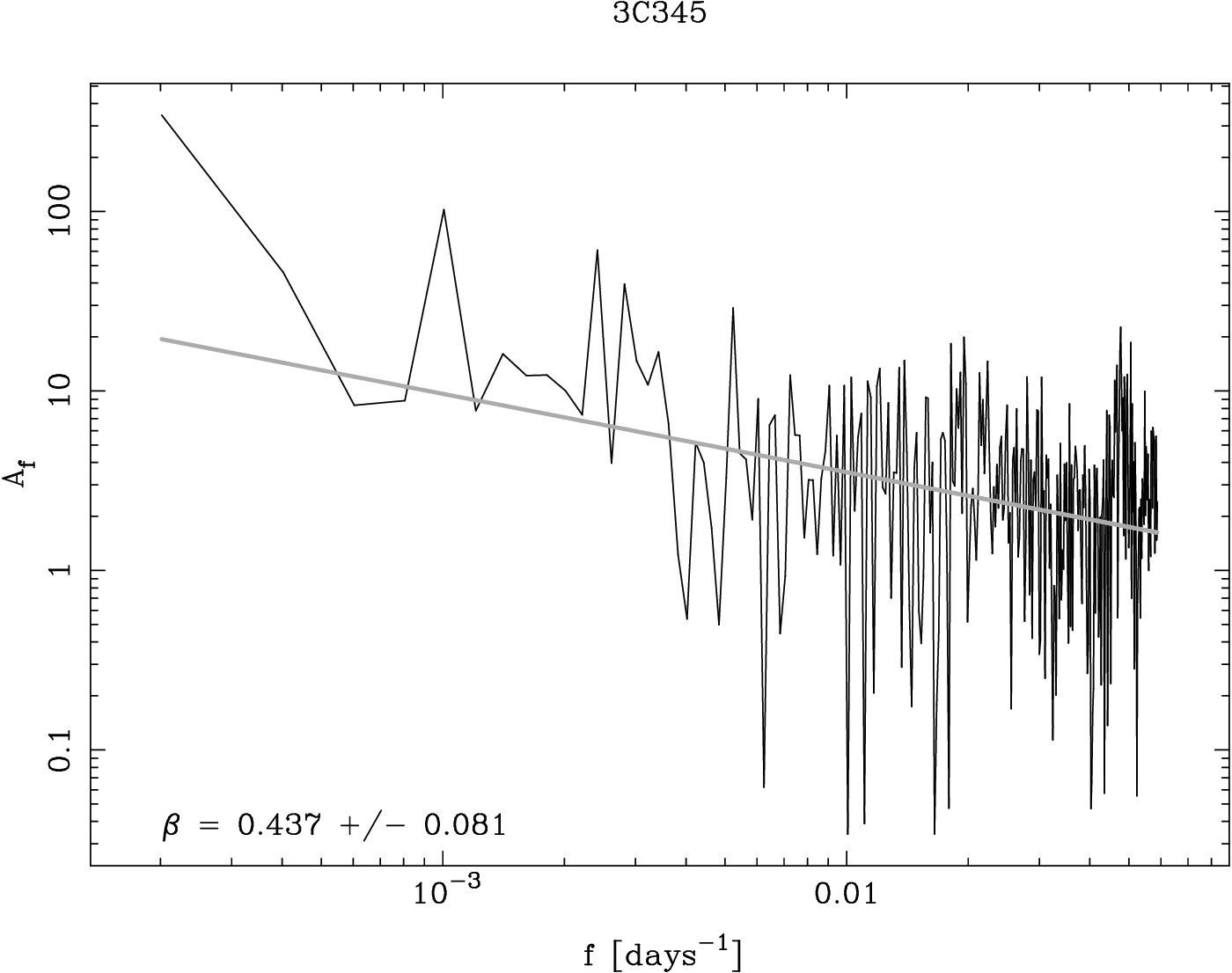} \\
\vspace{5mm}
\includegraphics[height=8.5cm,angle=-90]{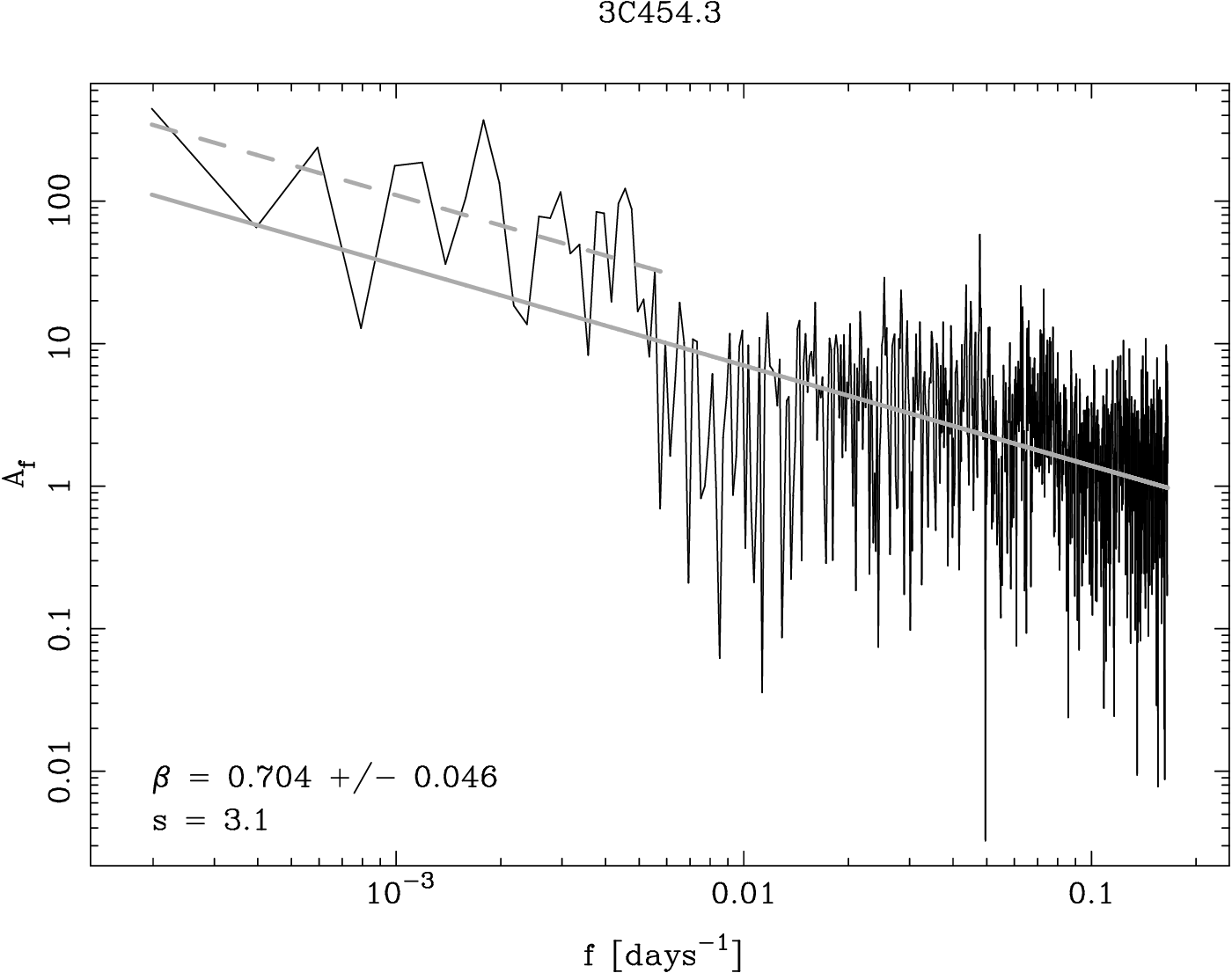}
\hspace{5mm}
\includegraphics[height=8.5cm,angle=-90]{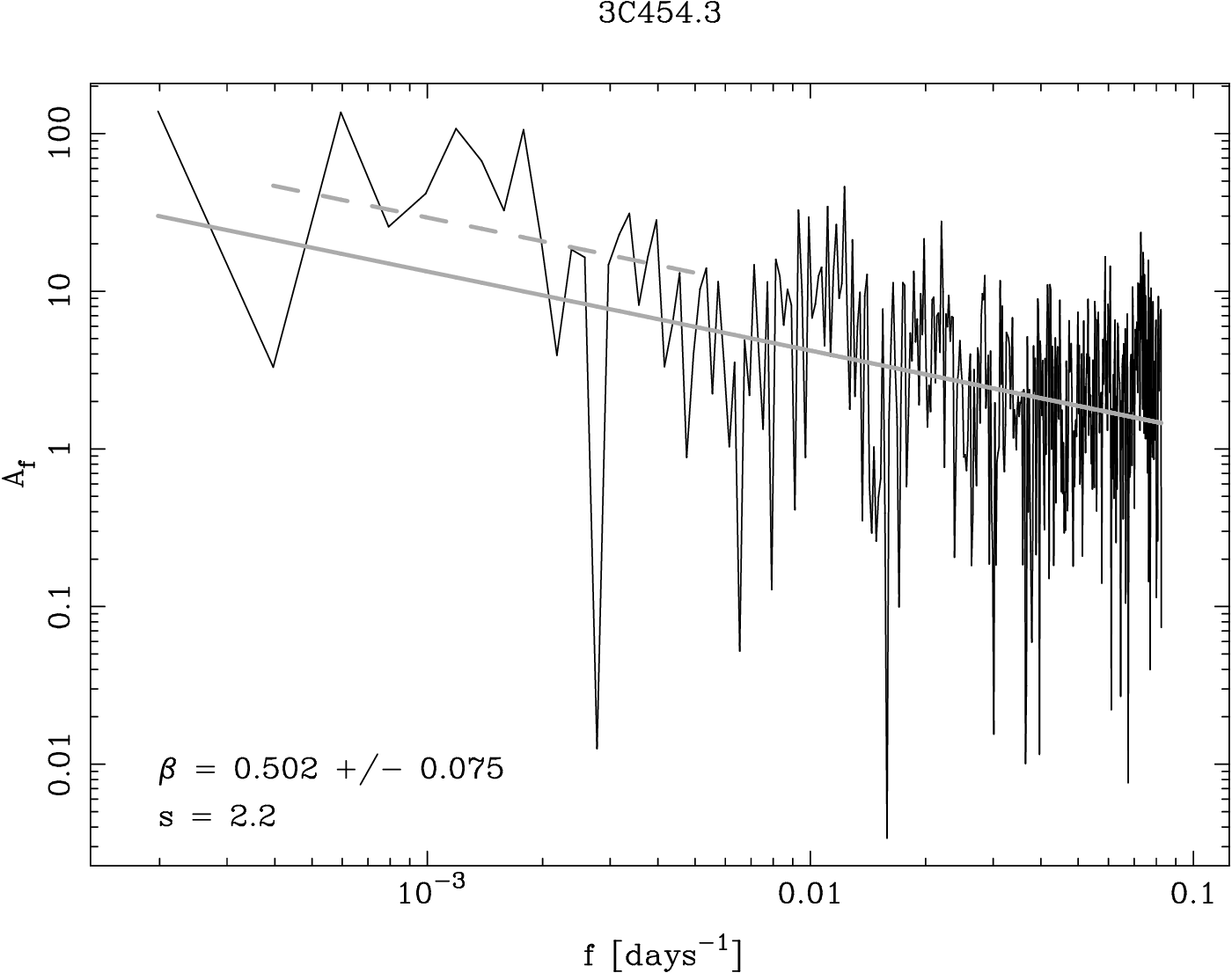} \\
\vspace{5mm}
\includegraphics[height=8.5cm,angle=-90]{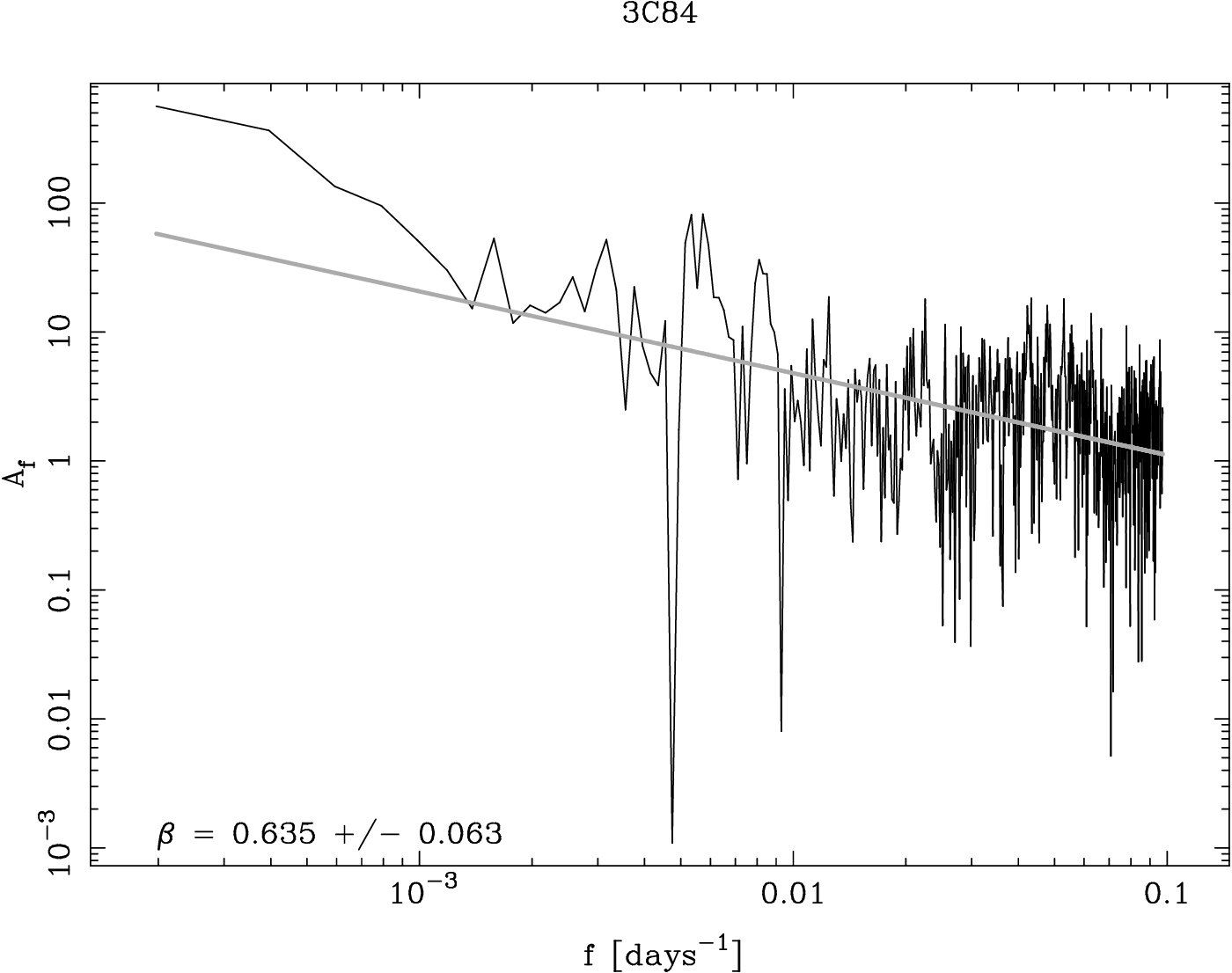}
\hspace{5mm}
\includegraphics[height=8.5cm,angle=-90]{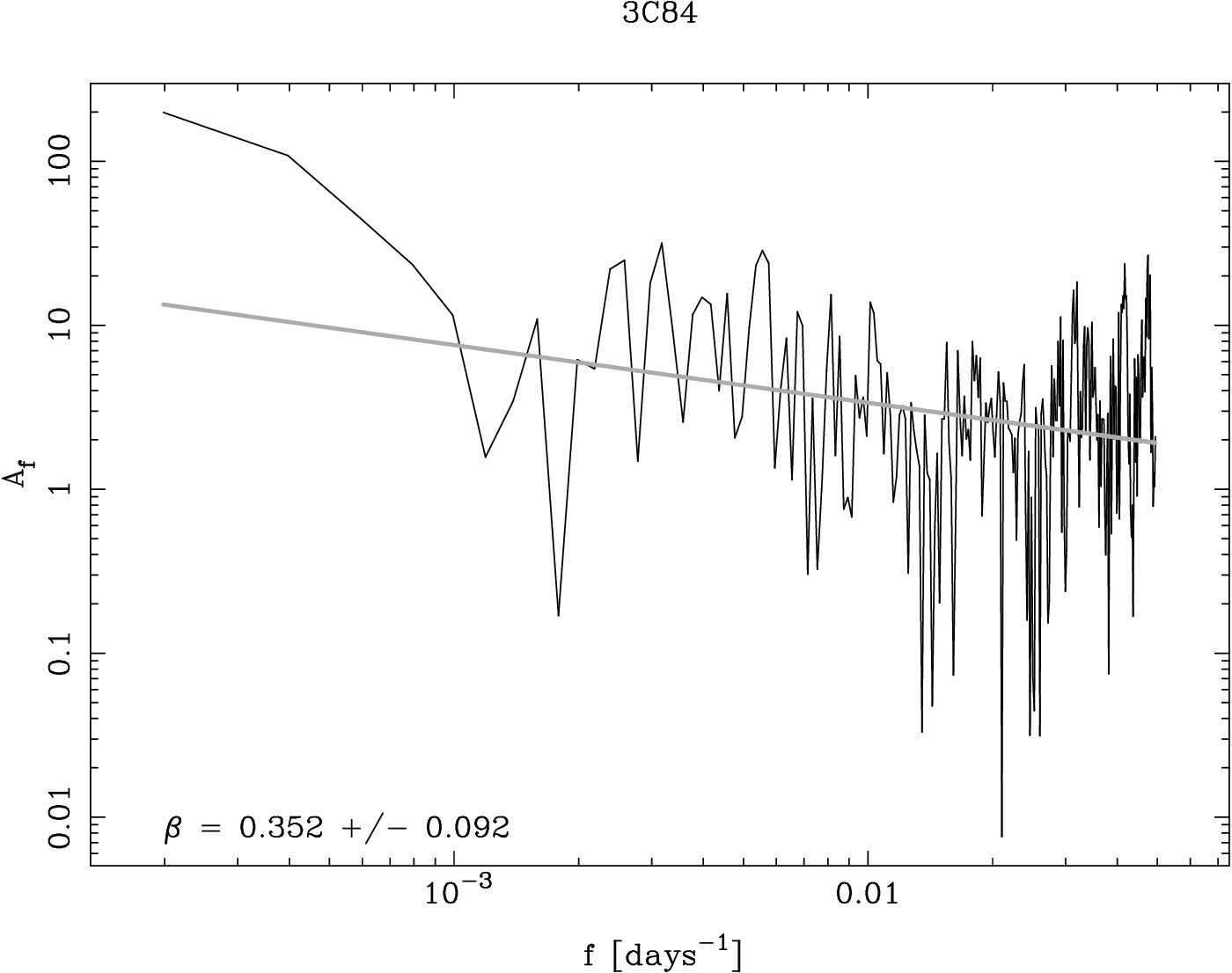}
\caption{Same as Fig.~\ref{fig_scargle1} for 3C~345, 3C~454.3, and 3C~84.}
\label{fig_scargle2}
\end{figure*}

In several cases we find that the low-frequency ends of the power spectra follow the average powerlaw slopes but are located systematically above the best-fitting powerlaw models. In those cases we approximated the affected spectral ranges as upscaled (by factors $\approx$2--3) versions of the overall power spectra. We present our results in Figs.~\ref{fig_scargle1} and \ref{fig_scargle2}.

\subsection{Time Offsets among Spectral Bands}

The long timeline of observations and the good time sampling of our flux data make it possible to test if the lightcurves of the various spectral bands are simultaneous or if time lags are present. In order to quantify the presence or absence of time offsets, we computed for each source a discrete correlation function (DCF) according to the scheme proposed by Edelson \& Krolik (\cite{edelson1988}). For two discrete datasets $\{a_i\}$, $\{b_j\}$ with averages $\bar a$, $\bar b$ one computes the {\it unbinned} discrete correlations

\begin{equation}
UDCF_{ij}(\Delta t_{ij}) = \frac{(a_i-\bar a)(b_j - \bar b)}{\left[\left(\sigma^2_a - \delta^2_a\right)\left(\sigma^2_b - \delta^2_b\right)\right]^{1/2}}
\label{eq_udcf}
\end{equation}

\noindent
for all pairs $(a_i, b_j)$ with time differences $\Delta t_{ij}$. The $\sigma^2_{a,b}$ are the variances, the $\delta_{a,b}$ are the mean measurement errors of $\{a_i\}$, $\{b_j\}$. Accordingly, the terms $\sigma^2_{a,b} - \delta^2_{a,b}$ are the variances corrected for artificial broadening due to finite measurement errors; this is necessary in order to preserve the correct normalization in noisy data. The actual DCF for a time offset $\tau$ is obtained via averaging over all $N$ $UDCF_{ij}(\Delta t_{ij})$ for which $\Delta t_{ij}$ is located within a selected bin $\Delta\tau$ (i.e., $\tau-\Delta\tau/2 \leq \Delta t_{ij} < \tau+\Delta\tau/2$):

\begin{equation}
DCF(\tau) = \frac{1}{N}\sum_{\tau-\Delta\tau/2}^{\tau+\Delta\tau/2} UDCF_{ij}(\Delta t_{ij}) 
\label{eq_dcf}
\end{equation}

\noindent
By definition, $DCF(\tau)$ is located in the range $[-1,+1]$, with $(-1)$ +1 indicating perfect (anti-)correlation and 0 corresponding to the absence of any correlation. The position of the maximum of the DCF indicates the time offset between the lightcurves. The statistical uncertainty of $DCF(\tau)$ is given by the standard error of mean

\begin{equation}
\sigma_{DCF}(\tau) = \frac{1}{N-1}\left[\sum_{\tau-\Delta\tau/2}^{\tau+\Delta\tau/2}\left(UDCF_{ij}-DCF(\tau)\right)^2\right]^{1/2} 
\label{eq_errdcf}
\end{equation}

For each of our six target AGN, we examined ranges of $\tau$ sufficient for covering the largest time offsets to be expected, either $\tau\pm365$~days or $\tau\pm730$~days. In the special case of 3C~454.3, we additionally probed two subsets of data with $\tau\pm180$~days (see Sect.~4.5). In order to preserve high time resolution while excluding the risk of creating oversampling artefacts, we used sampling times $\Delta\tau=T_{\rm 1mm}/N_{\rm 1mm}$, with $T_{\rm 1mm}$ ($N_{\rm 1mm}$) being the total observing time (number of measurements) obtained in the 1.3-mm band which is for all sources the wavelength band with the sparser sampling (see Table~\ref{tab_journal}). We note that the \emph{effective} resolution depends only weakly on the choice of bin sizes: smaller $\Delta\tau$ increase the formal resolution but also increase the errors as they cover smaller numbers of $UDCF_{ij}$ per bin (see Eq.~\ref{eq_errdcf}), and vice versa.

\begin{figure*}
\centering
\includegraphics[height=8.5cm,angle=-90]{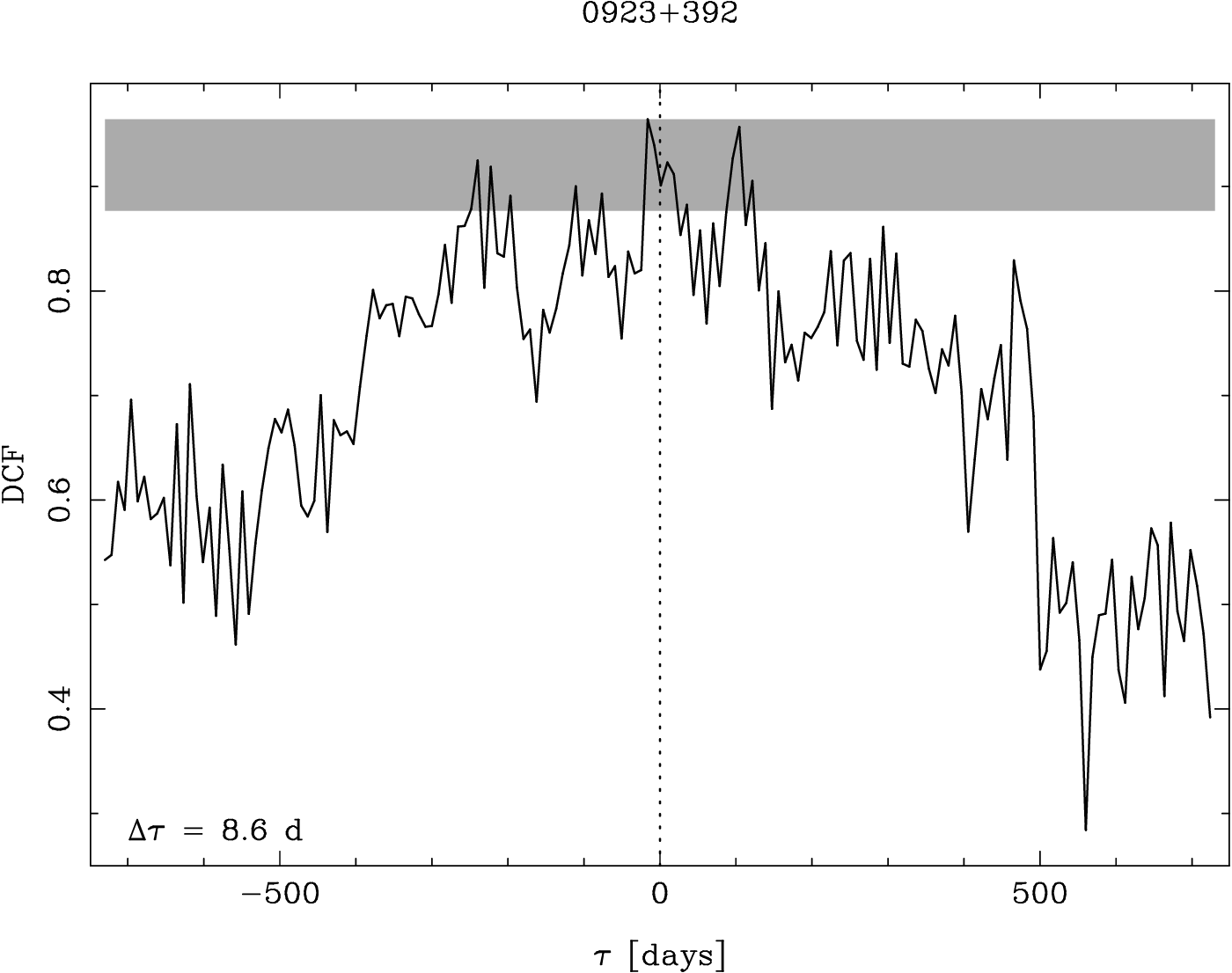}
\hspace{5mm}
\includegraphics[height=8.5cm,angle=-90]{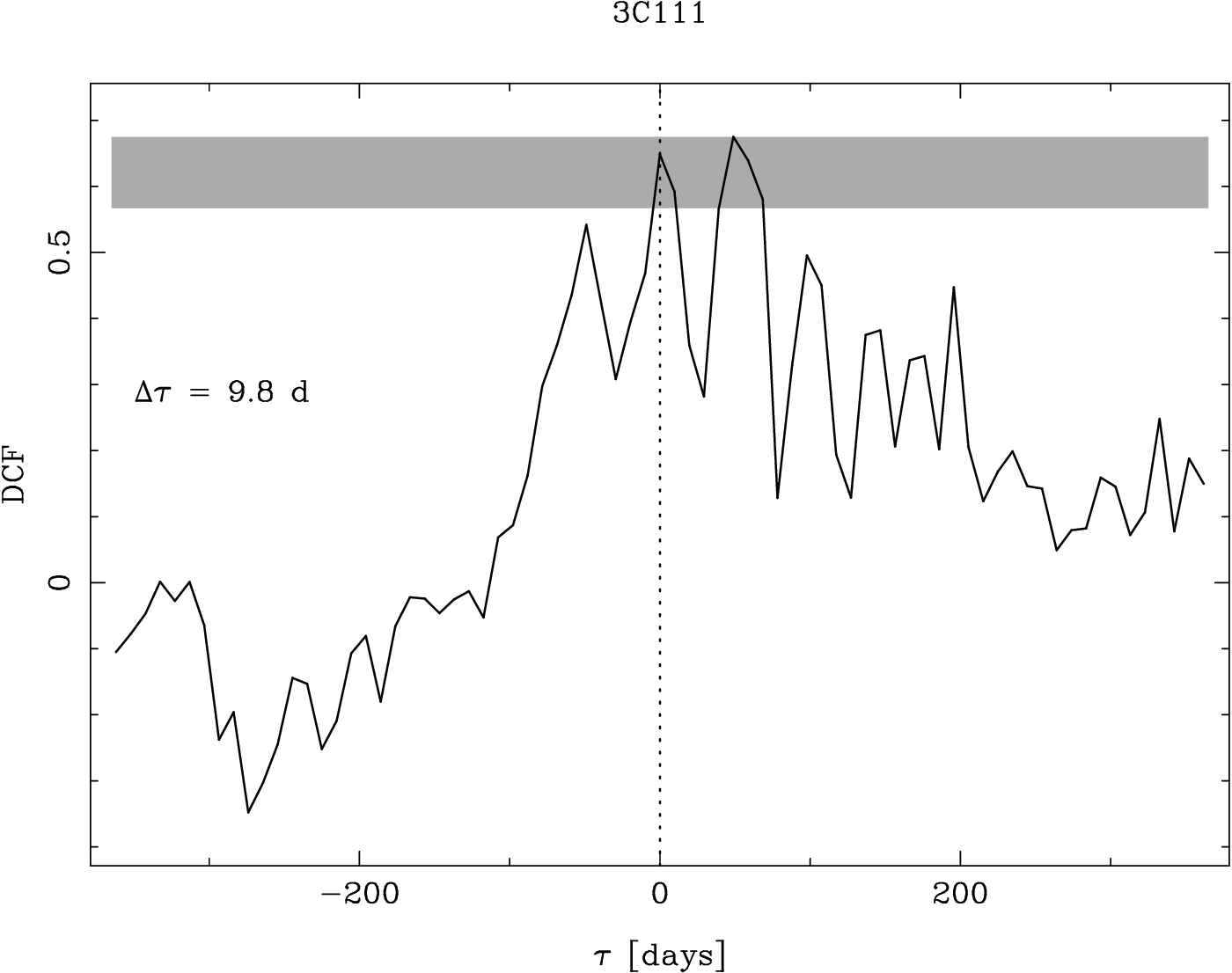} \\
\vspace{5mm}
\includegraphics[height=8.5cm,angle=-90]{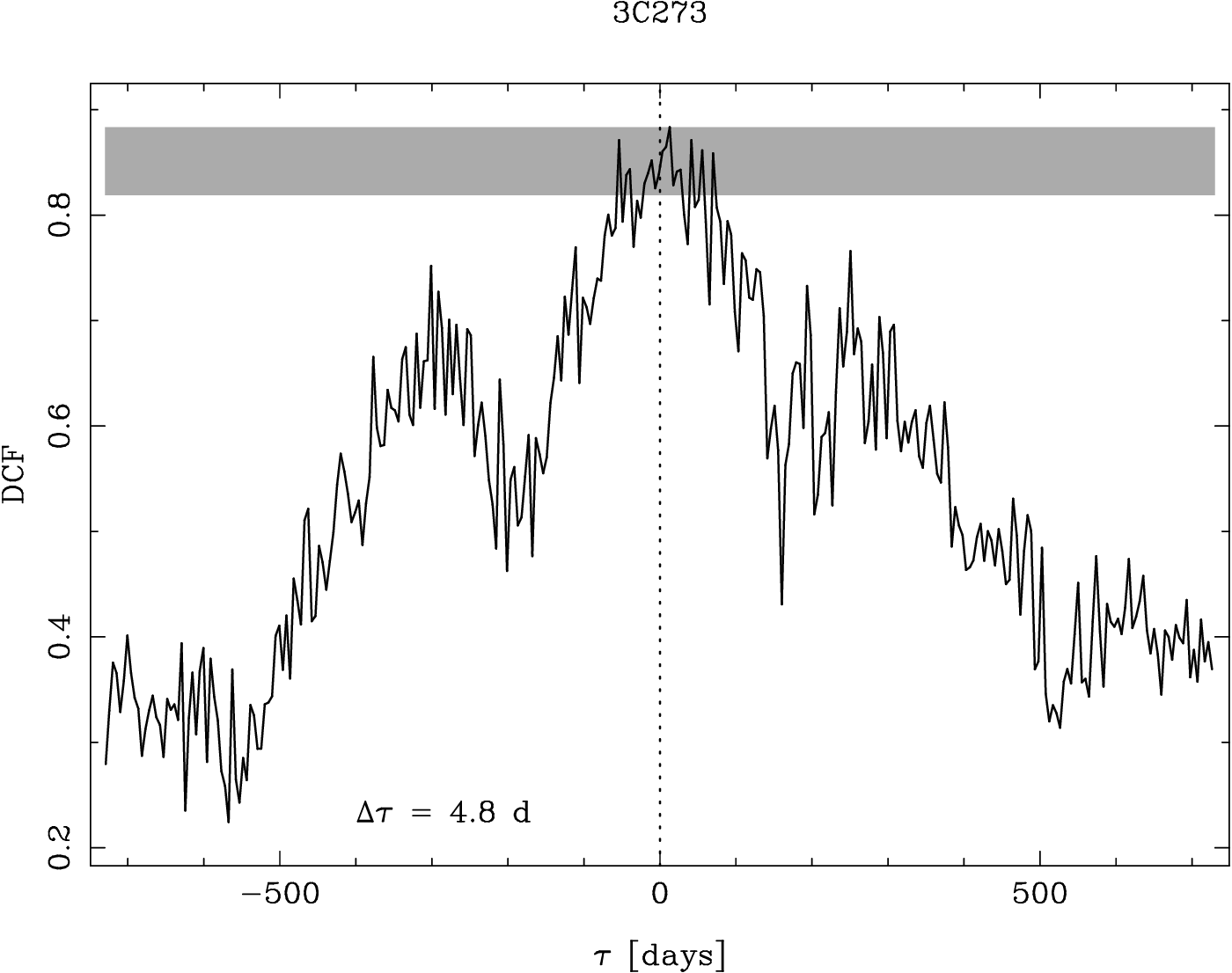}
\hspace{5mm}
\includegraphics[height=8.5cm,angle=-90]{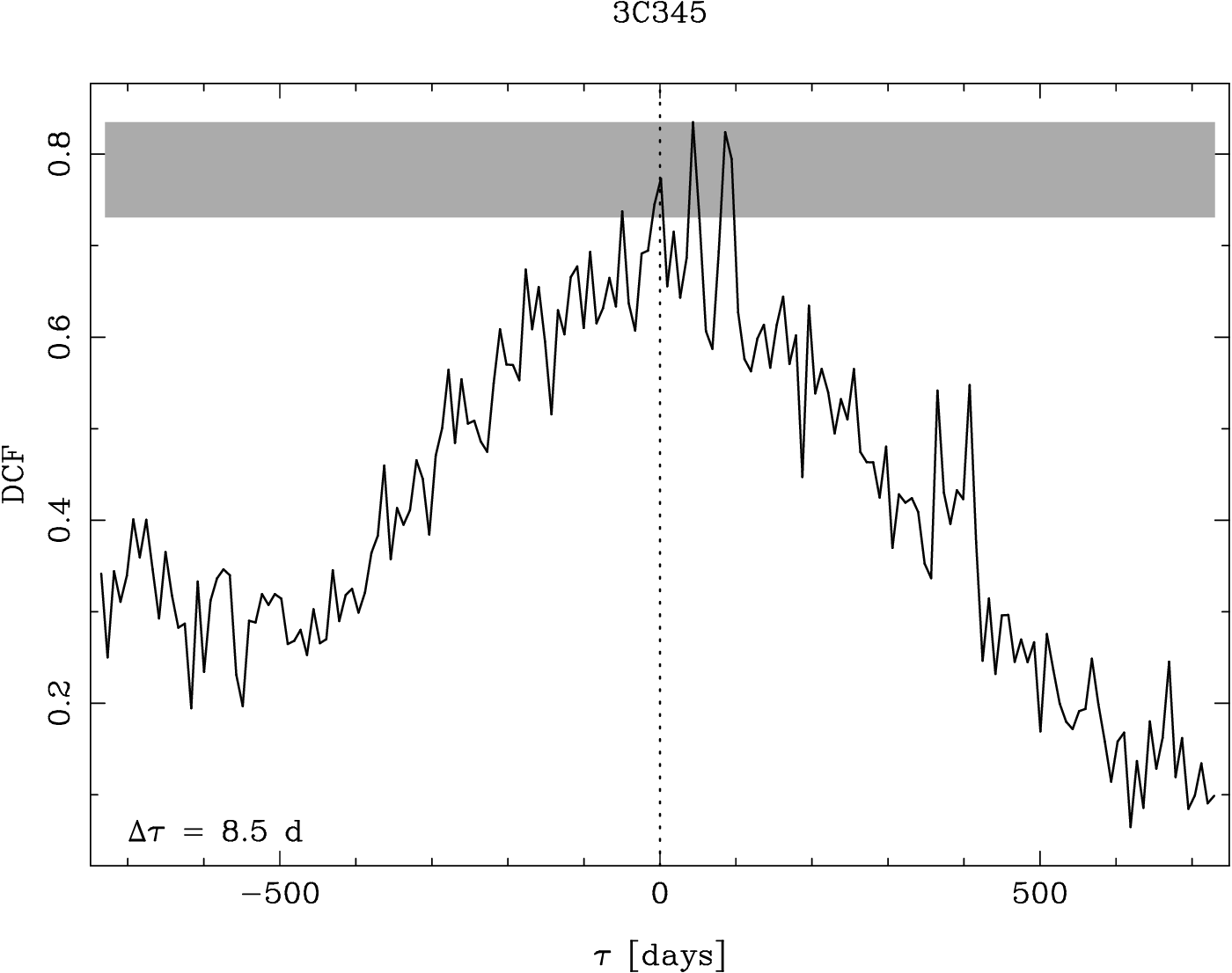} \\
\vspace{5mm}
\includegraphics[height=8.5cm,angle=-90]{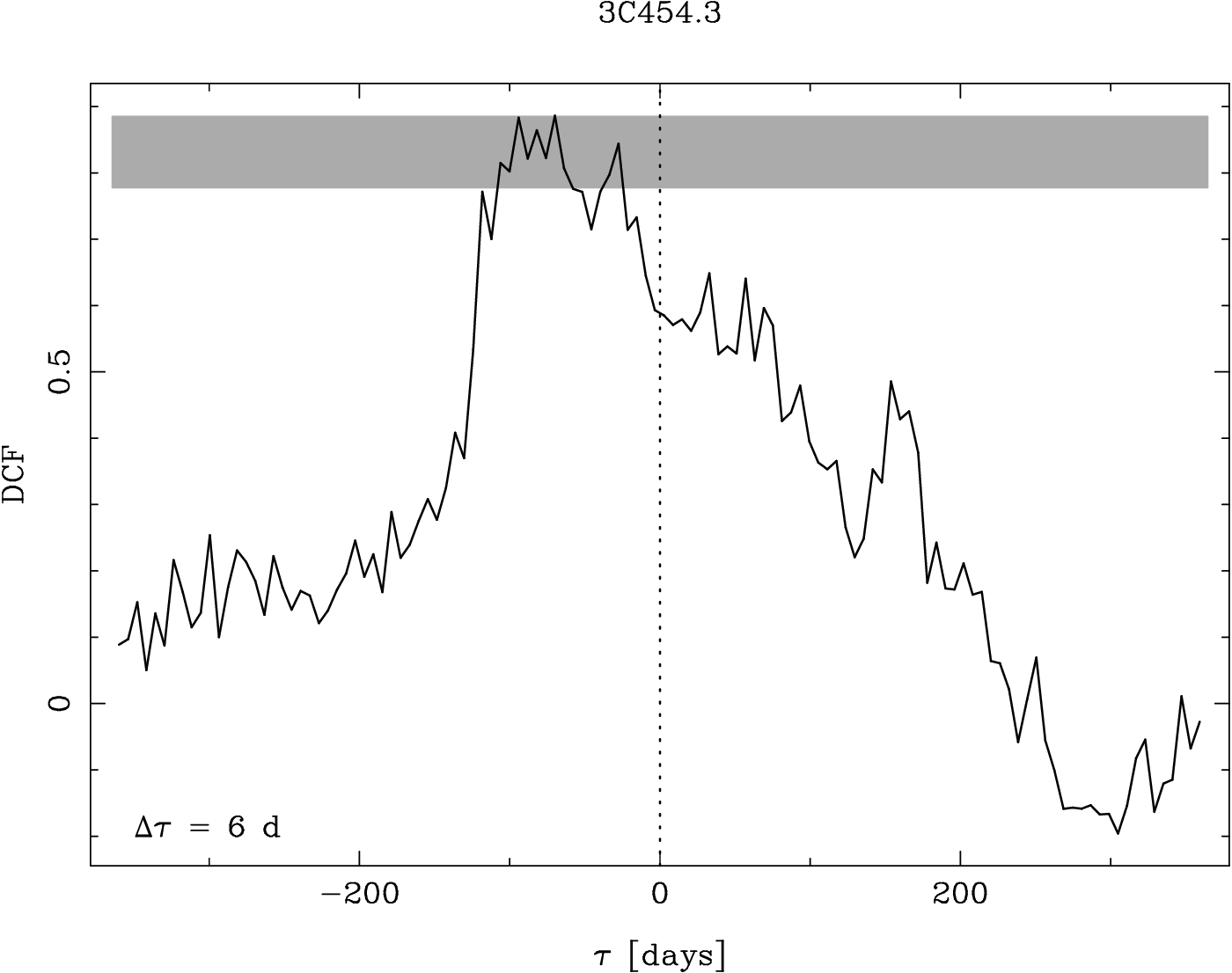}
\hspace{5mm}
\includegraphics[height=8.5cm,angle=-90]{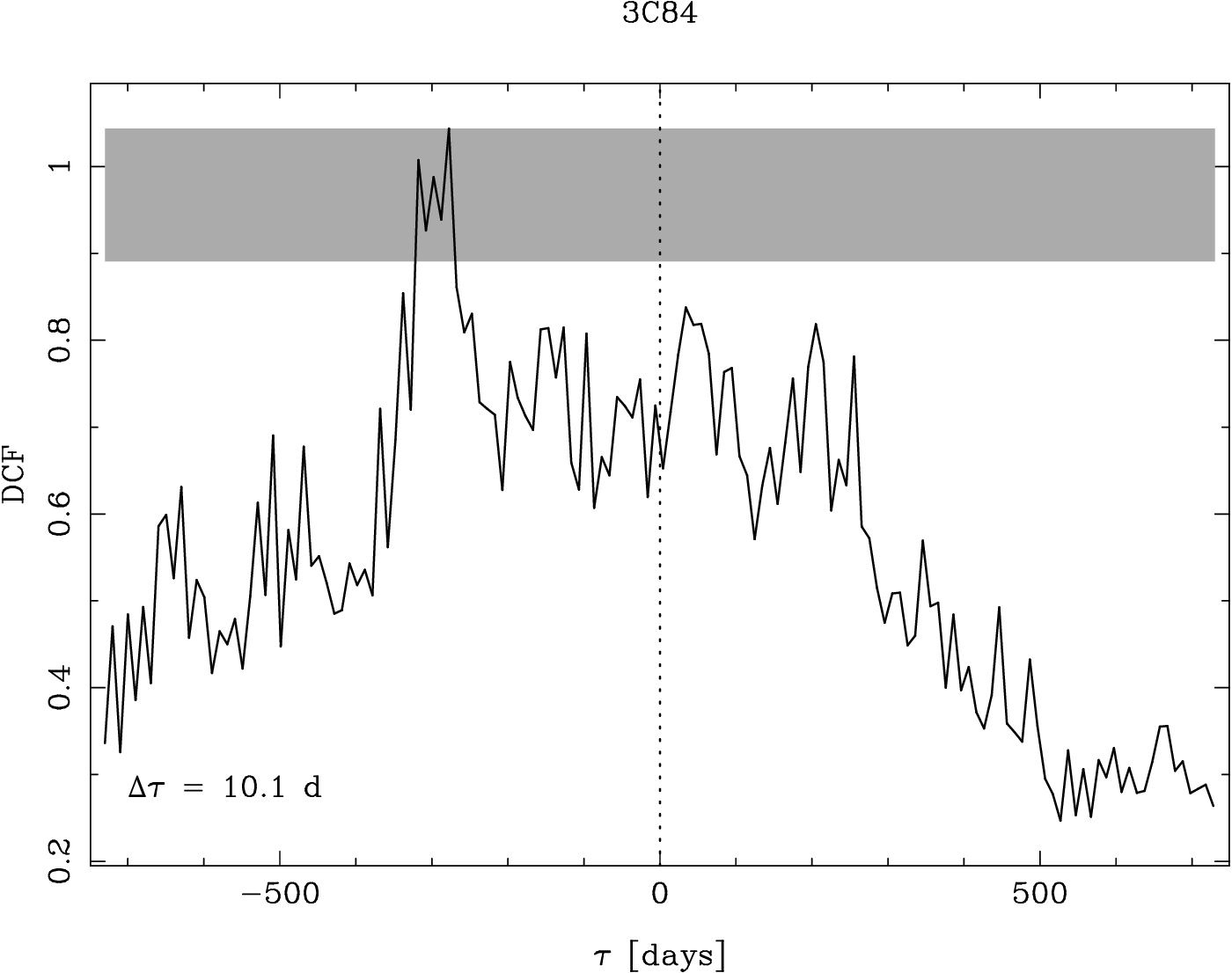}
\caption{Discrete correlation functions (DCF) as functions of time lags $\tau$ between the 1.3-mm and 3-mm lightcurves (see Fig.~\ref{fig_lightcurve}) of each AGN. The black graphs indicate the measured DCF, bin sizes $\Delta\tau$ are indicated in the corresponding diagrams. The maxima of the curves correspond to the most probable time lag between the spectral bands. Grey shaded areas mark the DCF ranges $[max(DCF), max(DCF)-3\sigma]$ with $\sigma$ being the statistical error of $max(DCF)$. Thus all values located within the shaded area are consistent with the maximum of the DCF curve within $3\sigma$ confidence. A positive (negative) time lag means that the 3-mm lightcurve preceeds (lags behind) the 1.3-mm lightcurve. Vertical dashed line mark the null positions of the time lag axes, corresponding to the absence of time lags.}
\label{fig_timelag}
\end{figure*}

We present the results of our analysis in Fig.~\ref{fig_timelag}. In our convention, a positive (negative) time lag means that the 3-mm lightcurve preceeds (follows) the 1.3-mm lightcurve. In each diagram, we indicate the ranges (as grey-shaded areas) covered by the range $\left[ max(DCF), max(DCF)-3\sigma_{max(DCF)} \right]$, thus covering all values of $DCF$ that are in agreement with the maximum of the DCF on a $3\sigma$ confidence level. We probe the null hypothesis ``the 1.3-mm and 3-mm lightcurves are simultaneous''. Accordingly, we regard the lightcurves as simultaneous as long as $DCF(\tau=0) \geq max(DCF)-3\sigma_{max(DCF)}$.

\section{Results}

\subsection{0923+392 (4C +39.25)}

The 3-mm (1.3-mm) flux densities of this flat spectrum radio quasar (FSRQ) dropped from $\approx$6~Jy ($\approx$3~Jy) in 1996 to $\approx$3~Jy ($\approx$1.5~Jy) in 2003 and have been rising again since then to present-day levels of $\approx$4.5~Jy ($\approx$3~Jy). For the flux density decrease from 1996 to 2003 we find characteristic (exponential) timescales of $\approx$5--7 years. The 3-mm flux distribution shows a double peak profile that can be described by two Gaussian components separated by $\approx$1.6~Jy. The 1.3-mm flux distribution shows an asymmetry indicative of a second component but no significant secondary peak. We therefore attempted to describe the histogram by a log-normal distribution according to Eq.~\ref{eq_lognormal}. However, we did not find a satisfactory solution, with the best fitting curve showing a $\chi^2/dof = 13.8$. 

The spectral index $\alpha$ is $\approx$0.8 in the first years of monitoring before 2000 and steepens to $\approx$1 around 2002.  The index then flattens again until reaching $\approx$0.6 in the time window 2007--2010. There appears to be no correlation between spectral index and 3-mm flux. Even though we find the steepest spectral indices for the lowest source fluxes, we find similar variations in $\alpha$ (by $\approx$0.2) within the same flux ranges.

The 1.3-mm and 3-mm lightcurves are simultaneous within the $3\sigma$ confidence interval. The power spectra of the 3-mm and 1.3-mm flux measurements can be well described as simple red-noise spectra with fairly flat powerlaw indices $\beta\approx0.4$. Within the errors, the values for $\beta$ are in agreement for both frequency bands.

\subsection{3C 111}

This Seyfert~1 galaxy shows episodic strong variability on timescales of years. The source seems to spend most of the time in a low-activity or ``quiescent'' state with fluxes in the range $\approx$1--3~Jy. In addition, we can identify at least five pronounced spikes in the lightcurves in 1997, 2001, 2006, 2008, and 2009. These outbursts reach flux densities of $\approx$6--15~Jy, thus exceeding the quiescent level by factors 3--6. The outbursts appear to rise and decay on fast ($\lesssim$1~yr) timescales; unfortunately, the irregular sampling of the lightcurves prevents a reliable quantitative analysis. The 3-mm (1.3-mm) flux distributions can be described by smooth, asymmetric profiles composed of two Gaussian components for fluxes $\lesssim$8~Jy ($\lesssim$5~Jy) plus sparsely populated tails ranging out to $\approx$14~Jy ($\approx$10~Jy). Except for the tails, it is also possible to describe the histograms with (unimodal) log-normal distributions. In both cases we find best-fit solutions with $\chi^2/dof \approx 1.6$. Although this is a relatively high value, the log-normal distribution appears to be an acceptable description of the data.
Within errors, the 1.3-mm and 3-mm lightcurves are simultaneous.

For most of the monitoring time the spectral index remains at $\alpha\approx0.5$ (within errors). The only exception appears around 1997 when the spectral slope is $\alpha\approx0.8$; this phase coincides with the first outburst. In the flux--$\alpha$ diagram the various points do not indicate a trend.

The power spectrum of the 3-mm (1.3-mm) flux data globally follows a red-noise powerlaw spectrum with $\beta\approx1.0$ ($\beta\approx0.6$). The low-frequency end ($f\approx0.002-0.01$~days$^{-1}$) of the 3-mm power spectrum is located systematically above the best-fitting overall powerlaw; this region can be described as an upscaled (by a factor $\approx$2) version of the global red-noise spectrum.

\subsection{3C 273}

At the beginning of the monitoring, the 3-mm (1.3-mm) lightcurves of this FSRQ rose fast towards a strong outburst peaking in 1997 at levels of $\approx$38~Jy ($\approx$30~Jy). From then on, the flux densities steadily decreased down to minimum values of $\approx$6~Jy ($\approx$4~Jy) in 2004 with an exponential decay timescale of $\approx$3~yr (measured for the decrease in 1997--2000). Since about 2006, 3C~273 has been in a state of high variability, with fluxes fluctuating in ranges $\approx$10--27~Jy ($\approx$6--20~Jy). The flux distributions follow complex, double-peaked profiles; they can be approximated by superpositions of three Gaussian distributions peaking at $\approx$8~Jy, $\approx$18~Jy, and $\approx$28~Jy ($\approx$4~Jy, $\approx$10~Jy, and $\approx$16~Jy). Within errors, the 1.3-mm and 3-mm lightcurves are simultaneous.

The spectral index varies substantially over the 14 years of observation, starting at $\alpha\approx0.3$ around 1997, steepening up to $\alpha\approx0.8$ in 2000--2003 and flattening again to $\alpha\approx0.4$ after 2007. There is a global trend towards flatter indices at higher flux levels, though no clear correlation: on the one hand $\alpha$ remains at similar levels over wide flux ranges, on the other hand strong variations in $\alpha$ occur within the same flux ranges.

The power spectra of the 3-mm and 1.3-mm lightcurves follow red-noise powerlaws with identical (within errors) indices $\beta\approx0.45$. For frequencies $f\lesssim0.04$~days$^{-1}$ the spectral power is located systematically above the best-fitting overall powerlaw; these ranges can be described as varieties of the overall power spectra upscaled by factors $\approx$3 (3~mm) and $\approx$2 (1.3~mm), respectively.

\subsection{3C 345}

The 3-mm (1.3-mm) lightcurves of this FSRQ start at flux densities of $\approx$4~Jy ($\approx$2~Jy) in 1997 and reach their highest values of $\approx$7~Jy ($\approx$4~Jy) in mid-1998. From then the fluxes decrease overall down to $\approx$3~Jy ($\approx$2~Jy) in the time range 2004--2008. Around 2009 an outburst occurs, reaching flux levels up to $\approx$6.5~Jy ($\approx$4.5~Jy) and slowly decreasing since then. The 2009 outburst raises in 2008--2009 with an exponential timescale of $\approx$1~yr. The 3-mm (1.3-mm) flux distributions follow double-peaked profiles we approximated as superpositions of two Gaussians peaking at $\approx$3~Jy and $\approx$5.5~Jy ($\approx$2~Jy and $\approx$3.5~Jy). Within errors, the 1.3-mm and 3-mm lightcurves are simultaneous.

Within errors, the spectral index remains on very similar levels in the range $\alpha\approx0.6-0.8$ for most of the monitoring time. After 2007 -- the time frame of the most recent outburst -- the spectrum flattens somewhat towards $\alpha\approx0.4$. We note that this somewhat steep index at mm-wavelengths is consistent with the classification of 3C~345 as FSRQ: the classification as FSRQ is based on cm-observations, whereas at mm-wavelengths the spectrum bends towards $\alpha>0$. In a $S_{\nu}-\alpha$ diagram the data points move with time on an almost circular trajectory, leaving no room for a correlation between flux and spectral index. 

The power spectra can be described as simple red-noise spectra, with powerlaw indices $\beta\approx0.5$ for the 3-mm and $\beta\approx0.4$ for the 1.3-mm lightcurve, respectively.

\subsection{3C 454.3}

This flat spectrum radio quasar shows the most spectacular lightcurves among the sources of our sample. For the first nine years (1996-2005) of monitoring the activity of this source was restricted to variability on levels of $\approx$2--8~Jy. This changed dramatically in 2005 when the first of at least three major outbursts occured, peaking at flux densities of $\approx$35--40~Jy. The series of adjacent outbursts has continued until the present day, making 3C~454.3 one of the brightest extragalactic mm/radio sources currently observed (see also, e.g., Jorstad et al. \cite{jorstad2010}). All outbursts show short exponential rise and decay times in the range $\approx$0.1--0.5 years. The 3-mm (1.3-mm) flux distribution is dominated by a Gaussian peak centered at $\approx$6~Jy ($\approx$4~Jy) with an extended tail ranging out to $\approx$40~Jy ($\approx$35~Jy). 

The spectral index has been fluctuating around $\alpha\approx0$, with the most extreme values being $\alpha\approx-0.4$ and $\alpha\approx0.5$, respectively. A somewhat steep value of $\alpha\approx0.5$ can be observed for the time 1996--2000. Notably, the most extreme change (from $\alpha\approx-0.4$ to $\alpha\approx0.4$) occurs at the time of the peak of the first bright outburst in 2005. The data almost fill the entire $S_{\nu}-\alpha$ area observed, leaving the impression that flux and spectral index are uncorrelated.

The 3-mm (1.3-mm) power spectrum globally follows a red noise powerlaw with index $\beta\approx0.7$ ($\beta\approx0.5$). However, for frequencies $f\lesssim5\times10^{-3}$~days$^{-1}$ the power spectrum seem to be an upscaled -- by a factor $\approx$3 ($\approx$2) -- version of the average powerlaw.

\begin{figure*}
\centering
\includegraphics[height=8.5cm,angle=-90]{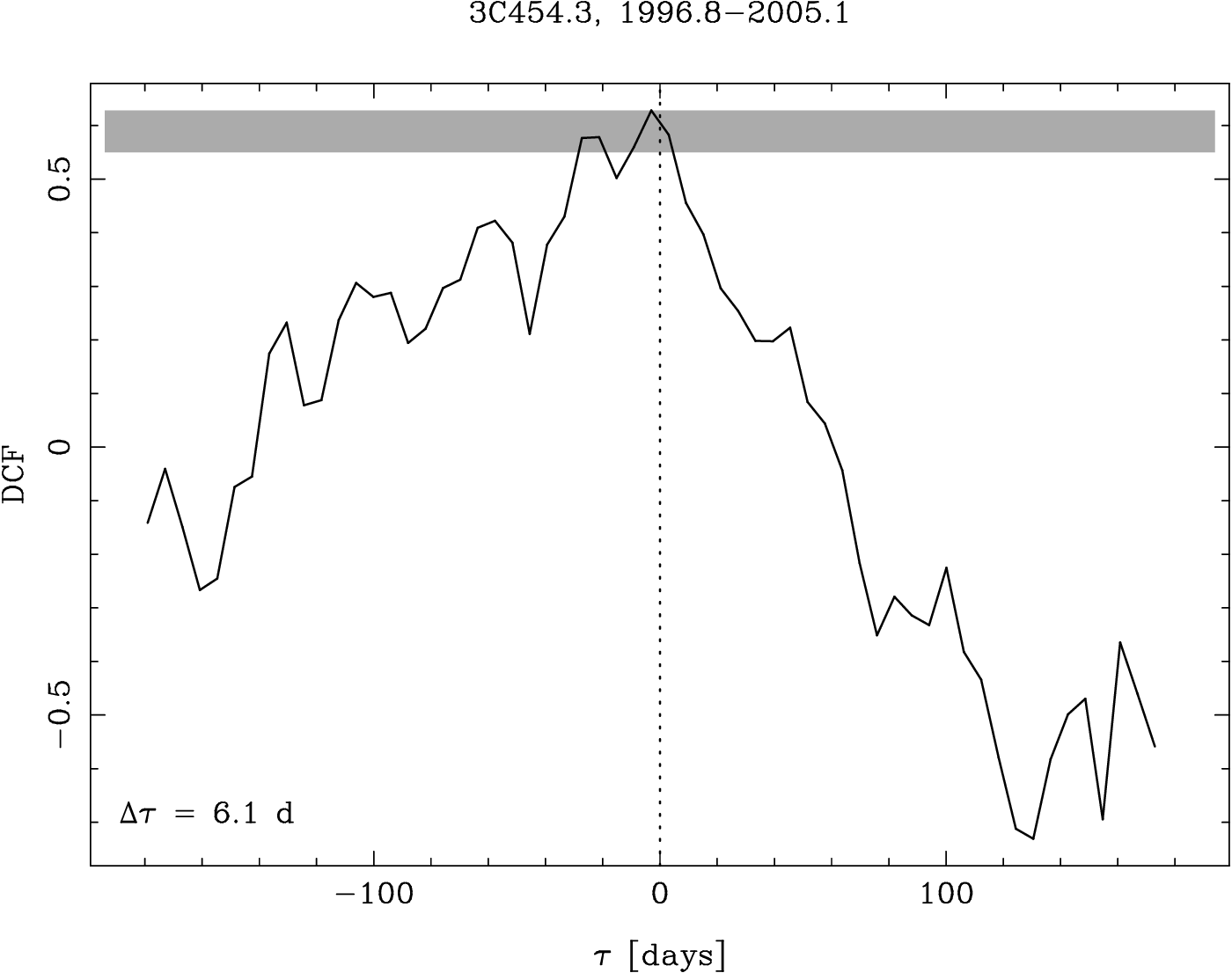}
\hspace{5mm}
\includegraphics[height=8.5cm,angle=-90]{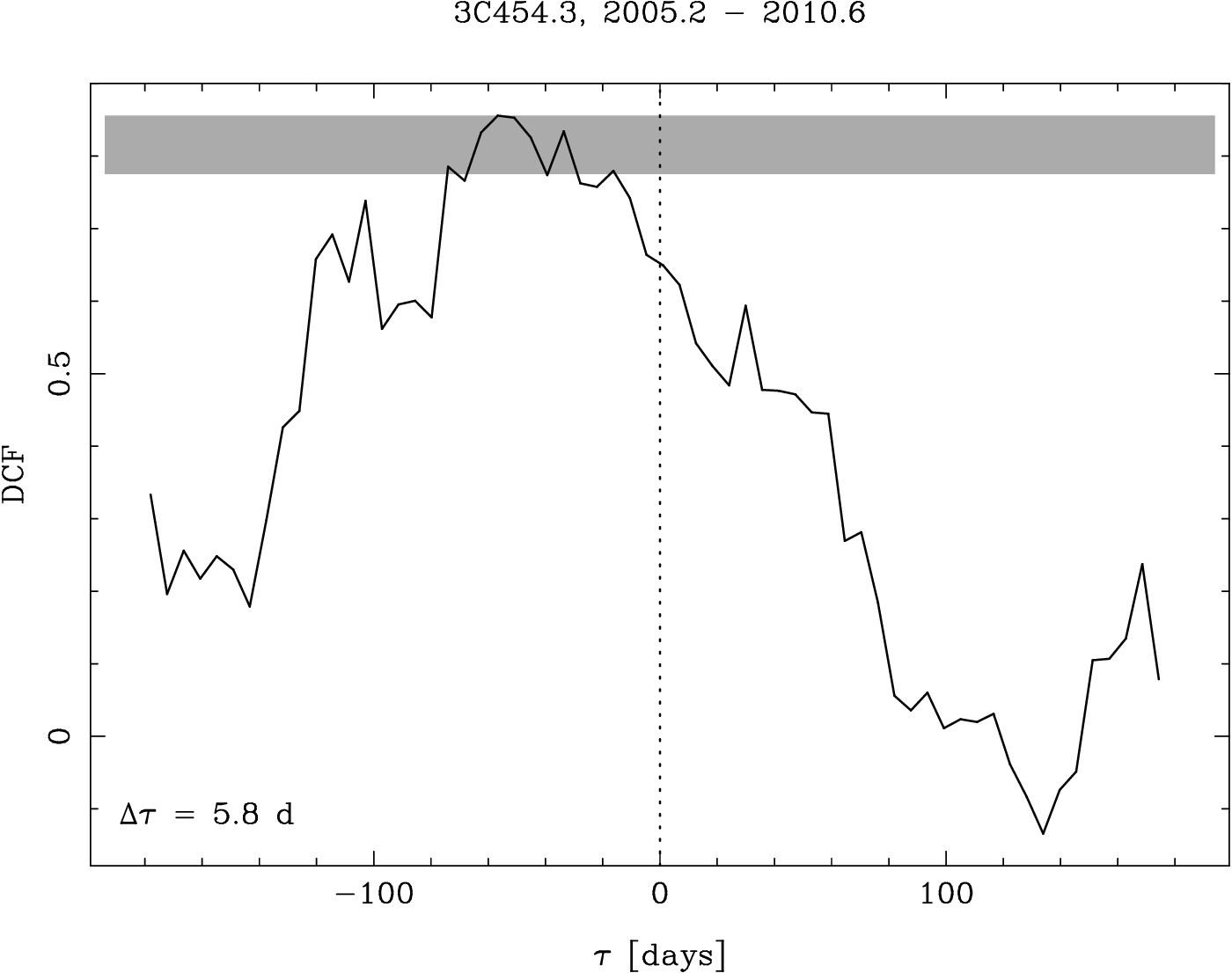}
\caption{Like Fig.~\ref{fig_timelag}, for 3C~454.3 before 2005.2 (left panel) and after 2005.2 (right panel). During the ``quiescent'' phase before 2005.2, the peak of the DCF is in good agreement with a time delay of zero. We note the low absolute value of the correlation, $DCF(0)\approx0.6$. For the ``flaring'' phase after 2005.2, we find a significant time delay in the range (as defined by the $3\sigma$ confidence interval) of $\sim-15...-80$ days, peaking at $\tau\approx-55$ days. The negative delay indicates that the 1.3-mm lightcurve is ahead in time of the 3-mm lightcurve.}
\label{fig_dcf3C454.3}
\end{figure*}

The discrete correlation function points to time delays $\tau$ significantly different from zero in terms of the $3\sigma$ confidence threshold, located in the range $\tau \approx -20 ... -110$ days. The negative delay indicates that the 1.3-mm emission preceeds the 3-mm emission. However, as 3C~454.3 shows a dramatic transition from ``quiescent'' to ``flaring'' around epoch 2005.2, the curve presented in Fig.~\ref{fig_timelag} (bottom left) might be misleading as the timing of the emission might be different in the two states. We therefore computed in addition one DCF for each of the two activity phases, one covering all data from 1996.8 to 2005.1, one including all measurements obtained between 2005.2 and 2010.6. The results are summarized in Fig.~\ref{fig_dcf3C454.3}. We note that for the ``quiescent'' phase (before 2005.2) the most probable time delay -- on a low absolute level, $DCF(0)\approx0.6$ -- is consistent with zero, whereas we find for the ``flaring'' phase (after 2005.2) $\tau\approx -15 ... -80$ days, with the maximum of the DCF located at $\tau\approx-55$ days. Again, the negative delay indicates that the 1.3-mm emission preceeds the 3-mm emission.

\begin{figure*}
\centering
\includegraphics[height=8.5cm,angle=-90]{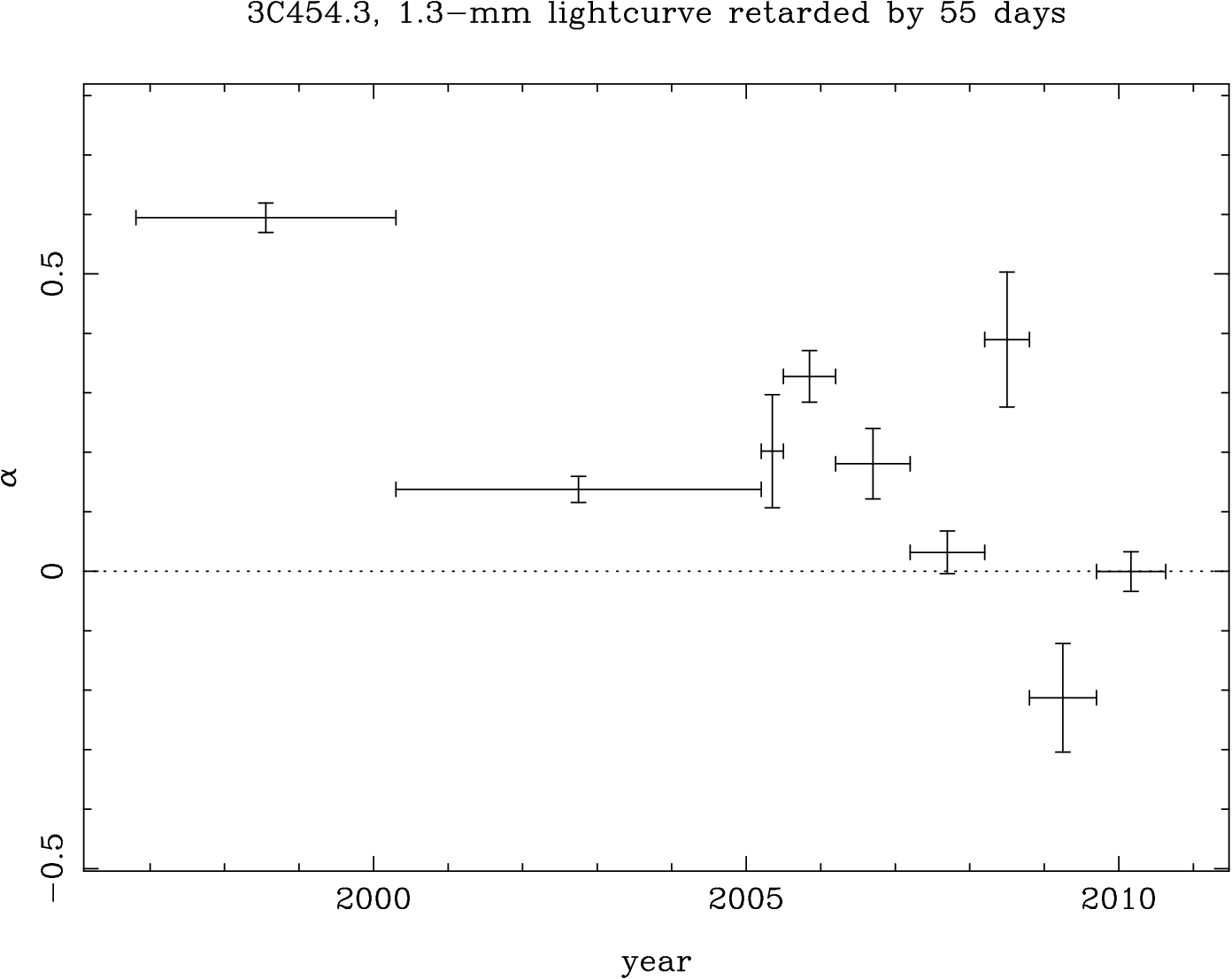}
\hspace{5mm}
\includegraphics[height=8.5cm,angle=-90]{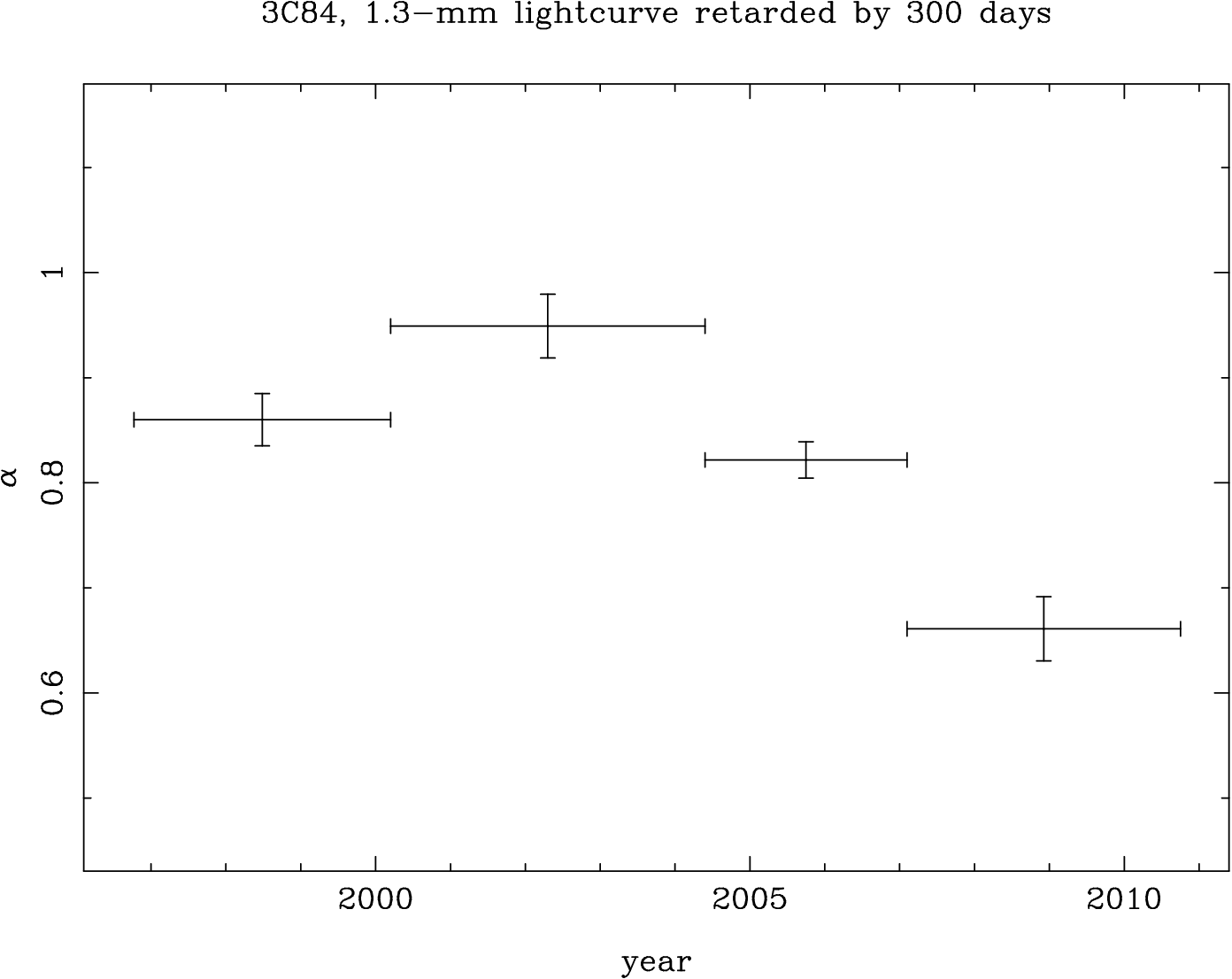}
\caption{Spectral index $\alpha$ vs. time based on 1.3-mm and 3-mm flux data only, with the 1.3-mm lightcurves artificially retarded. \emph{Left panel:} 3C~454.3, using a spectral time delay $\tau=-55$ days. \emph{Right panel: } 3C~84, using a spectral delay $\tau=-300$ days. The time delays are given by the correlation analysis illustrated in Fig.~\ref{eq_udcf}. This figure should be compared to Fig.~\ref{fig_specindex}.}
\label{fig_specindex_timeshift}
\end{figure*}

Finally, we probed the impact of the spectral time delay on the calculation of the spectral index $\alpha$. For this, we repeated the calculation outlined in Sect.~3.3 using 1.3-mm and 3-mm flux density data only. We compensated the spectral time delay by artificially retarding the 1.3-mm lightcurve by 55 days. We present the resulting modified spectral index evolution in Fig.~\ref{fig_specindex_timeshift} (left panel). Time bins are the same as in Fig.~\ref{fig_specindex} (bottom left panel) which should be compared to Fig.~\ref{fig_specindex_timeshift} (left panel). We note that retarding the 1.3-mm data removes the inverted values for $\alpha$ that are visible in Fig.~\ref{fig_specindex} (bottom left panel) during the outburst 2005/2006.

\subsection{3C 84}

The 3-mm (1.3-mm) lightcurve of this Seyfert~2 galaxy varies smoothly during the entire monitoring time between $\approx$3.5~Jy ($\approx$2~Jy) in 2002 and $\approx$11~Jy ($\approx$7~Jy) in 2010. Since about 2004 the fluxes raise continuously with exponential rise time scales of $\approx$6 years ($\approx$4 years). The flux distributions show somewhat complicated double-peak profiles that can be approximated as superpositions of three Gaussians peaking at $\approx$4~Jy, $\approx$5.5~Jy, and $\approx$9~Jy ($\approx$2~Jy, $\approx$4~Jy, and $\approx$6~Jy).

The spectral index is steep, with $\alpha\approx0.9$ (within errors) until about 2005 and then drops smoothly to $\alpha\approx0.5$ after 2007. We find the flattest spectral index for the highest flux value; however, as this is based on a single data point that occupies a large range in flux, this observation does not establish a significant trend.

The power spectra can be described as simple red-noise spectra, with powerlaw indices $\beta\approx0.7$ for the 3-mm and $\beta\approx0.4$ for the 1.3-mm lightcurve, respectively.

\begin{figure*}
\centering
\includegraphics[height=8.5cm,angle=-90]{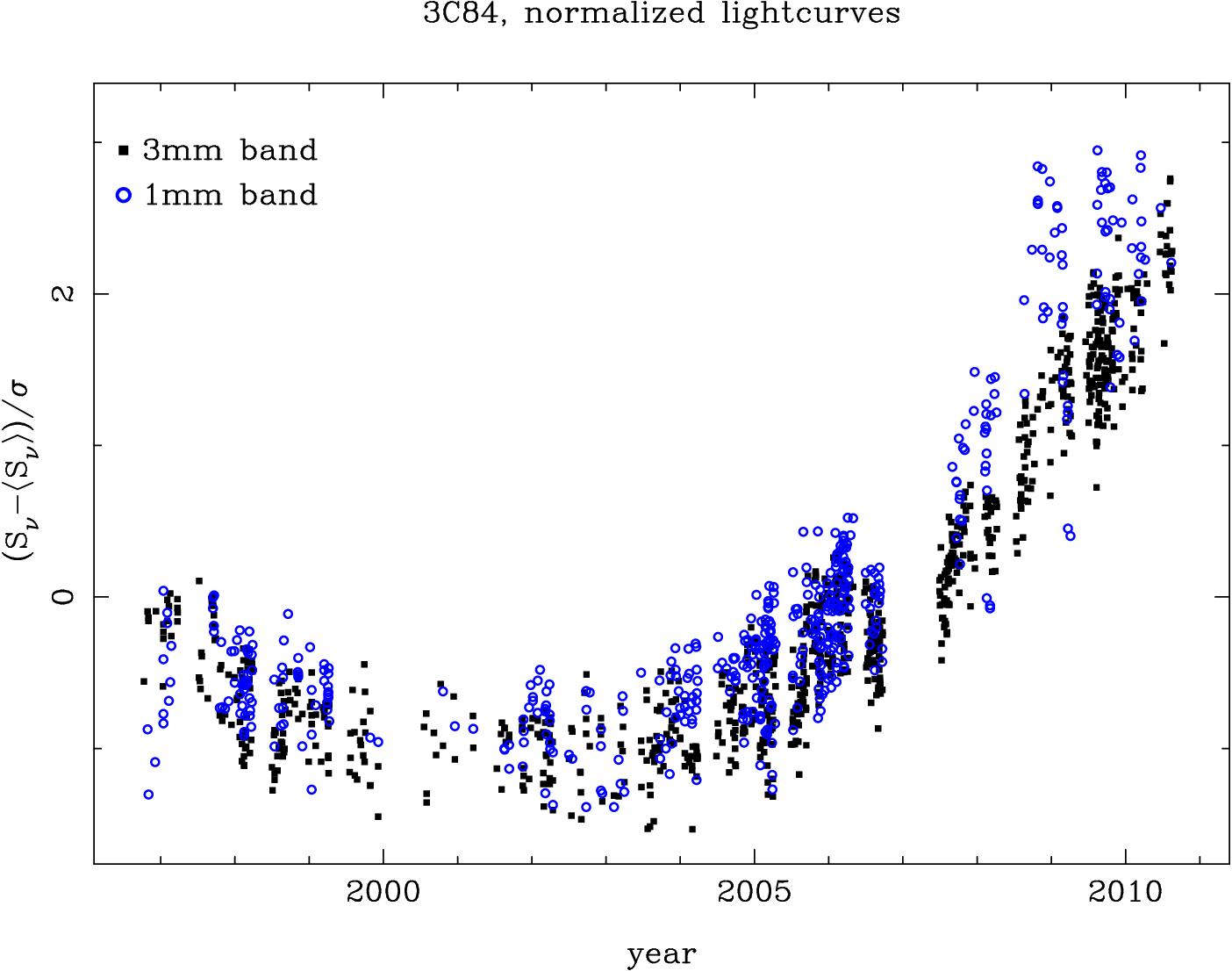}
\hspace{5mm}
\includegraphics[height=8.5cm,angle=-90]{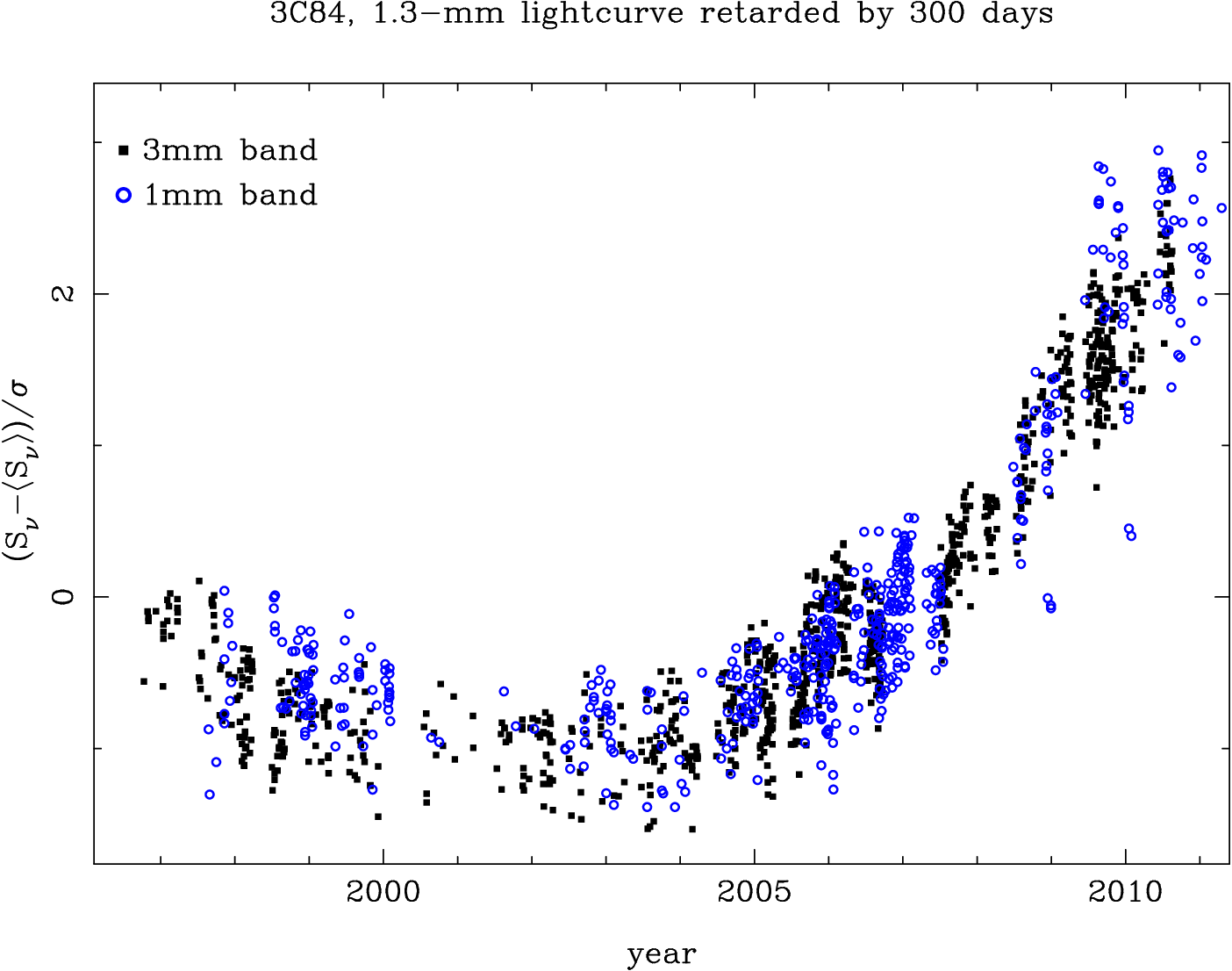}
\caption{A visualization of the time delay between the 1.3-mm and 3-mm lightcurves of 3C~84. \emph{Left panel:} The 1.3-mm and 3-mm lightcurves. Each lightcurve is normalized to zero mean and unity standard deviation, in analogy to the calculation of the unbinned discrete correlations according to Eq.~\ref{eq_udcf}. Here $S_{\nu}$ indicates the flux density, $\langle S_{\nu}\rangle$ is the average flux density, and $\sigma$ denotes the standard deviation of each lightcurve. \emph{Right panel:} Same as left panel, but with the 1.3-mm lightcurve artificially retarded by 300 days. Comparison of the two panel shows improved agreement of the two lightcurves when taking into account the time delay $\tau\approx-300$ days found from the discrete correlation function; see also Fig.~\ref{eq_udcf}, bottom right panel.}
\label{fig_3C84shift}
\end{figure*}

The discrete correlation function computed from the 1.3-mm and 3-mm lightcurves finds a spectral time delay significantly different from zero. The most probable time delay is located at $\tau\approx -270 ... -330$ days, meaning that the 1.3-mm emission preceeds the 3-mm emission by almost one year. We note the high absolute value of the maximum correlation, $DCF(\tau=-277{\rm d})=1.04\pm0.05$, which is in agreement with the theoretical maximum value of 1. However, as already $DCF(0)\approx0.7$, one might wonder how large the difference between $\tau\approx0$ and $\tau\approx-300$ actually is in \emph{absolute} terms. We therefore tested the impact of the time offset graphically. For this we normalize the 1.3-mm and 3-mm lightcurves to zero mean and unity standard deviation, in analogy to the computation of the unbinned discrete correlations according to Eq.~\ref{eq_udcf}. In Fig.~\ref{fig_3C84shift} we accordingly present the quantity $(S_{\nu}-\langle S_{\nu}\rangle)/\sigma$ for each of the two frequency bands; here $S_{\nu}$ is the flux density, $\langle S_{\nu}\rangle$ denotes the time average of the flux density, and $\sigma$ indicates the standard deviation of the flux density. In Fig.~\ref{fig_3C84shift} we present the normalized data separately for the two cases $\tau=0$ days (left panel) and $\tau=-300$ days days (right panel). Indeed the case $\tau=-300$ days shows a better agreement between the 1.3-mm and 3-mm lightcurves than the case  $\tau=0$ days. This agrees with the result obtained from the correlation analysis. However, we note that the lightcurves show clear trends that are only partially covered by our observations. This can mislead a correlation analysis, meaning that the time delay we find has to be taken with care.

Like for 3C~454.3, we probed the impact of the spectral time delay on the calculation of the spectral index. We compensated the spectral time delay by artificially retarding the 1.3-mm lightcurve by 300 days. We present the resulting modified spectral index evolution in Fig.~\ref{fig_specindex_timeshift} (right panel). Time bins are the same as in Fig.~\ref{fig_specindex} (bottom right panel) which should be compared to Fig.~\ref{fig_specindex_timeshift} (right panel). Comparison of the two diagrams indicates that removing the spectral time delay systematically shifts $\alpha$ towards steeper values by $\approx0.1$.

\section{Discussion}

\subsection{Lightcurves and Flux Distributions}

Our analysis reveals substantial -- by factors $\approx$2--8 -- flux variability on timescales of several years in all sources of our sample. Even though all lightcurves (see Fig.~\ref{fig_lightcurve}) show strong variations, the characteristic (exponential rise/decay) timescales differ by more than one order of magnitude, ranging from $\lesssim$0.5 years (3C454.3) to $\approx$7 years (0923+392). Accordingly, we find lightcurves with smooth morphologies during the entire $\approx$14-year monitoring time (0923+392, 3C~84) as well as lightcurves that are ``spiky'' (3C~111, 3C~454.3).

The flux distributions (Figs. \ref{fig_fluxdist1}, \ref{fig_fluxdist2}) of five out of six (the exception being 3C~111) AGN show bi- or even multimodality. This indicates that the flux variability of the concerned sources cannot be described by uniform stochastic processes:

\begin{itemize}

\item  In case of Gaussian (white) noise processes the distribution of amplitudes would be (obviously) Gaussian.

\item  In case of red noise processes (see also Sect. 3.4), the distribution of amplitudes is Gaussian for a given sampling frequency interval $[f, f+df]$ (e.g. Timmer \& K\"onig \cite{timmer1995}). The distribution of amplitudes resulting when summing over all $f$ would thus a priori be a superposition of Gaussians of varying widths (i.e. a unimodal, symmetric distribution). However, for physical lightcurves one needs to take into account that fluxes are constraint to positive values. This introduces a bias towards higher flux values, meaning that the resulting flux histogram shows an extended wing towards high values.

\item  In case of multiplicative stochastic processes the distribution of amplitudes would follow a unimodal log-normal distribution, i.e. an asymmetric distribution with an extended wing towards high values (see Uttley et al. \cite{uttley2005} for a discussion of AGN X-ray lightcurves).

\end{itemize}

\noindent
The flux distributions of 3C~111 are (at least marginally) consistent with log-normal distributions, pointing towards an underlying multiplicative process. Such processes can be understood as chains of events where each event is triggered by the previous one with a certain probability $<1$ (e.g., Montroll \& Shlesinger \cite{montroll1982}). Candidates are outbursts that trigger secondary emission, or shocks in jets that trigger secondary shocks or fragmentation of emission regions.

For the remaining five sources, it appears to be necessary to distinguish separate states of source activity like ``quiescent'' and (maybe multiple) ``flare states'' where each state is characterized by an individual flux distribution. The final, observed flux distribution would then be a superposition of the individual distributions.

\subsection{Spectral Indices}

All sources of our sample show (mostly) fairly steep spectral indices within ranges $\alpha\approx0.5-1$ (Fig.~\ref{fig_specindex}). All sources show a significant, smooth variability of their indices on timescales of years. Only 3C~454.3 provides an exception from both of the aforegiven statements: it shows a faster variability on timescales of $\approx$1--2 years and reaches flat and even inverted indices $\alpha\approx-0.5-0$. As our sample is composed of very luminous AGN, we can safely assume that their mm-emission is dominated by synchrotron radiation. We can thus classify the sources of our sample (with the possible exception of 3C~454.3) as outflow (i.e. halo or jet) dominated sources for which typically $\alpha\approx0.5-1$ (e.g. Krolik \cite{krolik1999}). This is consistent with the fact that for all sources luminous halos or jets on kilo-parsec scales have been observed (see Table~\ref{tab_journal}).

The variations of the spectral indices point towards variations of intrinsic source properties on timescales of years. A priori, there are three potential mechanisms:

\begin{enumerate}

\item  Variations of the optical depth. Synchrotron emission from optically thin emission regions shows steep spectral indices with $\alpha\gtrsim0.5$; strong absorption in emission regions (i.e. optically thick emission) leads to approximately flat  ($\alpha\approx0$) or even inverted, complex spectra (see, e.g., Kembhavi \& Narlikar \cite{kembhavi1999} for a review). Accordingly, variations of the optical depth can lead to variations of the spectral index inbetween these extreme cases. 

\item  Variations of the relative luminosities of different source components. Emission from (optically thick) nuclei shows usually fairly flat spectra ($\alpha\approx0$) whereas the emission from (optically thin) outflows (jets, lobes) typically shows steep spectra ($\alpha\approx0.5-1$; e.g. Krolik \cite{krolik1999}). Variations in the relative luminosities of the various components therefore modify the spectral index of the source-integrated emission. The spectrum becomes steeper if the relative contribution by outflows increases, it becomes flatter if the relative contribution by the nucleus increases.

\item  For \emph{optically thin} synchrotron emission from an ensemble of electrons with a powerlaw energy spectrum with index $\gamma$ the spectral index of the radiation is given by $\alpha=(\gamma-1)/2$ (e.g. Ginzburg \& Syrovatskii \cite{ginzburg1965}); for $\alpha=0.5-1$ this corresponds to $\gamma=2-3$. Accordingly, variations in the electron energy spectrum express themselves as variations of the spectral index.

\end{enumerate}

For three of our six target sources (namely 3C~111, 3C~273, and 3C~454.3) the spectral indices are -- at least occasionally -- significantly flatter than the range permitted for the assumption of optically thin emission (i.e. we observe $\alpha<0.5$) for parts of the monitoring time. This suggests that scenario 3 cannot be applied regularly as this scenario is valid only for optically thin emission.

Additional information is provided by the relation between source fluxes and their spectral indices (Fig.~\ref{fig_alpha_vs_flux}). Especially, the $S_{\nu}-\alpha$ relation reveals that scenario 2 can play only a minor role for the variability of the spectral index: if the $\alpha$ variations were due to (or at least dominated by) luminosity increases or decreases of parts of the sources, there should be a tight correlation between source fluxes and spectral indices -- which we do not observe. From the combined information we therefore conclude that variations of the optical depth with time (scenario 1) are the most likely mechanism for variations of the spectral index.

\subsection{Power Spectra}

\begin{table}
\caption{True spectral indices $\beta_{\rm true}$ vs. observed indices $\beta_{\rm obs}$. Observed indices are given for three cases: no white (Gaussian) noise added to the lightcurve ($\sigma=0$), Gaussian noise with $\sigma$ of 10\% of the flux values added ($\sigma=0.1S_{\nu}$), Gaussian noise with $\sigma$ of 16\% of the flux values added ($\sigma=0.16S_{\nu}$). These values are obtained from Monte Carlo simulations. Errors indicate the Gaussian widths of the distributions.}
\label{tab_rednoise}
\centering
\begin{tabular}{c c c c}
\hline\hline
$\beta_{\rm true}$ & $\beta_{\rm obs}(\sigma=0)$ & $\beta_{\rm obs}(\sigma=0.1S_{\nu})$ & $\beta_{\rm obs}(\sigma=0.16S_{\nu})$ \\
\hline
$2.0$ & 1.84$\pm$0.11 & 1.46$\pm0.23$ & 1.33$\pm$0.25 \\
$1.5$ & 1.51$\pm$0.16 & 1.30$\pm$0.14 & 1.17$\pm$0.17 \\
$1.0$ & 0.99$\pm$0.13 & 0.95$\pm$0.11 & 0.85$\pm$0.13 \\
$0.5$ & 0.48$\pm$0.15 & 0.44$\pm$0.13 & 0.39$\pm$0.10 \\
\hline
\end{tabular}
\end{table}

The power spectra of our sources (Fig. \ref{fig_scargle1}, \ref{fig_scargle2}) globally follow red noise distributions, meaning powerlaws with spectral indices $\beta>0$ (cf. Section~4.3). This, as well as the apparent absence of periodic signals, shows that the emission processes are intrinsically stochastic. The power-spectral indices $\beta\approx0.4-1$ are located between those of Gaussian (white) noise ($\beta=0$) and flicker noise ($\beta=1$) processes (e.g. Timmer \& K\"onig \cite{timmer1995}). They are significantly flatter than the indices found from optical and cm/radio (e.g. Hufnagel \& Bregman \cite{hufnagel1992}) as well as X-ray (e.g. Lawrence \& Papadakis \cite{lawrence1993}) studies of AGN emission which are located in the range 1--2.

When discussing the power spectral indices we have to take into account that we observe lightcurves that are affected by white (Gaussian) measurement noise. This noise is dominated by a systematic scatter that amounts to $\approx$10\% for the 2-mm and 3-mm fluxes and $\approx$16\% for the 1.3-mm flux data (see Sect.~2). The presence of white measurement noise leads to the observation of power-spectral indices $\beta$ that are actually flatter (i.e. closer to the case of white noise, $\beta=0$) than the indices intrinsic to the source emission. In order to estimate the intrinsic spectral indices, we performed Monte Carlo tests using simulated red-noise lightcurves with known ideal spectral indices. For this test we applied the algorithm by Timmer \& K\"onig (\cite{timmer1995}). We calculated artificial lightcurves corresponding to red noise lightcurves with $\beta=2$, $\beta=1.5$, $\beta=1$, and $\beta=0.5$, respectively. For each case, we examined three scenarios: the absence of measurement noise, Gaussian measurement noise with $\sigma=0.1S_{\nu}$ (i.e. a Gaussian width corresponding to 10\% of the observed fluxes), and Gaussian measurement noise with $\sigma=0.16S_{\nu}$ (i.e. a Gaussian width corresponding to 16\% of the observed fluxes). We simulated 100 lightcurves with 500 data points each for each of the 12 cases examined.
In order to assess the impact of gaps in the data, we applied a sampling approximately corresponding to the worst case actually encountered which is the sampling of the 1.3-mm lightcurve of 3C~111 (compare Fig.~\ref{fig_lightcurve}, top right panel). For this, we inserted  -- by simply leaving out data -- a total of eleven artificial gaps, each covering $\approx$2\% to $\approx$9\% of the total observing time.
For each artificial lightcurve we calculated a Scargle periodogram (Eq.~\ref{eq_periodogram}) and derived the powerlaw index. We present the results in Table~\ref{tab_rednoise}. Due to the stochastic nature of the noise signals, the ``observed'' values for $\beta$ show substantial scatter around the averages; the $1\sigma$ widths of the corresponding distributions are quoted as uncertainties in Table~\ref{tab_rednoise}.

From our test we can draw various conclusions:

\begin{itemize}

\item  The presence of Gaussian measurement noise flattens the power-spectral indices to $\beta_{\rm obs}<1.5$ even if $\beta_{\rm true}\approx2$.

\item  As $\beta_{\rm obs}\approx0.4-0.7$ for all sources except 3C~111, the intrinsic power-spectral indices should be flatter than 1 in general.

\item  Increasing the amplitude of measurement noise from $0.1S_{\nu}$ (like for our 3-mm flux data) to $0.16S_{\nu}$ (like for our 1.3-mm flux data) leads to a further flattening of the power-spectral index by about 0.1. This agrees with our observations: we find a systematic trend towards flatter (by $\approx$0.1) indices for the 1.3-mm data compared to the 3-mm power spectra of the corresponding sources.

\end{itemize}

This also shows that the difference between our results and those reported by Hufnagel \& Bregman (\cite{hufnagel1992}) or Uttley et al. (\cite{uttley2002}) cannot be attributed to measurement noise effects. Instead, we have to attribute this difference to the different frequency regime (and thus emission region) we observe. This indicates that at millimeter wavelengths, adjacent AGN emission events tend to be less correlated than in other regimes.

The fact that intrinsically $\beta<1$ (except for 3C~111) points towards stochastic emission processes located between the cases of white noise and flicker noise, with 3C~111 being closer to the case of random walk noise $(\beta=2)$. Formally, the various cases are connected by different levels of correlation among adjacent data. In case of white noise $(\beta=0)$ adjacent data are completely uncorrelated. Random walk noise $(\beta=2)$ corresponds to the integral of white noise over an infinite interval in time, meaning that adjacent data are fully correlated. Intermediate cases $(0<\beta<2)$ can be expressed as ``half'' integrals of white noise (again, over infinite intervals in time). Such a ``half'' integration is performed by convolving a white noise time series with a powerlaw kernel $G \propto t^{-\xi}$ with $\xi>0$ and $t$ being the time. For the specific case of flicker noise $(\beta=1)$ this was discussed in detail by Press (\cite{press1978}), leading to $\xi=1/2$. From this, one may a priori expect to find any $0\leq\beta\leq2$ in AGN time series driven by stochastic emission processes\footnote{Although there seems to be some confusion in the literature; e.g., Do et al. (\cite{do2009}) conclude that the infrared emission from Sagittarius~A* follows a red noise power spectrum with $\beta\approx2.5$, without providing a physical interpretation of this unexpected result.}.

Even though there is a straightforward \emph{mathematical} recipe for the creation of red noise (in the range $0\leq\beta\leq2$) time series, there is not yet a consistent understanding which \emph{physical} mechanisms create those time series. This is a problem of importance not only in astrophysics but in many fields of physics, and various approaches towards a general explanation have been proposed (e.g., Press \cite{press1978}; Pang et al. \cite{pang1995}; Makse et al. \cite{makse1996}). More recently, Kaulakys \& Alaburda (\cite{kaulakys2009}) and Kelly et al. (\cite{kelly2011}) proposed time series models based on superpositions of exponentially decaying outbursts occuring at random times and with random amplitudes. Both studies are able to reproduce the occurence of powerlaw power spectra. However, as already noted by Press (\cite{press1978}) who discussed (and eventually rejected) similar proposals, those approaches might only shift the problem from the question ``What process creates those power spectra?'' to the question ``What process generates those outburst sequences?'' without actually solving the problem.

Although the power spectra globally follow red noise laws, three out of six sources (3C~111, 3C~273, and 3C~454.3) are hardly consistent with a single stochastic emission process. Instead, we observe excess (by factors $\approx$2--3) spectral power with respect to the average powerlaw for sampling frequencies $f\lesssim0.005-0.01$~days$^{-1}$. In combination with the information from the flux distributions (Sect. 5.1; Fig. \ref{fig_fluxdist1}, \ref{fig_fluxdist2}) this strongly suggests that ``quiescent'' high-frequency and ``flaring'' low-frequency emission have to be treated as separate source states. This conclusion deviates from the more simple picture drawn by other studies (e.g., Hufnagel \& Bregman \cite{hufnagel1992}; Uttley et al. \cite{uttley2002}) that treat AGN time series and power spectra as uniform red-noise distributions. However, a physical dichotomy of quiescent vs. flaring source states has been suggested for the near-infrared emission observed from the Galactic nuclear black hole, Sagittarius~A* (e.g. Gillessen et al. \cite{gillessen2006}; Trippe et al. \cite{trippe2007}; Dodds-Eden et al. \cite{doddseden2011}), and has been hotly debated since then (compare, e.g., Meyer et al. \cite{meyer2007} vs. Meyer et al. \cite{meyer2008}).

We also note that a segregation of distinct source states is well-known for the case of Galactic microquasars. These sources mirror many properties of AGN (taking into account scaling with the black hole mass) and their emission shows distinct states that are ususally classified as ``hard'' and ``soft'' according to their X-ray spectral properties (e.g., Tananbaum et al. \cite{tananbaum1972}; Mirabel \& Rodriguez \cite{mirabel1998,mirabel1999}; Takeuchi \& Mineshige \cite{takeuchi1998}). A possible explanation for this transition (Klein-Wolt et al. \cite{kleinwolt2002}) is the difference between the ejection of fairly localized knots from the black hole vicinity (leading to X-ray and radio flares) and the ejection of quasi-continuous, extended jets that are initially optically thick (leading to ``plateaus'' in the X-ray and radio lightcurves). For micro-quasars, these events happen on timescales from minutes (minimum time between the ejection of knots) to tens of days (maximum length of ``plateau'' phases observed); for AGN, corresponding phenomena would have to occur on timescales of years -- thus deviating strongly from a simple linear black-hole mass scaling relation -- in order to be consistent with our observations.

Notably, the ``flare'' and ``high-frequency'' regimes of the power spectra are consistent with having the same powerlaw index though on different (by factors $\approx$2--3 in spectral power) energy levels. This indicates that both emission regimes, although they appear to be physically disconnected, are caused by intrinsically stochastic processes with similar correlation properties. Remarkably, the three sources in question -- 3C~111, 3C~273, and 3C~454.3 -- are also those for which we find the (temporarily) flattest spectral indices, indicating optically thick emission.

\subsection{Spectral Time Delays}

We tested the presence or absence of time lags between the 1.3-mm and 3-mm fluxes by means of a correlation analysis (cf. Eq.~\ref{eq_dcf}). We can exclude the presence of significant -- referring here to a $3\sigma$ confidence level -- offsets in time for four of our targets, namely 0923+392, 3C~111, 3C~273, and 3C~345 (see Fig.~\ref{fig_timelag}). This statement holds within accuracies (referring to the ranges of timelags consistent with a zero offset within the $3\sigma$ confidence band) of few tens of days. Our results are consistent with those obtained by the time delay analyses of Tornikoski et al. (\cite{tornikoski1994}) and Hovatta et al. (\cite{hovatta2008}) in the cm-radio range. Hovatta et al. (\cite{hovatta2008}) conclude that those limits on time delays, both in the cm- and mm-regimes, can be understood in the frame of the semi-empiric ``generalized shock model'' by Valtaoja et al. (\cite{valtaoja1992}).

For the remaining two sources, 3C~454.3 (after 2005.2; cf. Fig.~\ref{fig_dcf3C454.3}) and 3C~84, we do find indication that the 1.3-mm emission preceeds the 3-mm emission by $\approx55$ days (3C~454.3) and $\approx300$ days (3C~84). We note that the result for 3C~454.3 is consistent with the results by Tornikoski et al. (\cite{tornikoski1994}) and Hovatta et al. (\cite{hovatta2008}) who do not find a time delay in their radio lightcurves; the data of Hovatta et al. (\cite{hovatta2008}) end in 2005.3, just at the beginning of the ``flare phase'' of 3C~454.3.

For 3C~454.3, we suspect variations of the optical depth with frequency as the principal cause for the time delay. Adiabatically expanding plasmas become optically thin at higher energies first, introducing a systematic time delay between two lightcurves obtained at two frequencies with the high-frequency emission leading. Details depend on various parameters, notably the electron energy distribution, the optical depth at a reference frequency, the magnetic field strength, the electron density, and the emission region size (van der Laan \cite{vanderlaan1966}). This mechanism has been incorporated into the shock-in-jet model by Marscher \& Gear (\cite{marscher1985}); proper quantitative treatment takes into account the combination of Compton, synchrotron, and adiabatic cooling (e.g., Fromm et al. \cite{fromm2011}, and references therein). Our interpretation is supported by the fact that the spectral index shows fast variability from inverted to almost optically thin spectral indices in 2005--2007 and remains in the optically thick regime after 2007\footnote{This observation changes only marginally when artificially removing the spectral time delay; see Fig.~\ref{fig_specindex_timeshift}. The most notable difference between the two calculations is the time when inverted indices are observed.}. This is in agreement with the results by Jorstad et al. (\cite{jorstad2010}) who find time delays up to $\approx$215 days between 37~GHz and 230~GHz lightcurves, with the 230~GHz emission leading.

For 3C~84, the answer is less clear. High-resolution radio-interferometric observations let Asada et al. (\cite{asada2006}) conclude that the lightcurve, as observed until 2001, is in agreement with adiabatic cooling of an expanding radio lobe. The rise of the emission around 2005 nicely agrees with observations of the ejection of a new radio jet component in 2005 (Nagai et al. \cite{nagai2010}). Those observations are a priori in agreement with optical depth variations in the context of the shock-in-jet model. However, we note the unusually large time delay of $\approx$300 days -- assuming the result of the DCF analysis is correct. Such a large delay suggests additional travel time effects given by the geometry of the emission region. If the emission is composed of various components emitted from spatially separate zones, according time lags may occur. Given the observed time offsets of $\approx$300 days, this would suggest a spatial extension of the emission region of the order of one light-year.

\section{Conclusions}

We have analyzed the long-term activity of six luminous active galactic nuclei using mm/radio lightcurves spanning $\approx$14 years in time. From our analysis we draw the following principal conclusions:

\begin{enumerate}

\item  All sources show substantial, non-periodic variability by factors $\approx$2--8 on timescales of years. This is in good agreement with the results reported by Hovatta et al. (\cite{hovatta2007,hovatta2008}) and shows that source monitoring times on scales of decades are necessary in order to obtain a realistic overview on the source activity. We also note that at least for the special case of Sagittarius~A* both observational (e.g. Koyama et al. \cite{koyama2008}) and theoretical (e.g. Cuadra et al. \cite{cuadra2008}) evidence suggests activity timescales of several hundred years. An even more extreme example is provided by recent observations of ``Hanny's Voorwerp'' which suggest flux variations of the galactic nucleus of IC~2497 by factors $>$100 on timescales $<$70,000 years (e.g., Schawinski et al. \cite{schawinski2010}). Therefore one should be aware that AGN monitoring studies, like our work, might observe some of their targets at special times even when covering timelines of decades.

\item  The flux density distributions of five out of six sources show clear signatures of bi- or even multimodality. This indicates that the emission processes cannot be treated as uniform stochastic processes. Instead, we have to assume a segregation of ``quiescent'' and (maybe multiple) ``flare'' source states that correspond to different physical regimes. Each state would be characterized by an individual (probably red-noise type) flux distribution.

\item  Our sources show mostly steep ($\alpha\approx0.5-1$) spectral indices that indicate outflow (halos, jets) dominated synchrotron emission. The spectral indices show substantial variability that is not correlated with the source fluxes. From the combined information we conclude that optical depth variations are the dominant mechanism for the spectral variability.

\item  All sources show power spectra that are globally consistent with red noise (i.e. $\beta>0$). We find that the intrinsic (i.e. measurement noise corrected) power-spectral indices are flatter than 1 for five of our sources, placing their emission between the cases of flicker noise ($\beta=1$) and white noise ($\beta=0$). The only exception is 3C~111 whose power spectrum is consistent with being located between flicker noise and random walk noise ($\beta=2$). For three sources we find that the low-frequency ends of their power spectra appear to be upscaled in spectral power by factors $\approx$2--3 with respect to the overall powerlaws, presenting another indication towards a segregation between ``quiescent'' and ``flaring'' source states.

\item  For four of our targets -- 0923+392, 3C~111, 3C~273, and 3C~345 -- we find that the 1.3-mm and 3-mm lightcurves are simultaneous within $3\sigma$ confidence levels. This is in agreement with shock-models like the ``generalized'' model by Valtaoja et al. (\cite{valtaoja1992}); given the measurement accuracies of tens of days however these constraints are fairly loose. For 3C~454.3 (in its ``flare state(s)'' after 2005.2) and 3C~84 we find spectral time delays of $\approx$55 days and $\approx300$ days, respectively, with the 1.3-mm emission leading. The behaviour of 3C~454.3 can be traced back to variations of the optical depth with time and frequency, in agreement with previous work. In case of 3C~84, we suspect a combination of optical depth and emission region geometry effects as cause for the time lag.

\item  We do not see a correlation between the observed source properties -- variability, spectral index, power spectrum -- with physical source parameters -- redshift, black hole mass, kilo-parsec morphology -- as given in Table~\ref{tab_journal}.

\end{enumerate}

The discovery of distinct source emission states challenges the common assumption that AGN emission can be described by uniform stochastic processes. A promising road for further studies is indicated by the cases of microquasars, Sagittarius~A*, and IC~2497 where similar segregations of emission states have been seen. Long-term, multi-wavelength monitoring of AGN appears to be the most important tool towards a deeper understanding of this phenomenon.

\begin{acknowledgements}

We are grateful to the entire IRAM and PdBI staff whose hard work over many years made this study possible. We would like to thank Katie Dodds-Eden (MPE Garching) for pointing out the power of flux distribution analyses. Our work has made use of the NASA/IPAC Extragalactic Database (NED) which is operated by the Jet Propulsion Laboratory, California Institute of Technology, under contract with the National Aeronautics and Space Administration. We have also made use of data from the MOJAVE database that is maintained by the MOJAVE team (Lister et al. \cite{lister2009}). Last but not least, we are grateful to an anonymous referee whose comments helped to improve the quality of this paper.

\end{acknowledgements}

\clearpage

\clearpage

\clearpage

\clearpage

\clearpage

\clearpage

\clearpage

\clearpage

\clearpage

\end{document}